\documentclass[letterpaper,twoside,journal]{IEEEtran}
\usepackage[compatibility=false]{caption}
\usepackage{cite}
\usepackage{amsmath,amssymb,amsfonts}
\usepackage{mathtools}
\usepackage{algpseudocode}
\usepackage{algorithm}
\usepackage{textcomp}
\usepackage{graphicx}
\usepackage{enumitem}
\usepackage{bbm}
\usepackage{cases}
\usepackage[T1]{fontenc}
\usepackage{booktabs}
\usepackage{subcaption}
\usepackage{balance}
\usepackage{acro}
\usepackage[mathscr]{euscript}
\usepackage{colortbl}
\usepackage{physics}
\usepackage{url}
\usepackage{dsfont}
\usepackage{etoolbox}

\definecolor{rowblue}{rgb}{0.90,0.90,1}
\appto\appendix{\counterwithin{equation}{section}}

\newcommand{\mc}[1]{\mathcal{#1}}

\DeclareMathOperator*{\argmax}{arg\,max}

\DeclareAcronym{CPTP}{
  short = CPTP,
  long = completely positive and trace-preserving
}

\DeclareAcronym{CSI}{
  short = CSI,
  long = channel state information
}

\DeclareAcronym{DV}{
  short = DV,
  long = discrete variable
}

\DeclareAcronym{DoF}{
  short = DoF,
  long = degrees of freedom
}

\DeclareAcronym{FSO}{
  short = FSO,
  long = free-space optical
}

\DeclareAcronym{MIMO}{
  short = MIMO,
  long = multiple-input multiple-output
}

\DeclareAcronym{QuMIMO}{
  short = QuMIMO,
  long = quantum multiple-input multiple-output
}

\DeclareAcronym{QEC}{
  short = QEC,
  long = quantum error correction
}

\DeclareAcronym{SDP}{
  short = SDP,
  long = semidefinite program
}

\DeclareAcronym{SISO}{
  short = SISO,
  long = single-input single-output
}

\def\BibTeX{{\rm B\kern-.05em{\sc i\kern-.025em b}\kern-.08em
    T\kern-.1667em\lower.7ex\hbox{E}\kern-.125emX}}

\begin{document}

\title{QuMIMO Diversity over Discrete-Variable Free-Space Optical Channels}
\author{Shehbaz~Tariq, Junaid~ur~Rehman, and Symeon~Chatzinotas,~\IEEEmembership{Fellow, IEEE}%
\thanks{Interdisciplinary Centre for Security, Reliability and Trust (SnT), University of Luxembourg, Luxembourg (e-mails: shehbaz.tariq@uni.lu; symeon.chatzinotas@uni.lu).}%
\thanks{Department of Electrical Engineering, the Center for Advanced Quantum Computing, and the Center for Communication Systems and Sensing, King Fahd University of Petroleum and Minerals (KFUPM), Dhahran 31261, Saudi Arabia (e-mail: junaid.urrehman@kfupm.edu.sa).}%
\thanks{J.~ur~Rehman would like to acknowledge the support from KFUPM through the Ibn Battuta Global Scholarship grant with number ISP2651. The work of Shehbaz~Tariq and Symeon~Chatzinotas was supported by the project LUQCIA funded by the European Union~-- Next Generation EU, with the collaboration of the Department of Media, Connectivity and Digital Policy of the Luxembourgish Government in the framework of the RRF program. }%
\thanks{Corresponding author: Shehbaz~Tariq (email: shehbaz.tariq@uni.lu).}}
%The source code for reproducing the results is available at \url{https://github.com/ShehbazTariq/}.

\markboth
{QuMIMO Diversity over Discrete-Variable Free-Space Optical Channels}
{Tariq \textit{et al.}: QuMIMO Diversity over Discrete-Variable Free-Space Optical Channels}

\maketitle

\begin{abstract}
\Ac{FSO} links carry quantum states without a continuous fiber path, but
diffraction, pointing error, and atmospheric turbulence
couple spatial modes and reduce end-to-end state fidelity. Classical
\ac{MIMO} uses spatial channels to increase rate through multiplexing or
reliability through diversity. For unknown quantum states, however, the
no-cloning theorem prevents classical-style replication. Therefore, to study quantum
transmission over multiple spatial modes, we formulate an \ac{FSO}
\ac{QuMIMO} link for an unknown polarization qubit as a single \ac{CPTP} map.
It combines passive field transfer, bosonic pure loss lifted to Fock space, a
receiver map to erasure-augmented polarization qubits, and effective per-port
polarization noise. The plane-wave Rytov variance parameterizes turbulence,
while an internal-state Gram matrix captures partial photon distinguishability
and its effects on path coherence and multiphoton interference. We use
Haar-averaged state fidelity to compare direct transmission, fixed \ac{QEC},
approximate quantum cloning, coherent path superposition, and channel-adapted
encoder and recovery maps, while distinguishing \ac{CSI} from endpoint
availability. With Full \ac{CSI} on the two-rail channel, asymmetric
cloning and coherent path superposition exceed the fixed \ac{SISO} baseline in average fidelity. Adding rails, together with the photon-number
sectors they admit, enlarges the attained general-\ac{CPTP} gain over the fixed \ac{SISO} baseline, with the
widest margin at moderate turbulence. We find that benchmarks such as fixed
stabilizer encoders perform poorly as turbulence makes rail survival unequal and
produces errors unlike erasures at known code positions. Thus we show that \ac{QuMIMO}
realizes channel-adapted spatial diversity without requiring multiple copies
of the logical qubit.
\end{abstract}

\bstctlcite{ResearchBrain:BSTcontrol}

\section{Introduction}
\label{sec:introduction}

\IEEEPARstart{Q}{uantum} links transmit quantum states, distribute quantum
correlations between remote nodes, and provide optical connections for
distributed quantum information processing and the quantum
Internet~\cite{WEH:18:Sci}. Because
long-distance quantum links rely on optical propagation, attenuation imposes a
fundamental tradeoff between transmission rate and distance in the absence of
repeaters~\cite{PLOB:17:NC,Koudia2024_ACUR6M78}. \Ac{FSO} links support
satellite-to-ground and air-to-ground quantum
communication~\cite{TCFKST:17:NPh,NMRFHFW:13:NPh}. Such links can also employ
multiple apertures and spatial modes~\cite{GA:06:JLT}.
However, diffraction and aperture truncation reduce the collected optical
power, pointing errors displace the beam from the receiver, and atmospheric
turbulence produces random amplitude and phase fluctuations that couple
spatial modes~\cite{GA:06:JLT,ANV:08:AO,VSV:16:PRL}.

To mitigate the resulting degradation in link reliability, classical transmit
diversity sends the same information over multiple propagation paths and
combines the resulting observations at the
receiver~\cite{A:02:IEEE_J_JSAC}. However, this form of diversity cannot be
applied directly to quantum communication, because an unknown quantum state
cannot be copied perfectly~\cite{WZ:82:Nat}. Approximate cloning instead
produces correlated, imperfect clones of the input state. Sending the clones
through different spatial rails\footnote{The spatial mode labels the transmit
rail, while polarization carries the logical qubit; no additional internal
degree of freedom is reserved for source-rail indexing. After spatial mixing,
the receiver is not assumed to know a photon's source rail. Any temporal or
spectral distinguishability is modeled separately by the internal-state Gram
matrix.} gives the receiver multiple opportunities to recover the logical
qubit, at the cost of lower marginal fidelity for each clone~\cite{GM:97:PRL}.
Previous \ac{QuMIMO} studies have investigated
spatial diversity and multiplexing, correlated pure-loss channels,
crosstalk-resilient encoding, and \ac{CSI}-adaptive cloning and recovery
\cite{urRehman2025_39RDN6DM,Rehman2025_HGQ2Z98S,
Koudia2025_JJ3WBISQ,Tariq2026_878BK5NE}. Taken together, these studies do not
provide a common end-to-end \ac{FSO} model for comparing cloning, fixed
stabilizer \ac{QEC}, coherent path superposition, and general \ac{CPTP}
designs.

Accordingly, comparing these \ac{QuMIMO} approaches under a common
\ac{FSO} channel model requires separating the effects that alter the spatial mode,
photon number, and polarization. Spatial crosstalk occurs when diffraction,
pointing displacement, or turbulence transfers part of the optical field from
an intended transmit mode to other receive modes. This transfer is governed by
complex coupling amplitudes rather than by power alone, so it can produce
coherent interference at the
receiver~\cite{ANV:08:AO,BGL:20:JOSA}. Photon loss can produce an erasure when no usable single-photon polarization
qubit reaches the recovery stage~\cite{BDS:97:PRL_2,BMT:08:PRL_2}. A
multiphoton collision occurs when photons launched through different transmit
modes arrive at the same receive port. Their detection statistics depend on
bosonic interference and on the partial distinguishability of the photon
internal states. We therefore compare two receiver models: the squashing
receiver maps unresolved collisions to erasure, whereas the mode-resolving
receiver sorts resolvable internal modes before recovery, allowing additional
collision outcomes to be retained~\cite{HOM:87:PRL_2,T:15:PRA,S:15:PRA}.
Polarization noise is treated separately from the spatial effects above. An
isotropic depolarizing channel acting after spatial collection represents
unresolved polarization transformations together with imperfections in
polarization preparation and analysis. Atmospheric turbulence primarily
perturbs the spatial mode and wavefront phase~\cite{BGL:20:JOSA}.
Polarization can remain largely preserved on space-to-ground paths, although
measurable depolarization can arise depending on geometry and
weather~\cite{Peng2026_3AU6T4M9}.

This work considers one-way optical transmission of an unknown polarization
qubit over multiple spatial rails without preshared entanglement, using
Haar-averaged input-output state fidelity as the primary figure of merit. An
entanglement-based repeater instead establishes entanglement over elementary
links and extends it through heralding and entanglement swapping; some
architectures also use quantum memories and purification before teleporting
the unknown state~\cite{WEH:18:Sci}. Because repeater performance depends on
end-to-end rate, success probability, latency, and intermediate-node resources,
a quantitative comparison with the single-hop transceiver studied in this work
requires a separate resource-matched repeater model.

The contributions of this manuscript are as follows.
\begin{enumerate}[leftmargin=*,label=\arabic*)]
    \item We model an \ac{FSO} \ac{QuMIMO} channel as a \ac{CPTP} map
    that combines a passive complex field-transfer matrix, a Stinespring dilation
    of the associated bosonic pure-loss channel lifted to Fock space, a Gram
    matrix representing the partial distinguishability of photon internal
    states, a receiver map to erasure-augmented polarization qubits, and
    effective per-port polarization noise. We instantiate the model on split-step
    phase-screen realizations indexed by the plane-wave Rytov variance and
    evaluate three power-\ac{CSI} descriptors: mean single-photon survival,
    detector-plane crosstalk, and normalized rail-survival heterogeneity.
    \item We place five communication strategies, namely direct transmission,
    fixed stabilizer \ac{QEC}, general \ac{CPTP} encoder and recovery maps,
    asymmetric cloning, and single-photon coherent path superposition, in a
    common framework with Haar-averaged fidelity as the primary metric. For
    each strategy, we tabulate its design-parameter count, Choi rank, required
    \ac{CSI} granularity, and emitted photon number. We compare the five
    strategies at four and five rails under Full \ac{CSI}.
    \item We distinguish the estimated \ac{CSI} from its availability at
    each terminal, identify the measurements supplying detector-plane
    \ac{CSI} and the polarization and internal-state calibrations, state when
    an independent receiver-loss calibration is required, separate power-only
    from phase-sensitive requirements, and fix the minimal parameterization in
    which propagation and receiver-side loss are identifiable only jointly. We
    measure the fidelity gain of each availability case over a fixed \ac{SISO}
    baseline on a $2\times2$ channel with independently swept
    receive-mode scrambling.
    \item We identify the photon-number sector each strategy occupies and the
    two mechanisms by which the internal-state Gram matrix acts, damping
    single-photon path coherence and weighting the permutation terms of
    multiphoton interference. We quantify the resulting survival gain as a function of
    distinguishability and turbulence for both receiver maps.
    \item We evaluate how the attained Full \ac{CSI} general-\ac{CPTP}
    diversity gain changes as the square channel grows from two to five rails.
    For each realized channel we obtain encoder and recovery maps by alternating
    optimization and compare their attained margin over the fixed \ac{SISO}
    baseline across rail counts and turbulence strengths. The admitted nonvacuum
    photon-number sectors grow with the rail count, so this study scales the
    spatial and photon-number resources together.
\end{enumerate}

\textit{Notation:} In this manuscript, finite counts and dimensions are denoted by uppercase Roman
letters such as $N$, $M$, and $D$; abstract systems are identified explicitly
by labels such as $X$ and $Y$. Lowercase Roman and Greek
letters denote scalars or indices unless their functional dependence is shown.
The range of each index is stated at first use. Fonts separate the three roles a
subscript can play: an italic letter is a running mathematical index, an upright
Roman letter is a descriptive tag such as the success branch $\mathrm{s}$, and a
sans-serif letter is a label drawn from a finite named set, namely the strategy
$\mathsf a$, the availability case $\mathsf v$, and the probe basis $\mathsf b$.
Fidelity is denoted by the uppercase symbol $F$. Matrices and vectors are written in bold
uppercase and bold lowercase fonts, respectively, for example $\pmb{A}$ and
$\pmb{a}$. Finite-dimensional Hilbert spaces are written as $\mc H_X$,
single-photon spaces as $\mathfrak h_X$, and bosonic Fock spaces as
$\mc F_{\mathrm s}(\mathfrak h_X)$. Other calligraphic letters denote sets,
linear field operators, or quantum maps, as specified below. For a Hilbert
space $\mc H_X$, $\mathrm{L}(\mc H_X)$ is its operator space,
$\pmb{I}_{\mc H_X}$ is its identity operator, and
$\operatorname{id}_{\mc H_X}$ is the identity channel. A quantum state is
represented by a density operator $\pmb{\rho}$. The trace is $\Tr[\cdot]$;
the partial trace and partial transpose over system $X$ are $\Tr_X[\cdot]$ and
$(\cdot)^{\mathsf{T}_X}$, respectively.

\section{Related Works}
\label{sec:related-works}

\subsection{Transmitter}
\label{subsec:related-transmitter}

A classical transmitter (Tx) obtains diversity by replicating one symbol
across several spatial paths, as in space-time block
coding~\cite{A:02:IEEE_J_JSAC}; in the quantum case, diversity is instead
realized by the encoding map $\mc{E}$ applied at the Tx, which takes the logical
qubit into a joint state of several transmit rails. The
no-cloning theorem~\cite{WZ:82:Nat} constrains the admissible maps, and existing
constructions differ in the structure they impose: how the logical information
is encoded across the joint transmit state, how many photons are emitted, and
which channel parameters the map is allowed to depend on.

Universal symmetric $1\to M$ cloners produce $M$ correlated approximate clones
of the input state and attain the optimal state-independent marginal
fidelity under the symmetry assumptions of universal
cloning~\cite{GM:97:PRL,Wer:98:PRA}. \ac{QuMIMO} studies accept this
marginal-fidelity cost to provide multiple transmission opportunities across
lossy rails~\cite{urRehman2025_39RDN6DM,Rehman2025_HGQ2Z98S}.

Imbalanced rails motivate asymmetric universal cloning, which trades
fidelity among the clone marginals while preserving covariance with respect
to the input state. Complete positivity restricts the achievable fidelity
vectors, and the general $1\to M$ region is more involved than the
$1\to2$ case~\cite{C:00:JMO,NPR:23:LMP,kay_optimal_2016}. Previous adaptive
\ac{QuMIMO} work optimized clone asymmetry and receiver purification
under controlled-SWAP crosstalk and local depolarization for different
\ac{CSI} locations~\cite{Tariq2026_878BK5NE}. That study found that assigning
a clone to every rail need not maximize end-to-end fidelity, but its
comparison covered only direct transmission and cloning with purification.
It did not include coherent single-photon encoding, fixed stabilizer codes, or
general
\ac{CPTP} encoders under a complex field-transfer model.

Moreover, stabilizer \ac{QEC} embeds the logical qubit isometrically into a code
space and targets a correctable error set rather than high-fidelity clone
marginals~\cite{GBP:97:PR,KO:26:NQI,LMPZ:96:PRL,KL:97:PR}.
Channel-adapted techniques optimize an encoder and recovery for a
specified noise channel and can exploit its error
structure~\cite{FSW:08:PRA,TKL:10:IEEE_J_IT}. Joint design can be approached
through alternating convex subproblems, but the resulting bilinear problem is
not globally convex~\cite{RW:05:PRL_2,KL:09:QIP_2}. Single-photon coherent path
encoding places a photon in a superposition of spatial paths; its
performance can depend on relative phases and the physical channel
implementation~\cite{abbott_communication_2020,rubino_experimental_2021}.

\begin{figure}[t]
    \centering
    \includegraphics[width=\columnwidth]
    {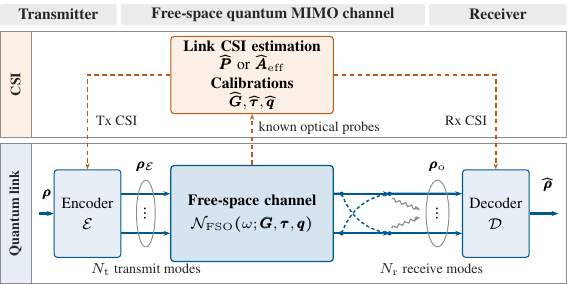}
    \caption{End-to-end model of the $N_{\mathrm{t}}\times N_{\mathrm{r}}$
    \ac{FSO} \ac{QuMIMO} link, from the logical qubit $\pmb\rho$ through the
    encoder $\mc E$, the free-space channel
    $\mc N_{\mathrm{FSO}}(\omega;\pmb G,\pmb\tau,\pmb q)$ and the decoder
    $\mc D$ to the recovered state $\widehat{\pmb\rho}$. Solid, crossing, and
    wavy paths denote direct transfer, spatial-mode coupling, and leakage out
    of the collected subspace. Dashed red links indicate the encoder and
    decoder maps that may depend on the available link estimates and
    calibrations.}
    \label{fig:free-space-qmimo-system}
\end{figure}

\begin{figure}[t]
    \centering
    \includegraphics[width=\columnwidth]
    {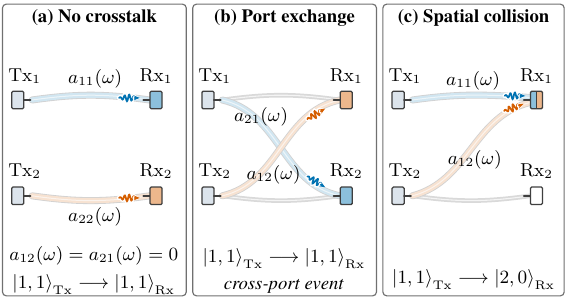}
    \caption{Two-rail spatial outcomes at the detector plane, before the
    receiver map: (a) no crosstalk,
    $\ket{1,1}_{\mathrm{Tx}}\rightarrow\ket{1,1}_{\mathrm{Rx}}$; (b) port
    exchange, with the same output occupation as (a); and (c) a spatial
    collision, $\ket{1,1}_{\mathrm{Tx}}\rightarrow\ket{2,0}_{\mathrm{Rx}}$.
    Colours distinguish launch-rail contributions to the transfer amplitudes
    for illustration; they are not persistent photon labels or receiver
    measurement records. The labelled amplitudes $a_{ji}(\omega)$ are entries
    of the field-transfer matrix $\pmb A(\omega)$.}
    \label{fig:qumimo-crosstalk-distinguishability}
\end{figure}

\subsection{QuMIMO Channels}
\label{subsec:qumimo-channel}

An $N_{\mathrm{t}}\times N_{\mathrm{r}}$ \ac{FSO} \ac{QuMIMO} link couples
$N_{\mathrm{t}}$ transmit spatial modes to $N_{\mathrm{r}}$ receive spatial
modes; an occupied mode carries the
quantum state in an internal degree of freedom such as polarization. We refer
to each labeled spatial mode as a rail. These rails can be used for diversity,
multiplexing, path selection, or
coherent joint transmission and reception. Early multi-aperture studies used
spatial diversity to improve the reliability of free-space quantum key
distribution~\cite{GA:06:JLT}. In turbulent propagation, projection onto
prescribed transmit and receive mode bases produces a complex transfer matrix.
Its off-diagonal elements are coherent mode-coupling amplitudes that include
optical phase~\cite{ANV:08:AO,BGL:20:JOSA}. Physical-layer
treatments distinguish diversity, multiplexing, crosstalk control, and
\ac{CSI}-assisted adaptation~\cite{Koudia2024_ACUR6M78}.

\ac{QuMIMO}
studies analyze the fidelity-rate tradeoff, symmetric cloning under correlated
erasures, and \ac{CSI}-adaptive cloning and recovery
schemes~\cite{urRehman2025_39RDN6DM,Rehman2025_HGQ2Z98S,Tariq2026_878BK5NE}.
Gottesman-Kitaev-Preskill encodings can preserve logical information for
specified passive mode-mixing transformations~\cite{Koudia2025_JJ3WBISQ},
while a composite turbulent-polarization channel has been used to derive
measurement-device-independent quantum key-distribution security
expressions~\cite{Peng2026_3AU6T4M9}.

\subsection{Receiver}
\label{subsec:related-receiver}

Recovery is the corresponding decoding map $\mc{D}$ applied at the receiver
(Rx), taking the joint state of the receive ports back to one logical qubit. A
classical \ac{MIMO} combiner
processes numerical observations from several ports, using rules such as
maximal-ratio combining, zero forcing, or selection
combining~\cite{PGN:04:PoIEEE}. Existing quantum constructions differ in how
much of the received state
they use: whether correlations across ports are retained, whether the channel
is fixed or adapted to a channel estimate, and whether the output is accepted
deterministically or through postselection. Single-port selection is the
simplest choice and discards correlations and signal components at the other
ports. Prior \ac{QuMIMO} diversity studies used selection-based recovery:
an orthogonal erasure measurement identifies a surviving clone without
measuring its logical state; when both clones survive, one is retained and the
other is discarded~\cite{urRehman2025_39RDN6DM,Rehman2025_HGQ2Z98S}. A
fidelity-based ranking of surviving outputs would instead be computed from the
available \ac{CSI} and the assumed channel model; it cannot be obtained by
measuring a single unknown logical state while preserving that state. For a specified encoder and
noise channel, the recovery that maximizes entanglement fidelity can instead
be obtained through an \ac{SDP}~\cite{FSW:07:PR}.

Heralded joint recovery provides another receiver primitive. In standard
purification protocols, a joint operation on several noisy systems trades
success probability against the fidelity of the accepted
output~\cite{CEM:99:PRL,yao_protocols_2025}. These protocols usually assume
identical or independently degraded inputs under a specified noise model.
Asymmetric cloning followed by multiport propagation instead produces a
correlated state with erasure flags and possible multiphoton collisions. A
recovery rule derived for independent copies is not generally optimal for this
state; the recovery must act on the output of the composite channel.

\section{System Model and Problem Formulation}
\label{sec:system-model-problem}

\subsection{FSO QuMIMO Channel}
\label{subsec:free-space-channel}

In this work, we model the \ac{FSO} link as a single \ac{CPTP} map that combines
coherent spatial transfer, photon loss, partial photon distinguishability,
receiver handling of multiphoton arrivals, and polarization noise.
Let $\mc H_{\mathrm{src}}$ denote the admissible source Hilbert space, a finite
subspace of the input Fock space, and let $\mc H_{\mathrm{rx}}$ denote the
receiver Hilbert space of erasure-augmented polarization qubits. For a fixed realization $\omega$, the \ac{FSO} channel is
\begin{equation}
\begin{aligned}
    \mc{N}_{\mathrm{FSO}}(\omega;\pmb G,\pmb\tau,\pmb q)
    &\colon
    \mathrm{L}(\mc H_{\mathrm{src}})
    \longrightarrow
    \mathrm{L}(\mc H_{\mathrm{rx}}),
    \\
    \mc{N}_{\mathrm{FSO}}(\omega;\pmb G,\pmb\tau,\pmb q)
    &=
    \mc{N}^{\mathrm{pol}}_{\pmb{q}}
    \circ
    \mc{S}
    \circ
    \Tr_{\mathrm{E}}
    \circ
    \mc{N}^{\mathrm{bos}}_{\pmb{W},\pmb{G}} .
\end{aligned}
    \label{eq:composite-fso-channel}
\end{equation}
The bosonic map $\mc{N}^{\mathrm{bos}}_{\pmb{W},\pmb{G}}$ is generated by the passive Stinespring isometry $\pmb W$
and the source internal-state Gram matrix $\pmb G$. The environment trace
removes inaccessible loss modes, $\mc S$ maps the optical output to local
erasure-augmented polarization qubits, and $\mc N^{\mathrm{pol}}_{\pmb q}$ applies
effective per-port polarization noise. Its placement after $\mc S$ specifies
the factorization of the effective channel rather than the physical origin of
the polarization impairment. The vector $\pmb\tau$ collects the fixed
post-collection transmissivities of the receive modes and enters $\pmb W$
through the corresponding attenuation. The remainder of this subsection
defines these factors in their order of action.
Figure~\ref{fig:free-space-qmimo-system} shows the end-to-end model together
with the two terminals at which an estimate may inform the design:
the transmitter, where it conditions the encoder, and the receiver, where it
conditions the recovery. The encoder $\mc E$ maps the logical qubit
$\pmb\rho$ onto $N_{\mathrm t}$ transmit rails as $\pmb\rho_{\mc E}$, the
channel carries those rails to $N_{\mathrm r}$ receive rails and produces
$\pmb\rho_{\mathrm o}$, and the decoder returns $\widehat{\pmb\rho}$. \Ac{CSI} estimation uses known optical probes to obtain detector-plane
estimates of the power matrix $\widehat{\pmb P}$ or the complex field-transfer
matrix $\widehat{\pmb A}_{\mathrm{eff}}$ within the assumed structured channel
model. Separate receiver and source calibrations provide
$\widehat{\pmb\tau}$, $\widehat{\pmb q}$, and $\widehat{\pmb G}$, with
$\pmb G$ a source calibration rather than link \ac{CSI}.

\subsubsection{Spatial Field Transfer}
\label{subsubsec:field-transfer}

For each realization $\omega$, classical propagation defines the single-photon
transverse-field propagation operator $\mc K_{\mathrm{prop}}(\omega)$ on
$L^2(\mathbb R^2)$. It includes diffraction, atmospheric turbulence, pointing
error, extinction, and receive aperture truncation. Aperture truncation and
extinction make $\mc K_{\mathrm{prop}}(\omega)$ a contraction; for normalized
inputs, the decrease in squared norm is the probability that the photon is not collected. See Appendix~\ref{app:phase-screens} for the stochastic
paraxial model, split-step construction, and sampling assumptions used
here~\cite{BGL:20:JOSA,KS:23:PRA,C:20:JOSA}. Each realization $\omega$ is one draw of the atmospheric and pointing state.
The plane-wave Rytov variance $\sigma_R^2$ fixes the ensemble from which that
draw is taken, so realizations sharing a common $\sigma_R^2$ still differ in
their instantaneous modal transfer matrix.

With orthonormal transmit modes
$\{\ket{u_i^{\mathrm{tx}}}\}_{i=1}^{N_{\mathrm{t}}}$ and receive modes
$\{\ket{u_j^{\mathrm{rx}}}\}_{j=1}^{N_{\mathrm{r}}}$ of the
$N_{\mathrm{t}}\times N_{\mathrm{r}}$ \ac{FSO} \ac{QuMIMO} channel, we define
the isometries that embed the finite mode spaces into the transverse-field
space by
$\mc{U}_{\mathrm{tx}}\pmb{e}_i=\ket{u_i^{\mathrm{tx}}}$ and
$\mc{U}_{\mathrm{rx}}\pmb{e}_j=\ket{u_j^{\mathrm{rx}}}$, where $\pmb{e}_i$
and $\pmb{e}_j$ are standard coordinate vectors in their respective finite
mode spaces. The complex field-transfer matrix collects the matrix elements of
$\mc K_{\mathrm{prop}}(\omega)$ between the selected transmit and receive
modes,
\begin{equation}
    \begin{aligned}
        \pmb{A}(\omega)
        &\coloneqq
        \mc{U}_{\mathrm{rx}}^{\dagger}
        \mc K_{\mathrm{prop}}(\omega)
        \mc{U}_{\mathrm{tx}}
        \in\mathbb{C}^{N_{\mathrm{r}}\times N_{\mathrm{t}}},
        \\
        a_{ji}(\omega)
        &\coloneqq
        [\pmb{A}(\omega)]_{ji}
        =
        \big\langle u_j^{\mathrm{rx}}
        \big|
        \mc K_{\mathrm{prop}}(\omega)
        \big|
        u_i^{\mathrm{tx}}
        \big\rangle .
    \end{aligned}
    \label{eq:field-transfer}
\end{equation}
The scalar $a_{ji}(\omega)$ is the probability
amplitude for a photon launched in transmit mode $i$ to be collected in receive
mode $j$ before receiver-side attenuation. The operator
$\mc K_{\mathrm{prop}}(\omega)$ acts on the complete transverse-field space,
whereas $\pmb A(\omega)$ retains only the selected transmit-to-receive modal
amplitudes. We adopt the standard scalar paraxial turbulence
model~\cite{BGL:20:JOSA,KS:23:PRA}, under which the propagation effects included
in $\mc K_{\mathrm{prop}}(\omega)$ act identically on the two polarization
components. Accordingly, the same $\pmb A(\omega)$ applies to both components;
polarization changes outside this scalar spatial-transfer model are represented
separately by $\mc N^{\mathrm{pol}}_{\pmb q}$.

Collection into the selected receive modes and spatial crosstalk are quantified
by the modal power-coupling coefficients
\begin{equation}
    \eta_{ji}(\omega)=\left|a_{ji}(\omega)\right|^2 .
    \label{eq:power-transfer}
\end{equation}
For a photon prepared in transmit mode $i$, $\eta_{ji}(\omega)$ is the
probability of collection in receive mode $j$ before receiver-side
attenuation. When the transmit and receive bases are paired, the diagonal
coefficients describe coupling to the corresponding receive modes, while the
off-diagonal coefficients describe spatial crosstalk.

Passivity implies
$\pmb{A}^{\dagger}(\omega)\pmb{A}(\omega)\preceq
\pmb{I}_{N_{\mathrm{t}}}$, so the leakage probability is
\begin{equation}
    p_i^{\mathrm{leak}}(\omega)
    =
    1-\sum_{j=1}^{N_{\mathrm{r}}}\eta_{ji}(\omega)
    \geq 0 .
    \label{eq:spatial-leakage}
\end{equation}
Thus $p_i^{\mathrm{leak}}(\omega)$ is the probability that a photon launched in transmit
rail $i$ reaches no modeled receive mode. It includes explicit extinction,
aperture truncation, coupling outside the selected receive-mode subspace, and
mismatch to the receive basis. It therefore quantifies the
pre-receiver selected-mode leakage of transmit rail $i$. Its
complement, $1-p_i^{\mathrm{leak}}(\omega)=\sum_j\eta_{ji}(\omega)$, is the total fraction
collected in the modeled receive modes. The coefficients
$\{\eta_{ji}(\omega)\}$ specify how this collected power is distributed among
those modes, while the phases of $\pmb{A}(\omega)$ are discarded. Those
relative phases govern coherence between spatial modes and affect multiphoton
interference, so two field realizations with identical power-coupling
coefficients can induce different quantum channels.

\subsubsection{Receiver-Side Loss and Stinespring Dilation}
\label{subsubsec:loss-dilation}

After spatial collection, each receive mode undergoes an effective
receiver-side pure-loss channel. Let
$\tau_j^{\mathrm{rx}}\in[0,1]$ denote its transmissivity for receive mode $j$.
This factor is the same for both polarization components and is fixed
across atmospheric realizations. It represents the downstream insertion,
filtering, coupling, and detection inefficiencies that remain after the modal
selection defining $\pmb{A}$. We collect the receiver transmissivities into
\begin{equation}
    \pmb{\tau}
    =
    \begin{bmatrix}
        \tau_1^{\mathrm{rx}} & \cdots & \tau_{N_{\mathrm{r}}}^{\mathrm{rx}}
    \end{bmatrix}^{\mathsf{T}} .
    \label{eq:receiver-transmissivity}
\end{equation}
Mode $j$ therefore retains a fraction $\tau_j^{\mathrm{rx}}$ of its collected
power and loses the complement $1-\tau_j^{\mathrm{rx}}$ after spatial
collection, which is accounted separately from the selected-mode
leakage $p_i^{\mathrm{leak}}(\omega)$.

We represent each efficiency by a pure-loss beamsplitter immediately before an
ideal detector, and define the receiver attenuation matrix
\begin{equation}
    \pmb{D}_{\tau}
    \coloneqq
    \operatorname{diag}
    \left(
        \sqrt{\tau_1^{\mathrm{rx}}},
        \ldots,
        \sqrt{\tau_{N_{\mathrm{r}}}^{\mathrm{rx}}}
    \right).
    \label{eq:receiver-attenuation-matrix}
\end{equation}
The effective field transfer is then
\begin{equation}
\begin{aligned}
    \pmb{A}_{\mathrm{eff}}(\omega)
    &=
    \pmb{D}_{\tau}\pmb{A}(\omega),
    \\
    a_{\mathrm{eff},ji}(\omega)
    &\coloneqq
    [\pmb{A}_{\mathrm{eff}}(\omega)]_{ji}
    =
    \sqrt{\tau_j^{\mathrm{rx}}}\,a_{ji}(\omega).
    \label{eq:effective-field-transfer}
\end{aligned}
\end{equation}
The corresponding detector-plane power-coupling coefficients and matrix are
\begin{equation}
\begin{aligned}
    p_{ji}(\omega)
    &\coloneqq
    \left|a_{\mathrm{eff},ji}(\omega)\right|^2
    =
    \tau_j^{\mathrm{rx}}\eta_{ji}(\omega),
    \\
    \pmb{P}(\omega)
    &\coloneqq
    \pmb{A}_{\mathrm{eff}}(\omega)
    \odot
    \pmb{A}_{\mathrm{eff}}^{*}(\omega)
    =
    \left[p_{ji}(\omega)\right]
    \in\mathbb{R}_{+}^{N_{\mathrm{r}}\times N_{\mathrm{t}}},
\end{aligned}
    \label{eq:power-csi}
\end{equation}
where $\odot$ denotes the Hadamard product. Thus, receiver-side attenuation
scales the field amplitude in receive mode $j$ by $\sqrt{\tau_j^{\mathrm{rx}}}$ and its
modal power by $\tau_j^{\mathrm{rx}}$. Section~\ref{sec:csi} defines the observations
available under power and field \ac{CSI}. Since $\pmb A(\omega)$ and
$\pmb D_{\tau}$ are contractions,
$\pmb A_{\mathrm{eff}}^{\dagger}(\omega)
\pmb A_{\mathrm{eff}}(\omega)\preceq\pmb I_{N_{\mathrm t}}$.

The contraction $\pmb{A}_{\mathrm{eff}}(\omega)$ describes only the amplitudes
that remain in the modeled receiver modes. To dilate this contraction to an
isometry, we introduce inaccessible environment modes and define
\begin{equation}
\begin{aligned}
    \pmb{L}(\omega)
    &=
    \sqrt{
        \pmb{I}_{N_{\mathrm{t}}}
        -
        \pmb{A}_{\mathrm{eff}}^{\dagger}(\omega)
        \pmb{A}_{\mathrm{eff}}(\omega)
    }
    \succeq
    \pmb{0},
    \\
    \pmb{W}(\omega)
    &=
    \begin{bmatrix}
        \pmb{A}_{\mathrm{eff}}(\omega)\\
        \pmb{L}(\omega)
    \end{bmatrix},
    \qquad
    \pmb{W}^{\dagger}(\omega)\pmb{W}(\omega)
    =
    \pmb{I}_{N_{\mathrm{t}}} .
\end{aligned}
    \label{eq:loss-dilation}
\end{equation}
The upper
block of $\pmb{W}(\omega)$ gives the amplitudes in the modeled receiver modes,
while $\pmb{L}(\omega)$ gives the amplitudes transferred to the environment.
Thus,
$\pmb{L}(\omega)\in\mathbb C^{N_{\mathrm t}\times N_{\mathrm t}}$ and
$\pmb{W}(\omega)\in
\mathbb C^{(N_{\mathrm r}+N_{\mathrm t})\times N_{\mathrm t}}$.
The matrix $\pmb{W}(\omega)$ is an isometry. Its orthonormal columns
can be extended to a unitary scattering transformation on the combined
receiver and environment modes.

Taking the environment inputs to be in vacuum, lifting this passive
transformation to Fock space, and tracing out the environment yields the
bosonic pure-loss channel associated with
$\pmb{A}_{\mathrm{eff}}(\omega)$~\cite{ISS:11:PRA,TRS:18:PRX}. The resulting
reduced map is \ac{CPTP} because it is obtained
from an isometric evolution followed by a partial trace.

The lower block $\pmb{L}(\omega)$ carries the amplitude that leaves
the modeled detector plane. It therefore combines the selected-mode
leakage already contained in $\pmb{A}(\omega)$ with the receiver-side loss
introduced by $\pmb{D}_{\tau}$. The factorization
$\pmb{A}_{\mathrm{eff}}=\pmb{D}_{\tau}\pmb{A}$ separates the two mechanisms
only when $\pmb{\tau}$ is independently calibrated or assumed; an end-to-end
field or power observation determines their product alone.

\subsubsection{Source and Receiver Spaces}
\label{subsubsec:spaces}

The logical Hilbert space is
$\mc H_{\mathrm L}\coloneqq\operatorname{span}\{\ket{0},\ket{1}\}
\cong\mathbb C^2$. The logical qubit is encoded in the polarization of occupied transmit
spatial modes. Each transmit mode contains at most one
photon at the channel input and has the local Hilbert space
\begin{equation}
    \mc H_{\mathrm{tx},i}^{(\leq 1)}
    =
    \operatorname{span}
    \left\{
        \ket{\mathrm{H}},\;
        \ket{\mathrm{V}},\;
        \ket{\mathrm{vac}}
    \right\}
    \cong
    \mathbb{C}^{3},
    \label{eq:transmit-rail-space}
\end{equation}
where $\ket{\mathrm{vac}}$ is the physical zero-photon state and
$\ket{\mathrm{H}}$ and $\ket{\mathrm{V}}$ are the single-photon states with
horizontal and vertical polarization, respectively. In an occupied mode we
identify $\ket{0}\equiv\ket{\mathrm{H}}$ and
$\ket{1}\equiv\ket{\mathrm{V}}$; each unused transmit mode is prepared in
$\ket{\mathrm{vac}}$. The joint transmit space is
$\mc H_{\mathrm{tx}}^{(\leq 1)}=
\bigotimes_{i=1}^{N_{\mathrm{t}}}\mc H_{\mathrm{tx},i}^{(\leq 1)}$.
A communication strategy
selects an admissible encoder-output space
\begin{equation}
    \mc H_{\mathrm{src}}
    =
    \bigoplus_{n\in\mathsf N_{\mathrm{src}}}\mc H_{\mathrm{src}}^{(n)}
    \subseteq
    \mc H_{\mathrm{tx}}^{(\leq 1)},
    \label{eq:source-space}
\end{equation}
where $\mc H_{\mathrm{src}}^{(n)}$ is the sector with total photon number $n$
and $\mathsf N_{\mathrm{src}}\subseteq\{0,\ldots,N_{\mathrm{t}}\}$ contains
the sectors used by the strategy. Under the single-photon-per-input-mode
assumption,
\begin{equation}
    \mc H_{\mathrm{src}}\cong\mathbb{C}^{D_{\mathrm{src}}},
    \qquad
    D_{\mathrm{src}}
    =
    \sum_{n\in\mathsf N_{\mathrm{src}}}
    \binom{N_{\mathrm{t}}}{n}2^{n} .
    \label{eq:source-space-dimension}
\end{equation}
The receiver map sends the local optical state of each accessible spatial mode
$j$ to an erasure-augmented polarization qubit,
\begin{equation}
    \mc H_{\mathrm{rx},j}
    =
    \operatorname{span}
    \left\{
        \ket{\mathrm{H}},\;
        \ket{\mathrm{V}},\;
        \ket{\mathrm{e}}
    \right\}
    \cong
    \mathbb{C}^{3}.
    \label{eq:receiver-erasure-augmented-qubit}
\end{equation}
The joint receiver space is
\begin{equation}
    \mc H_{\mathrm{rx}}
    =
    \bigotimes_{j=1}^{N_{\mathrm{r}}}
    \mc H_{\mathrm{rx},j}
    \cong
    \left(\mathbb{C}^{3}\right)^{\otimes N_{\mathrm{r}}}.
    \label{eq:receiver-space}
\end{equation}
The state
$\ket{\mathrm{e}}$ is orthogonal to the polarization subspace and flags mode
$j$ as carrying no usable single-photon polarization qubit. It is
distinct from $\ket{\mathrm{vac}}$: vacuum is a physical optical state before
the receiver map, whereas $\ket{\mathrm{e}}$ belongs to the receiver output
space. The squashing receiver applies a binary test to the total occupation of
receive port~$j$, summed over its polarization and internal modes: exactly one
photon versus any other occupation. The exactly-one-photon branch is mapped to
the polarization subspace, whereas zero local occupation and multiphoton
occupation are both mapped to $\ket{\mathrm{e}}$. The flag therefore identifies
only whether that port contains a usable single-photon qubit; it does not
distinguish no local arrival from a multiphoton collision.

\subsubsection{Bosonic Propagation and Distinguishability}
\label{subsubsec:multiphoton}

Bosonic propagation acts on the finite direct sum $\mc H_{\mathrm{src}}$ of the
photon-number sectors admitted by $\mathsf N_{\mathrm{src}}$, before the
receiver map. This common domain accommodates encoders with different emitted
photon numbers. Passive dilation preserves total photon number; tracing the
environment produces receiver photon-number mixtures, while multiphoton
interference redistributes photons within a fixed sector.

Let
$\mathfrak{h}_{\mathrm{in}}
\coloneqq\mathbb{C}^{N_{\mathrm{t}}}\otimes\mathbb{C}^{2}$,
$\mathfrak{h}_{\mathrm{rec}}
\coloneqq\mathbb{C}^{N_{\mathrm{r}}}\otimes\mathbb{C}^{2}$, and
$\mathfrak{h}_{\mathrm{E}}
\coloneqq\mathbb{C}^{N_{\mathrm{t}}}\otimes\mathbb{C}^{2}$ denote the
transmitter, collected-receiver, and environment single-photon spaces,
respectively, each carrying a polarization factor. The factor
$\mathbb{C}^{N_{\mathrm{t}}}$ in $\mathfrak{h}_{\mathrm{in}}$ is the existing
transmit spatial-mode degree of freedom, so assigning an input rail does not
consume an auxiliary source-index register. We set
$\mathfrak{h}_{\mathrm{out}}
\coloneqq\mathfrak{h}_{\mathrm{rec}}\oplus\mathfrak{h}_{\mathrm{E}}$.
For any single-photon space $\mathfrak h$, we write
$\mc{F}_{\mathrm{s}}(\mathfrak h)
\coloneqq\bigoplus_{n=0}^{\infty}\operatorname{Sym}^{n}(\mathfrak h)$ for its
bosonic Fock space. See Appendix~\ref{app:fock} for this common description and
the occupation-basis and second-quantization conventions used below. The occupation basis canonically identifies
$\mc H_{\mathrm{src}}^{(n)}$ with a subspace of
$\operatorname{Sym}^{n}(\mathfrak h_{\mathrm{in}})$; throughout, we use the
resulting isometric identification
$\mc H_{\mathrm{src}}
\subset\mc F_{\mathrm s}(\mathfrak h_{\mathrm{in}})$. Because the
spatial transformation is polarization-independent, its single-photon action is
the isometry
$\pmb{W}(\omega)\otimes\pmb{I}_2:
\mathfrak{h}_{\mathrm{in}}\rightarrow\mathfrak{h}_{\mathrm{out}}$.
Its second quantization is denoted by
$\Gamma(\pmb W(\omega)\otimes\pmb I_2)$ and maps
$\mc F_{\mathrm s}(\mathfrak h_{\mathrm{in}})$ to
$\mc F_{\mathrm s}(\mathfrak h_{\mathrm{out}})$ sector by sector. For an
$n$-photon input whose photons are
indistinguishable in their internal degrees of freedom, the second-quantized
transition amplitude is
\begin{equation}
    \left\langle \pmb{m}\middle|
    \Gamma\!\left(\pmb{W}(\omega)\otimes\pmb{I}_2\right)
    \middle|\pmb{n}\right\rangle
    =
    \frac{
        \operatorname{per}
        \left(
            \left[
                \pmb{W}(\omega)\otimes\pmb{I}_2
            \right]
            [\pmb{m}\mid\pmb{n}]
        \right)
    }{
        \sqrt{
            \prod_{i=1}^{2N_{\mathrm{t}}} n_i!\,
            \prod_{k=1}^{2(N_{\mathrm{r}}+N_{\mathrm{t}})} m_k!
        }
    },
    \label{eq:fock-lift}
\end{equation}
where $\pmb{n}\in\mathbb{N}_0^{2N_{\mathrm{t}}}$ and
$\pmb{m}\in\mathbb{N}_0^{2(N_{\mathrm{r}}+N_{\mathrm{t}})}$ are the
input and output occupation-number vectors over the rail-and-polarization
modes. The output vector includes both receiver and environment modes, so
$\pmb{n}$ and $\pmb{m}$ have
the same total photon number. The notation $[\pmb{m}\mid\pmb{n}]$ repeats
column $i$ of $\pmb{W}(\omega)\otimes\pmb{I}_2$ exactly $n_i$ times and row
$k$ exactly $m_k$ times. The factorials normalize repeated occupations, and
$\operatorname{per}(\cdot)$ denotes the matrix permanent~\cite{Scheel:08:APS}.
The blockwise direct sum over $n\in\mathsf N_{\mathrm{src}}$ gives the
propagation on $\mc H_{\mathrm{src}}$
and includes Hong--Ou--Mandel interference~\cite{HOM:87:PRL_2}.

Photons supplied to different rails can occupy different unresolved temporal
or spectral states. Relative emission-time or path-delay offsets,
carrier-frequency detuning, unequal spectral envelopes, and residual spectral
phase or dispersion can reduce their overlap
~\cite{HOM:87:PRL_2,LWHRK:04:PRL}. Their output probabilities are then
weighted sums over interfering many-particle
permutations~\cite{T:15:PRA,S:15:PRA}. For each occupied transmit rail
$i\in\{1,\ldots,N_{\mathrm{t}}\}$, let $\ket{\varphi_i}$ be the normalized
internal state of its photon. For a pure wave packet, let
$\widetilde{\varphi}_i(\omega_{\mathrm{opt}})$ be its normalized
angular-frequency amplitude before the relative delay and let $t_i$ be its
arrival-time offset. With
$\Delta t_{ii'}\coloneqq t_{i'}-t_i$, a differential free-space path length
$\Delta L_{ii'}$ contributes $\Delta t_{ii'}\simeq\Delta L_{ii'}/c$. The
internal-state Gram matrix
$\pmb{G}=[g_{ii'}]_{i,i'=1}^{N_{\mathrm{t}}}
\in\mathbb{C}^{N_{\mathrm{t}}\times N_{\mathrm{t}}}$ is defined by
\begin{equation}
\begin{aligned}
    g_{ii'}
    &=
    \langle\varphi_i|\varphi_{i'}\rangle
    =
    \int_{\mathbb{R}} d\omega_{\mathrm{opt}}\,
    \widetilde{\varphi}_i^*(\omega_{\mathrm{opt}})
    \widetilde{\varphi}_{i'}(\omega_{\mathrm{opt}})
    e^{-i\omega_{\mathrm{opt}}\Delta t_{ii'}},
    \\
    \zeta_{ii'}
    &=
    1-\left|g_{ii'}\right|^2
    \in[0,1].
\end{aligned}
    \label{eq:internal-state-distinguishability}
\end{equation}
For equal transform-limited Gaussian spectra with carrier angular frequencies
$\omega_i^{\mathrm{c}}$ and common rms angular-frequency bandwidth
$\sigma_\omega$ of the spectral intensity, set
$\Delta\omega_{ii'}^{\mathrm{c}}\coloneqq
\omega_{i'}^{\mathrm{c}}-\omega_i^{\mathrm{c}}$. Equation
\eqref{eq:internal-state-distinguishability} then gives
\begin{equation}
    \zeta_{ii'}
    =
    1-
    \exp\!\left[
        -\sigma_\omega^2\Delta t_{ii'}^2
        -\frac{\left(\Delta\omega_{ii'}^{\mathrm{c}}\right)^2}
        {4\sigma_\omega^2}
    \right].
    \label{eq:gaussian-wavepacket-distinguishability}
\end{equation}
The overlap integral also covers unequal spectral envelopes and residual
spectral phase. In this work, these wave-packet properties enter only through
the separately calibrated $\pmb G$; the phase-screen realization $\omega$
changes $\pmb A(\omega)$, not $\pmb G$. The Gram matrix satisfies
$\pmb{G}\succeq\pmb{0}$ and $g_{ii}=1$. We use the equicorrelated
one-parameter model
\begin{equation}
    g_{ii'}(\zeta)
    =
    \begin{cases}
        1, & i=i',\\
        \sqrt{1-\zeta}, & i\neq i',
    \end{cases}
    \qquad \zeta\in[0,1].
    \label{eq:equicorrelated-gram-model}
\end{equation}
This matrix assigns the same real, nonnegative overlap to every pair and is
positive semidefinite for $\zeta\in[0,1]$. It is a symmetric equal-overlap
comparison model rather than a function of $\sigma_R^2$; for
$N_{\mathrm{t}}>2$, its common $\zeta$ need not correspond to a unique set of
pairwise delays or detunings.
For three or more photons, a general Gram matrix can have complex
off-diagonal entries. Multiplying one internal state by a phase, which leaves
the physical state unchanged, changes the phase of an individual overlap
$g_{ii'}$, so only combinations invariant under such a rephasing are
observable. One such combination is the collective phase of a closed product
$g_{12}g_{23}g_{31}$, which can affect multiphoton probabilities. The scalar
$\zeta$ model therefore excludes this closed-loop phase degree of
freedom~\cite{T:15:PRA,S:15:PRA}.

For any Gram matrix, we choose normalized internal states
$\{\ket{\varphi_i}\}$ that realize $\pmb G$, and set
$\mathfrak h_{\mathrm{int}}\coloneqq
\operatorname{span}\{\ket{\varphi_i}\}$. The isometry
$\mc V_{\pmb G}=\bigoplus_{n\in\mathsf N_{\mathrm{src}}}
\mc V_{\pmb G}^{(n)}$ attaches
$\ket{\varphi_i}$ to the photon supplied by each occupied transmit rail $i$
and fixes the vacuum. See Appendix~\ref{app:internal-state-assignment} for its
occupation-basis construction. Tracing the internal states
multiplies the rail coherence
$\ket{i,\sigma}\!\bra{i',\sigma'}$ by
$\langle\varphi_{i'}|\varphi_i\rangle=g_{i'i}$.

We write $n_{\max}\coloneqq\max\mathsf N_{\mathrm{src}}$ and define the truncated
Fock space
$\mc F_{\leq n_{\max}}(\mathfrak h)
\coloneqq\bigoplus_{n=0}^{n_{\max}}\operatorname{Sym}^{n}(\mathfrak h)$.
Let $\Gamma_{\mathrm{ext}}(\omega)\coloneqq
\Gamma(\pmb{W}(\omega)\otimes\pmb{I}_{2}\otimes\pmb{I}_{\mathrm{int}})$ denote
the second-quantized action that leaves the internal states unchanged. The
bosonic propagation map, restricted to the admissible source space, is
$\mc N_{\pmb W,\pmb G}^{\mathrm{bos}}:
\mathrm L(\mc H_{\mathrm{src}})\rightarrow
\mathrm L\!\left(\mc F_{\leq n_{\max}}
(\mathfrak h_{\mathrm{out}}\otimes\mathfrak h_{\mathrm{int}})\right)$ and is given by
\begin{equation}
\begin{aligned}
    \mc{N}_{\pmb{W},\pmb{G}}^{\mathrm{bos}}(\pmb{\rho})
    &\coloneqq
    \Gamma_{\mathrm{ext}}(\omega)
    \mc{V}_{\pmb{G}}\pmb{\rho}\mc{V}_{\pmb{G}}^{\dagger}
    \Gamma_{\mathrm{ext}}^{\dagger}(\omega).
    \label{eq:partial-distinguishability-channel}
\end{aligned}
\end{equation}
The bosonic map is \ac{CPTP} because $\mc{V}_{\pmb{G}}$ and
$\Gamma_{\mathrm{ext}}(\omega)$ are isometries. Any two sets of
internal states that realize the same $\pmb{G}$ are related by an isometry
between their spans and therefore give equivalent maps up to an internal-mode
basis change. When the receiver does not resolve internal modes,
they are discarded, so the physical output depends on the internal states only
through $\pmb{G}$. The mode-resolving receiver instead requires a calibrated
sorter that maps a chosen orthonormal basis of the existing temporal or spectral
mode space to distinct output bins before detection. Multi-output quantum pulse
gates have experimentally demultiplexed five-dimensional single-photon temporal
modes and realized a complete decoder suitable for high-dimensional quantum key
distribution~\cite{SGSREBS:23:PQ}. Our receiver uses this sorting principle but
additionally assumes that the operation preserves the polarization qubit. The
sorter resolves existing internal modes; it does not create a source label or
perfectly identify nonorthogonal states $\ket{\varphi_i}$. If all occupied rail
components have the same internal state, the internal factor is common to every
photon and the spatial
and polarization factors reduce to the coherent bosonic transformation in
\eqref{eq:fock-lift}.

Figure~\ref{fig:qumimo-crosstalk-distinguishability} illustrates two limiting
spatial-transfer patterns. In panel (a), $a_{11}$ and $a_{22}$ couple
corresponding transmit and receive rails, whereas in panel (b), $a_{21}$ and
$a_{12}$ couple crossed rails. The panels have the same detected occupation;
their difference lies in the field-transfer matrix, not in a source-rail record
at the receiver. For identical internal states, the direct and exchange
permutations are indistinguishable probability amplitudes and interfere. An
event-by-event source-rail record would require a source-correlated temporal or
spectral state, which uses an additional physical degree of freedom. If these
states are orthogonal, a polarization-preserving sorter can resolve them without
measuring the logical $\mathrm{H}/\mathrm{V}$ state, but the resulting
which-path information removes the corresponding spatial coherence. For
nonorthogonal states, the reduction is set by their overlaps in $\pmb G$, and
the source rail cannot be identified perfectly~\cite{E:96:PRL}. In panel (c), a
collision means that two photons occupy the same receive rail, which the
receiver map of Section~\ref{subsubsec:receiver-maps} records as an erasure,
whereas a calibrated mode-resolving receiver can separate orthogonal internal
modes when $\zeta=1$.

\subsubsection{Receiver Maps and the Composite Channel}
\label{subsubsec:receiver-maps}

The receiver and environment Fock factors separate canonically. See
Appendix~\ref{app:receiver-map-realizations} for this factorization. After the
environment is traced out, the optical state lies in the finite receiver Fock space
\begin{equation}
    \mc F_{\mathrm{rec}}
    \coloneqq
    \mc F_{\leq n_{\max}}
    (\mathfrak h_{\mathrm{rec}}\otimes\mathfrak h_{\mathrm{int}}).
    \label{eq:receiver-fock-space}
\end{equation}
On the output support generated by $\mc H_{\mathrm{src}}$, this
factorization gives the
environment trace the type
$\Tr_{\mathrm E}:\mathrm L(\mc F_{\leq n_{\max}}
(\mathfrak h_{\mathrm{out}}\otimes\mathfrak h_{\mathrm{int}}))
\rightarrow\mathrm L(\mc F_{\mathrm{rec}})$.
The receiver considered here, referred to below as the squashing receiver,
applies a local coarse-graining map that tests whether each port contains
exactly one photon.
For each orthonormal internal mode $\gamma$, write
$\bra{\mathrm P,\gamma}\coloneqq\bra{\mathrm P}\otimes\bra{\gamma}$, with
$\mathrm P\in\{\mathrm H,\mathrm V\}$, on the local single-photon space at port
$j$. The corresponding Kraus operator is
\begin{equation}
    \pmb{K}^{\mathrm{sq},(1)}_{j,\gamma}
    =
    \ket{\mathrm{H}}\!\bra{\mathrm{H},\gamma}
    +
    \ket{\mathrm{V}}\!\bra{\mathrm{V},\gamma}.
    \label{eq:single-tag-squash-kraus}
\end{equation}
For every local Fock occupation-number vector $\pmb{n}^{\mathrm{loc}}_j$ with total
photon number
$|\pmb{n}^{\mathrm{loc}}_j|\coloneqq
\sum_{\mu}n^{\mathrm{loc}}_{j,\mu}\ne1$, where $\mu$ ranges over the
polarization and internal modes at port $j$, a separate Kraus operator
$\pmb{K}^{\mathrm{sq},(\mathrm{e})}_{j,\pmb{n}^{\mathrm{loc}}_j}
=\ket{\mathrm{e}}\!\bra{\pmb{n}^{\mathrm{loc}}_j}$ maps that occupation to erasure. These
operators satisfy the Kraus completeness relation on the truncated local Fock
space. Thus $\mc{S}=\bigotimes_{j=1}^{N_{\mathrm{r}}}\mc{S}_j$ is a \ac{CPTP}
map from $\mc{F}_{\mathrm{rec}}$ to $\mc H_{\mathrm{rx}}$. It preserves
$\mathrm{H}/\mathrm{V}$ coherence within one internal mode, but removes
coherence between internal modes and between distinct complete occupation
patterns. It also removes coherence between arrival at different ports. The
map models an ideal port-resolving, nondestructive photon-number-selective
receiver interface; coherent spatial processing must act before its which-port
record is created. A detector-only realization would instead require a
measurement-specific squashing model, and an arbitrary optical measurement
need not admit a qubit squashing model~\cite{BMT:08:PRL_2}. The loss of path
coherence is the usual consequence of retaining which-path
information~\cite{E:96:PRL}.

The vector
$\pmb{q}=[q_1,\ldots,q_{N_{\mathrm{r}}}]^{\mathsf{T}}$ contains the
effective per-port depolarization probabilities, with $q_j\in[0,1]$.
We define the local projectors
$\pmb{\Pi}_{\mathrm{pol}}=\ketbra{\mathrm{H}}{\mathrm{H}}+
\ketbra{\mathrm{V}}{\mathrm{V}}$ and
$\pmb{\Pi}_{\mathrm{e}}=\ketbra{\mathrm{e}}{\mathrm{e}}$. Every local state
$\pmb\rho_j$ in the image of $\mc S_j$ is block diagonal with
$\pmb\rho_j^{\mathrm{pol}}\coloneqq
\pmb\Pi_{\mathrm{pol}}\pmb\rho_j\pmb\Pi_{\mathrm{pol}}$ and
$p_{\mathrm e,j}\coloneqq\bra{\mathrm e}\pmb\rho_j\ket{\mathrm e}$. The
per-port polarization channel therefore acts as
\begin{equation}
    \mc{N}^{\mathrm{pol}}_{q_j}(\pmb{\rho}_j)
    =
    (1-q_j)\pmb\rho_j^{\mathrm{pol}}
    +q_j\Tr[\pmb\rho_j^{\mathrm{pol}}]\frac{\pmb I_2}{2}
    +p_{\mathrm e,j}\ketbra{\mathrm e}{\mathrm e}.
    \label{eq:local-polarization-depolarization}
\end{equation}
See Appendix~\ref{app:receiver-map-realizations} for a Kraus realization on all
of $\mathrm L(\mc H_{\mathrm{rx},j})$. Equation
\eqref{eq:local-polarization-depolarization} fixes the erasure flag. The
product map is
$\mc{N}^{\mathrm{pol}}_{\pmb{q}}
=\bigotimes_{j=1}^{N_{\mathrm{r}}}\mc{N}^{\mathrm{pol}}_{q_j}$. The
parameter $q_j$ quantifies effective isotropic depolarization at receive port
$j$ after spatial collection. It represents unresolved polarization
transformations induced by the atmospheric path or terminal optics, together
with equivalent imperfections in polarization preparation and analysis over
the observation interval, and is distinct from the receiver transmissivity
$\tau_j^{\mathrm{rx}}$.

For an $N_{\mathrm t}\times N_{\mathrm r}$ \ac{QuMIMO} channel, we separate
the atmospheric transfer from controlled unitary mixing of the receive modes.
Let $c_{\mathrm{rx}}\in[0,1]$, and assign each realization
$\omega$ a target unitary
$\pmb{U}_{\omega}\sim\operatorname{Haar}(\mathrm U(N_{\mathrm r}))$, independent
of $c_{\mathrm{rx}}$. Let $\pmb{H}_{\omega}$ be its principal Hermitian generator,
whose eigenvalues lie in $(-\pi,\pi]$, such that
$\pmb{U}_{\omega}=\exp(-\mathrm{i}\pmb{H}_{\omega})$, and define
\begin{equation}
    \pmb{U}_{\mathrm{scr}}(c_{\mathrm{rx}},\omega)
    \coloneqq
    \exp\!\left(-\mathrm{i}c_{\mathrm{rx}}\pmb{H}_{\omega}\right).
    \label{eq:controlled-receive-mode-scrambling}
\end{equation}
Hence,
$\pmb{U}_{\mathrm{scr}}(0,\omega)=\pmb I_{N_{\mathrm r}}$,
$\pmb{U}_{\mathrm{scr}}(1,\omega)=\pmb{U}_{\omega}$, and
$\pmb{U}_{\mathrm{scr}}^{\dagger}\pmb{U}_{\mathrm{scr}}
=\pmb I_{N_{\mathrm r}}$ for every $c_{\mathrm{rx}}$. The corresponding
field-transfer matrices are
\begin{equation}
\begin{aligned}
    \pmb{A}^{(c_{\mathrm{rx}})}(\omega)
    &\coloneqq
    \pmb{U}_{\mathrm{scr}}(c_{\mathrm{rx}},\omega)\pmb{A}(\omega),
    \\
    \pmb{A}_{\mathrm{eff}}^{(c_{\mathrm{rx}})}(\omega)
    &\coloneqq
    \pmb{D}_{\tau}\pmb{A}^{(c_{\mathrm{rx}})}(\omega).
    \label{eq:controlled-effective-field-transfer}
\end{aligned}
\end{equation}
Thus $c_{\mathrm{rx}}=0$ leaves the physical atmospheric realization
unchanged, whereas $c_{\mathrm{rx}}=1$ applies its assigned target unitary.
Because the mixing is unitary,
$\pmb{A}^{(c_{\mathrm{rx}})}(\omega)$ and $\pmb{A}(\omega)$ have identical
singular values. Thus $c_{\mathrm{rx}}$ changes the receive-mode basis while
preserving the squared singular values.

Let $\pmb{W}^{(c_{\mathrm{rx}})}(\omega)$ denote the passive Stinespring dilation
obtained from \eqref{eq:loss-dilation} using
$\pmb{A}_{\mathrm{eff}}^{(c_{\mathrm{rx}})}(\omega)$. The controlled composite
channel is
\begin{equation}
    \mc{N}_{\mathrm{FSO}}^{(c_{\mathrm{rx}})}
    (\omega;\pmb G,\pmb\tau,\pmb q)
    \coloneqq
    \mc{N}^{\mathrm{pol}}_{\pmb q}
    \circ\mc S
    \circ\Tr_{\mathrm E}
    \circ\mc{N}^{\mathrm{bos}}_{\pmb{W}^{(c_{\mathrm{rx}})}(\omega),\pmb G}.
    \label{eq:controlled-composite-fso-channel}
\end{equation}
The scalar $c_{\mathrm{rx}}$ is a model parameter controlling receive-mode
mixing, independent of \ac{CSI}. At fixed $c_{\mathrm{rx}}$, the \ac{CSI}-availability case determines
whether the encoder and recovery may depend on the realization $\omega$.

\subsection{System Model \& Communication Strategies}
\label{subsec:system-model-strategies}

The Tx transmits an unknown pure logical qubit
$\pmb{\rho}=\ketbra{\psi}{\psi}\in\mathrm{L}(\mc H_{\mathrm L})$ to the Rx
through the
\ac{FSO} channel in
\eqref{eq:composite-fso-channel}. The encoder
\begin{equation}
    \mc{E}:
    \mathrm{L}(\mc H_{\mathrm L})
    \longrightarrow
    \mathrm{L}(\mc H_{\mathrm{src}}),
    \qquad
    \mc H_{\mathrm{src}}\cong\mathbb{C}^{D_{\mathrm{src}}},
    \label{eq:generic-encoder}
\end{equation}
prepares a joint optical state in the admissible source space
\eqref{eq:source-space}. A rail to which the encoder assigns no photon is
prepared in
$\pmb{\rho}_{\mathrm{vac}}\coloneqq\ketbra{\mathrm{vac}}{\mathrm{vac}}$.
After propagation, a decoder
$\mc{D}:\mathrm{L}(\mc H_{\mathrm{rx}})
\rightarrow\mathrm{L}(\mc H_{\mathrm L})$ maps the receiver space in
\eqref{eq:receiver-space} back to the logical qubit. The resulting logical
channel and the states along the link are
\begin{equation}
\begin{aligned}
    \Lambda_{\mc{E},\mc{D},\omega}
    &=
    \mc{D}
    \circ
    \mc{N}_{\mathrm{FSO}}(\omega;\pmb G,\pmb\tau,\pmb q)
    \circ
    \mc{E},
    \\
    \pmb{\rho}_{\mc{E}}
    &=
    \mc{E}(\pmb{\rho}),
    \\
    \pmb{\rho}_{\mathrm{o}}
    &=
    \mc{N}_{\mathrm{FSO}}(\omega;\pmb G,\pmb\tau,\pmb q)
    (\pmb{\rho}_{\mc{E}}),
    \\
    \widehat{\pmb{\rho}}
    &=
    \mc{D}(\pmb{\rho}_{\mathrm{o}})
    =
    \Lambda_{\mc{E},\mc{D},\omega}(\pmb{\rho}).
\end{aligned}
    \label{eq:end-to-end-logical-map}
\end{equation}
Unless a heralded instrument is stated explicitly, both $\mc{E}$ and $\mc{D}$
are deterministic quantum channels (\ac{CPTP} maps). For a heralded
recovery, each instrument branch is completely positive and trace
nonincreasing, and the sum of the success and failure branches is \ac{CPTP}.
The unconditional performance measure includes both outcomes.

For the mode-resolving receiver, a calibrated temporal or spectral mode sorter
feeds the deterministic \ac{CPTP} mode-resolved decoder
$\mc{D}_{\mathrm{mr},\pmb{q}}^{(\varpi)}:\mathrm{L}(\mc{F}_{\mathrm{rec}})
\rightarrow\mathrm{L}(\mc H_{\mathrm L})$, where $\varpi$ fixes the
spatial-port priority. The polarization-preserving sorter resolves a
calibrated orthonormal basis of internal modes into separate bins. The decoder
scans the ports in the order $\varpi$ and the internal modes in their fixed
order. It selects the first bin containing exactly one photon, preserves its
$\mathrm{H}/\mathrm{V}$ coherence, applies the selected port's depolarization
parameter $q_j$, and traces all other bins. If no such bin exists, it outputs
$\pmb I_2/2$. This decoder acts on the resolved receiver Fock state after the
environment trace, before $\mc S$ and $\mc N^{\mathrm{pol}}_{\pmb q}$. The
composite channel retains both maps for receiver-space comparisons. When
$\zeta=0$, a multiphoton collision remains in one
unresolved internal bin; when $\zeta=1$, orthogonal internal states can be
separated. For $0<\zeta<1$, the result depends on the complete Gram matrix and
the calibrated sorter basis~\cite{T:15:PRA,S:15:PRA}.

We consider five communication strategies: direct transmission, fixed
\ac{QEC} encoders with recovery, general \ac{CPTP} encoder and recovery maps
obtained by alternating optimization, asymmetric cloning with recovery, and
single-photon coherent path superposition. Each strategy restricts the admissible
form of $\mc{E}$, $\mc{D}$, or both under the same physical channel
and logical input-output interface. A strategy is included only when the
specified source space contains its required rail count and photon-number
sectors. The objective and the available \ac{CSI} are defined next.
\subsection{Optimization Formalism}
\label{subsec:optimization-formalism}

For fixed $\mc{E}$, $\mc{D}$, and $\omega$, let
$\Lambda\coloneqq\Lambda_{\mc{E},\mc{D},\omega}$ denote the end-to-end logical
channel in \eqref{eq:end-to-end-logical-map}. All fidelities below refer to
this composite end-to-end map, whose output is $\widehat{\pmb{\rho}}$. Its
input and output spaces are
both $\mc H_{\mathrm L}\cong\mathbb{C}^{D}$, where $D=2$.
We write $\pmb\rho_{\mathrm{mix}}\coloneqq\pmb I_{\mc H_{\mathrm L}}/D$ for the
maximally mixed logical state.

Fidelities use the input-first, unnormalized Choi convention. See
Appendix~\ref{app:choi} for its definition. Let
$\pmb{\Phi}_D\coloneqq\ketbra{\Phi^{+}}{\Phi^{+}}$ denote the maximally
entangled state of the logical system with a reference
$\mc H_{\mathrm{L}_{\mathrm{ref}}}\cong\mc H_{\mathrm L}$, which for $D=2$ is a
Bell state, and let $\pmb{J}^{\Lambda}$ denote the Choi operator of
$\Lambda$, so that $\Tr\pmb{J}^{\Lambda}=D$ when $\Lambda$ is trace
preserving.

We compute the Haar-averaged input-output state fidelity through its exact
relation to entanglement fidelity. This relation avoids explicit integration
over the input ensemble: $F_{\mathrm{e}}$ is obtained from one joint
reference-output state and is linear in $\pmb{J}^{\Lambda}$, so the encoder and
decoder subproblems introduced below remain semidefinite programs.
For the maximally mixed logical input, the entanglement fidelity of $\Lambda$
relative to ideal logical transmission is
\begin{equation}
    F_{\mathrm{e}}(\Lambda)
    =
    \bra{\Phi^{+}}
    \left(\operatorname{id}_{\mc H_{\mathrm{L}_{\mathrm{ref}}}}
    \otimes\Lambda\right)
    \left(\pmb{\Phi}_D\right)
    \ket{\Phi^{+}}
    =
    \frac{
        \bra{\Phi^{+}}\pmb{J}^{\Lambda}\ket{\Phi^{+}}
    }{D}.
    \label{eq:entanglement-fidelity}
\end{equation}
For a pure logical input $\pmb{\psi}\coloneqq\ketbra{\psi}{\psi}$, the
state fidelity after transmission is $\Tr[\pmb{\psi}\Lambda(\pmb{\psi})]$.
Averaging over the normalized Haar measure and using the second-moment
identity of that measure gives
\begin{equation}
    F_{\mathrm{av}}(\Lambda)
    =
    \frac{DF_{\mathrm{e}}(\Lambda)+1}{D+1}
    =
    \frac{2F_{\mathrm{e}}(\Lambda)+1}{3},
    \qquad D=2 ,
    \label{eq:average-entanglement-fidelity-relation}
\end{equation}
which holds for every quantum channel (\ac{CPTP} map)~\cite{N:02:PL}.
See Appendix~\ref{app:haar-average} for derivation. We use
$F_{\mathrm{av}}$ as the primary figure of merit because it measures the mean
fidelity of an unknown logical qubit and is affine in the logical Choi
operator.

The five strategies solve one design problem under different restrictions. Let
$\widehat{\mc{I}}_{\mathrm{t}}$ and
$\widehat{\mc{I}}_{\mathrm{r}}$ denote the channel descriptions available to
the Tx and the Rx, as specified in
Section~\ref{subsec:csi-availability}, and let $\mathfrak{S}$ be the set of
encoder-recovery pairs admitted by a chosen strategy. The common objective is
to maximize the Haar-averaged state fidelity under the strategy constraints.
The \ac{CSI}-dependent encoder and recovery induce the logical channel
\begin{equation}
\begin{aligned}
    \Lambda(\omega,\widehat{\mc I}_{\mathrm t},\widehat{\mc I}_{\mathrm r})
    &\coloneqq
    \mc D(\widehat{\mc I}_{\mathrm r})
    \circ
    \mc N_{\mathrm{FSO}}(\omega;\pmb G,\pmb\tau,\pmb q)
    \\
    &\quad\circ
    \mc E(\widehat{\mc I}_{\mathrm t}).
    \label{eq:csi-adaptive-logical-channel}
\end{aligned}
\end{equation}
The design problem is
\begin{subequations}
\label{eq:design-problem}
\begin{alignat}{2}
    &\max_{\mc{E},\mc{D}}
    &\quad&
    \mathbb{E}_{\omega}
    \mathbb{E}_{\widehat{\mc I}_{\mathrm t},
    \widehat{\mc I}_{\mathrm r}\mid\omega}
    \left[
        F_{\mathrm{av}}
        \left(
            \Lambda(\omega,\widehat{\mc I}_{\mathrm t},
            \widehat{\mc I}_{\mathrm r})
        \right)
    \right]
    \label{eq:design-objective}
    \\
    &\operatorname{subject\ to}
    &&
    (\mc{E},\mc{D})\in\mathfrak{S},
    \label{eq:design-strategy}
    \\
    &&&
    \mc{E}:\mathrm{L}(\mc H_{\mathrm L})
    \rightarrow\mathrm{L}(\mc H_{\mathrm{src}}),
    \label{eq:design-encoder}
    \\
    &&&
    \mc{D}:\mathrm{L}(\mc H_{\mathrm{rx}})
    \rightarrow\mathrm{L}(\mc H_{\mathrm L}),
    \label{eq:design-decoder}
    \\
    &&&
    \mc H_{\mathrm{src}}
    =\bigoplus_{n\in\mathsf N_{\mathrm{src}}}\mc H_{\mathrm{src}}^{(n)}
    \subseteq\mc H_{\mathrm{tx}}^{(\leq 1)},
    \label{eq:design-source}
    \\
    &&&
    \mc{E}=\mc{E}(\widehat{\mc{I}}_{\mathrm{t}}),
    \quad
    \mc{D}=\mc{D}(\widehat{\mc{I}}_{\mathrm{r}}).
    \label{eq:design-csi}
\end{alignat}
\end{subequations}
The inner expectation in \eqref{eq:design-objective} includes finite-shot
estimator randomness conditional on the channel realization.
Equation~\eqref{eq:design-strategy} restricts each strategy to its code and
placement, cloning support and coefficients, selected rails and path weights,
or direct rail--port pair. For the general design, $\mathfrak S$ contains all
\ac{CPTP} encoder-recovery pairs on these spaces and hence every resource-matched
structured strategy admitted by $\mc H_{\mathrm{src}}$. Its optimum therefore
upper-bounds the corresponding structured optima.
Equation~\eqref{eq:design-source} is the physical resource constraint because
$\mathsf N_{\mathrm{src}}$ fixes the allowed photon-number sectors. A
fixed-sector strategy $\mathsf N_{\mathrm{src}}=\{n\}$ emits exactly $n$
photons. If $\mathsf N_{\mathrm{src}}$ contains several
sectors, the comparison must instead state the mean emitted photon number and
whether coherences between sectors are permitted. In the absence of an
explicit optical phase reference, the encoder output is dephased between
distinct total-photon-number sectors~\cite{MS:14:PRA}.
Equation~\eqref{eq:design-csi} is the information constraint: each map may
depend on $\omega$ only through the description available at the endpoint that
applies it. Thus a quantity absent from $\widehat{\mc I}_{\mathrm t}$ cannot
enter $\mc E$. Allowing both endpoints to depend on the complete realized
channel gives an upper benchmark for all four \ac{CSI}-availability cases.

Two consequences of this formulation are used throughout. First, the five
strategies are constrained subsets of the same problem rather than distinct
formulations. When their rail, photon-number, and \ac{CSI} resources are matched,
their comparison measures the cost of a structural restriction rather than a
difference in objective. Second, for fixed $\mc{E}$ and $\mc{D}$ the average
fidelity is affine in the channel, and therefore
$F_{\mathrm{av}}\!\left(
\mc D\circ
\mathbb E_{\omega}[\mc N_{\mathrm{FSO}}
(\omega;\pmb G,\pmb\tau,\pmb q)]\circ\mc E\right)
=\mathbb{E}_{\omega}\!
[F_{\mathrm{av}}(\mc{D}\circ
\mc{N}_{\mathrm{FSO}}(\omega;\pmb G,\pmb\tau,\pmb q)\circ\mc{E})]$.
The distinction is adaptation: optimizing one fixed pair for the ensemble is
generally different from averaging the fidelity of realization-adapted pairs.

A heralded decoder is represented by a quantum instrument. Each branch is
completely positive and trace nonincreasing, and the sum of all branches is
\ac{CPTP}. Its Haar-averaged success probability
$\overline{p}_{\mathrm{s}}$, the success-conditioned fidelity
$F_{\mathrm{av}}^{(\mathrm{s})}$, the success-weighted fidelity
$\overline f_{\mathrm{s}}=\overline p_{\mathrm{s}}
F_{\mathrm{av}}^{(\mathrm{s})}$, and the unconditional average fidelity are
defined there. See Appendix~\ref{app:heralded} for these definitions. We evaluate
a heralded design through $\overline f_{\mathrm{s}}$ together with its success probability. Direct
comparisons with deterministic recovery use the unconditional average fidelity,
which assigns a fixed logical state to rejected trials.

\subsubsection{Direct Transmission}
\label{subsubsec:single-channel-selection}

Direct transmission sends the logical qubit through one spatial path. The fixed
\ac{SISO} baseline uses a prescribed pair
$(i_0,j_0)\in\{1,\ldots,N_{\mathrm t}\}\times
\{1,\ldots,N_{\mathrm r}\}$ for every evaluation realization. For transmit
rail $i$, the Tx uses the isometric encoder
\begin{equation}
    \mc{E}_{i}^{\mathrm{dir}}(\pmb{\rho})
    =
    \pmb{\rho}_{\mathrm{vac}}^{\otimes(i-1)}
    \otimes
    \pmb{\rho}
    \otimes
    \pmb{\rho}_{\mathrm{vac}}^{\otimes(N_{\mathrm{t}}-i)} .
    \label{eq:direct-rail-encoder}
\end{equation}
The Rx decoder $\mc{D}_{j}^{\mathrm{dir}}$ selects one receive port $j$, traces
out the remaining ports, preserves the $\mathrm{H}/\mathrm{V}$ block, and maps
the local erasure flag to $\pmb{\rho}_{\mathrm{mix}}$. The baseline applies
$\mc E_{i_0}^{\mathrm{dir}}$ and $\mc D_{j_0}^{\mathrm{dir}}$ throughout; its
pair is independent of $\omega$ and instantaneous \ac{CSI}. We evaluate joint
multiport recovery as another decoder class and realization-adaptive rail or
port selection as separate strategies.

\subsubsection{Quantum Error Correction}
\label{subsubsec:quantum-error-correction}

A \ac{QEC} encoder is an isometry
$\pmb{V}_{\mathrm{QEC}}:\mc H_{\mathrm L}\rightarrow
\mc H_{\mathrm{code}}\subseteq\mc H_{\mathrm{src}}$ followed by a
recovery channel at the Rx.
For an error set $\{\pmb{E}_{a}\}$ and code projector
$\pmb{P}_{\mathrm{code}}$, exact correction is possible when the
Knill-Laflamme conditions
\begin{equation}
    \pmb{P}_{\mathrm{code}}
    \pmb{E}_{a}^{\dagger}\pmb{E}_{b}
    \pmb{P}_{\mathrm{code}}
    =
    c_{ab}\pmb{P}_{\mathrm{code}} .
    \label{eq:knill-laflamme}
\end{equation}
hold~\cite{KL:97:PR}. The orthogonal erasure flag in
\eqref{eq:receiver-erasure-augmented-qubit} allows the Rx to identify a receive
rail whose carrier is lost. A code with minimum distance $d_{\min}$ can therefore correct up to
$d_{\min}-1$ erasures when their code positions are known, whereas it corrects
$\lfloor(d_{\min}-1)/2\rfloor$ arbitrary errors~\cite{GBP:97:PR,KO:26:NQI}.
This bound assumes distinct erased code positions. With spatial crosstalk,
off-diagonal coupling can combine carriers from different code positions in
one receive port. The receiver then identifies erasures by receive port rather
than by the original code positions. The distance bound therefore applies
before spatial mixing and does not guarantee finite-crosstalk performance.
Where Rx \ac{CSI} is available, the recovery is optimized for the encoded
channel rather than inferred from code distance.

We use two fixed stabilizer encoders as benchmarks, written
$\left[\!\left[n_{\mathrm{c}},k,d_{\min}\right]\!\right]$ for a code of
length $n_{\mathrm{c}}$ carrying $k$ logical qubits with minimum distance
$d_{\min}$. Four physical qubits are necessary and sufficient to encode one
logical qubit while correcting one erasure at a known position; the
$\left[\!\left[4,1,2\right]\!\right]$ encoder used here is a
one-logical-qubit subspace of the
$\left[\!\left[4,2,2\right]\!\right]$ code~\cite{GBP:97:PR,KO:26:NQI}. The
$\left[\!\left[5,1,3\right]\!\right]$ perfect code has the minimum block
length for correcting an arbitrary single-qubit error and also corrects any
two erasures at known positions~\cite{LMPZ:96:PRL,KL:97:PR}. These are
therefore the shortest block codes for one- and two-erasure correction. More
generally, the quantum Singleton bound for an
$\left[\!\left[n_{\mathrm{c}},k,d_{\min}\right]\!\right]$ code is
$n_{\mathrm{c}}-k\geq2(d_{\min}-1)$~\cite{CC:97:PRA}. For $k=1$, a guarantee
of three correctable erasures requires $d_{\min}\geq4$ and hence
$n_{\mathrm{c}}\geq7$. Thus no six-qubit code provides a higher guaranteed
erasure order than the $\left[\!\left[5,1,3\right]\!\right]$ code; the next
distinct distance class requires at least seven occupied rails and seven
transmitted photons. Each encoder is a fixed isometry with Choi rank one, and
its construction
introduces no encoder optimization variables. A code is evaluated only when the
source space contains its required number of occupied rails. A placement
$\pi=(\pi_1,\ldots,\pi_{n_{\mathrm{c}}})$ is an ordered tuple of distinct
transmit rails, $\pi_r\in\{1,\ldots,N_{\mathrm{t}}\}$, that carries code
position $r$ on rail $\pi_r$. Write $\mc{C}_{c}$ for the channel of code $c$
and $\mc{P}_{\pi}(\cdot)\coloneqq\pmb{V}_{\pi}(\cdot)\pmb{V}_{\pi}^{\dagger}$
for the placement channel of the isometry
\begin{equation}
\begin{aligned}
    \pmb{V}_{\pi}
    \ket{x_1\cdots x_{n_{\mathrm{c}}}}
    &\coloneqq
    \bigotimes_{i=1}^{N_{\mathrm{t}}}
    \ket{\xi_i},
    \\
    \ket{\xi_{\pi_r}}\coloneqq\ket{x_r},
    \qquad
    &\ket{\xi_i}\coloneqq\ket{\mathrm{vac}}
    \quad (i\notin\pi),
\end{aligned}
    \label{eq:qec-placement-isometry}
\end{equation}
written on the polarization basis $\ket{x_1\cdots x_{n_{\mathrm{c}}}}$ of the
code positions, so that every rail outside $\pi$ is prepared in
$\pmb{\rho}_{\mathrm{vac}}$. The placed encoder is
\begin{equation}
    \mc{E}_{c,\pi}^{\mathrm{QEC}}
    \coloneqq
    \mc{P}_{\pi}\circ\mc{C}_{c},
    \qquad
    \pi\in\mc{Q}_{c} .
    \label{eq:qec-encoder-placement}
\end{equation}
The admissible set $\mc{Q}_{c}$ contains the increasing tuples, one for each
rail subset of size $n_{\mathrm{c}}$, so
$\lvert\mc{Q}_{c}\rvert=\binom{N_{\mathrm{t}}}{n_{\mathrm{c}}}$, which at
$N_{\mathrm{t}}=5$ gives five placements for
$\left[\!\left[4,1,2\right]\!\right]$ and one for
$\left[\!\left[5,1,3\right]\!\right]$. Reordering a tuple permutes the physical
qubits of the code against a channel whose rails carry unequal survival and
crosstalk, and such reorderings lie outside $\mc{Q}_{c}$. The Rx recovery
$\mc{D}_{c}^{\mathrm{QEC}}$ is optimized for the encoded channel.

\subsubsection{Iterative Optimization of CPTP Maps}
\label{subsubsec:iterative-cptp-optimization}

In this subsubsection, $\mc N_{\mathrm{FSO}}$ denotes the fixed channel used in the optimization
and its arguments are suppressed, and $(\mc{E}^{\mathrm{CPTP}},
\mc{D}^{\mathrm{CPTP}})$ denotes a pair restricted only by complete positivity
and trace preservation. For a fixed encoder and channel,
$F_{\mathrm{av}}$ is affine in the decoder
Choi operator. Write
$\langle\pmb X,\pmb Y\rangle
\coloneqq\Tr(\pmb X^{\dagger}\pmb Y)$ for the Hilbert--Schmidt inner product.
For the fixed channel and encoder, we define the Hermitian coefficient operator
$\pmb C_{\mc D}$ through the Hilbert--Schmidt inner product with the Choi
operator $\pmb J^{\mc D}$,
\begin{equation}
    \left\langle\pmb C_{\mc D},\pmb J^{\mc D}\right\rangle
    \coloneqq
    F_{\mathrm{av}}
    (\mc D\circ\mc N_{\mathrm{FSO}}\circ\mc E),
    \label{eq:decoder-objective-operator}
\end{equation}
for the decoder represented by $\pmb J^{\mc D}$. The globally optimal
deterministic decoder within the specified
\ac{CPTP} class is therefore obtained from
\begin{equation}
\begin{aligned}
    \underset{\pmb{J}^{\mc{D}}}{\operatorname{maximize}}
    \quad&
    \left\langle
        \pmb{C}_{\mc{D}},
        \pmb{J}^{\mc{D}}
    \right\rangle                                                     \\
    \operatorname{subject\ to}
    \quad&
    \pmb{J}^{\mc{D}}\succeq\pmb{0},
    \qquad
    \Tr_{\mc H_{\mathrm L}}\pmb{J}^{\mc{D}}
    =
    \pmb{I}_{\mc H_{\mathrm{rx}}}.
    \label{eq:decoder-sdp}
\end{aligned}
\end{equation}
The operator $\pmb C_{\mc D}$ is obtained by contracting the joint
reference-output state of $\mc N_{\mathrm{FSO}}\circ\mc E$ with the
maximally entangled projector in \eqref{eq:entanglement-fidelity}. For a
heralded success branch, trace preservation is
replaced by
$\Tr_{\mc H_{\mathrm L}}\pmb{J}^{\mc{D}_{\mathrm{s}}}
\preceq\pmb{I}_{\mc H_{\mathrm{rx}}}$; a desired value of
$\overline{p}_{\mathrm{s}}$ can be imposed through an additional linear
constraint. This is the standard channel-adapted recovery
\ac{SDP}~\cite{RW:05:PRL_2,FSW:07:PR}.

For a fixed decoder, we define the Hermitian coefficient operator
$\pmb C_{\mc E}$ analogously by
\begin{equation}
    \left\langle\pmb C_{\mc E},\pmb J^{\mc E}\right\rangle
    \coloneqq
    F_{\mathrm{av}}
    (\mc D\circ\mc N_{\mathrm{FSO}}\circ\mc E),
    \label{eq:encoder-objective-operator}
\end{equation}
for the encoder represented by $\pmb J^{\mc E}$. The transmitter subproblem is
\begin{equation}
\begin{aligned}
    \underset{\pmb{J}^{\mc{E}}}{\operatorname{maximize}}
    \quad&
    \left\langle
        \pmb{C}_{\mc{E}},
        \pmb{J}^{\mc{E}}
    \right\rangle                                                     \\
    \operatorname{subject\ to}
    \quad&
    \pmb{J}^{\mc{E}}\succeq\pmb{0},
    \qquad
    \Tr_{\mc H_{\mathrm{src}}}\pmb{J}^{\mc{E}}
    =
    \pmb{I}_{\mc H_{\mathrm L}},                                       \\
    &
    (\pmb{I}_{\mc H_{\mathrm L}}\otimes\pmb{\Pi}_n)
    \pmb{J}^{\mc{E}}
    (\pmb{I}_{\mc H_{\mathrm L}}\otimes\pmb{\Pi}_{n'})
    =
    \pmb{0},
    \quad
    n\ne n' .
    \label{eq:encoder-sdp}
\end{aligned}
\end{equation}
Here $\pmb{\Pi}_n$ projects the encoder output onto the $n$-photon sector
$\mc H_{\mathrm{src}}^{(n)}$. The last constraint expresses the photon-number
superselection
that applies when the transmitter holds no optical phase reference: coherences
between sectors of different total photon number are dephased and carry no
operational content, so the encoder Choi operator is block diagonal in
$n$~\cite{MS:14:PRA}. An available reference lifts the constraint. If
$\mathsf N_{\mathrm{src}}$ contains one sector, the constraint is vacuous.

The alternating design solves \eqref{eq:decoder-sdp} and
\eqref{eq:encoder-sdp} in turn. If both subproblems use the same unconditional
average-fidelity objective and constraints and are solved exactly, the previous
iterate remains feasible. The objective is therefore nondecreasing and, because
fidelity is bounded, converges in value. The joint problem is bilinear over two
convex \ac{CPTP} sets rather than jointly convex, so the attained value can
depend on initialization~\cite{RW:05:PRL_2,KL:09:QIP_2,TKL:10:IEEE_J_IT}.

Let $N_{\mathrm{init}}$ be the number of initial encoders, and let
$\{\mc E_u\}_{u=1}^{N_{\mathrm{init}}}$ include rail selection, cloning,
coherent path superposition, and fixed codes admitted by the source space. For
each initialization, we optimize $\mc D_u^\star$, alternate the two subproblems,
perform a final decoder update, and select the largest attained fidelity. This
value is a feasible lower bound on the global \ac{CPTP} optimum. When all
compared structured encoders are registered as initializations, monotonicity
also guarantees that it is no worse than the best registered strategy after
decoder optimization. See Appendix~\ref{app:multistart} for the argument.

Thus the attained general-design value should be interpreted carefully: it
certifies one feasible encoder-recovery pair, not the optimum of the joint
problem. Conversely, the unknown optimum over all admissible \ac{CPTP} pairs
remains an upper bound on every resource-matched structured strategy. These two
statements distinguish the numerical guarantee from the expressiveness of the
design class within the source space.

The general \ac{CPTP} search grows rapidly with the source dimension. Let
$d_n=\binom{N_{\mathrm{t}}}{n}2^n$ be the dimension of the $n$-photon sector.
A deterministic map from one logical qubit to a source space of dimension
$D_{\mathrm{src}}$ has
\begin{equation}
    \begin{cases}
        4D_{\mathrm{src}}^{2}-4,
        & \text{shared phase reference},
        \\[3pt]
        4\sum_{n\in\mathsf N_{\mathrm{src}}}d_{n}^{2}-4,
        & \text{no shared reference},
    \end{cases}
    \label{eq:general-encoder-dof}
\end{equation}
real \ac{DoF}, and its Choi rank, equivalently its minimal Kraus
rank, can be as large as
$2D_{\mathrm{src}}$. For a two-rail source with
$\mathsf N_{\mathrm{src}}=\{1,2\}$, the design dimension is $124$ without a
shared phase
reference and $252$ with one; the Choi-rank bound is $16$. By
comparison, the universal cloner below has $M-1$ real design variables and
Choi rank at most $M$; the coherent path encoder has $2M-2$ real variables and
Choi rank one; and a rail-selection or fixed stabilizer encoder also has Choi
rank one. Adding unused vacuum rails to an encoder does not change its
Choi rank. These restrictions reduce the optimization dimension and the
number of operators required by a minimal Kraus or Stinespring
realization.\footnote{However, Choi rank does not, by itself, determine optical
gate depth~\cite{NC:10:NoVenue}.}
Design complexity is not the only resource that separates these encoders. The
encoders for direct transmission and coherent path superposition emit one
photon per channel use, whereas an
    $M$-clone encoder emits $M$ photons and a
$\left[\!\left[n_{\mathrm{c}},1,d_{\min}\right]\!\right]$ stabilizer
encoder emits $n_{\mathrm{c}}$. A multiphoton encoder therefore gains
additional opportunities for a carrier to survive loss, although these need
not be statistically independent under crosstalk or correlated fading, at a
source cost that grows in proportion. Each structured encoder emits into a
single photon-number sector, so its photon number per channel use is fixed;
only the general \ac{CPTP} encoder can spread over several sectors of
$\mathsf N_{\mathrm{src}}$, for which the mean applies. Every strategy consumes one unknown logical qubit per transmission attempt.
Multiphoton encoders add locally prepared carriers while preserving this
single-input-state cost. We therefore distinguish the mean emitted photon
number, which is optical carrier overhead, from quantum-state consumption. For
statistically identical heralded attempts,
$1/\overline{p}_{\mathrm{s}}$ is the mean number of input states consumed per
successful output, while the encoder and recovery dimensions and Choi ranks
characterize complexity scaling.

\subsubsection{Asymmetric Cloning and Heralded Joint Recovery}
\label{subsubsec:asymmetric-cloning-purification}

The system-level cloning strategy is a union over the emitted clone number,
the occupied transmit-rail support, and the universal-cloner coefficients.
For $M\in\{1,\ldots,N_{\mathrm t}\}$, let
$S\subseteq\{1,\ldots,N_{\mathrm t}\}$ be a support of size $|S|=M$, and
define
\begin{equation}
\begin{aligned}
    \mathfrak B_M
    \coloneqq
    \biggl\{
        \pmb\beta\in\mathbb R_{+}^{M}:
        \left(\sum_{m=1}^{M}\beta_m\right)^2
        +\sum_{m=1}^{M}\beta_m^2=2
    \biggr\}.
\end{aligned}
    \label{eq:universal-cloner-coefficient-set}
\end{equation}
Let $\mc C_{\pmb\beta}^{(M)}$ be the universal asymmetric
$1\to M$ cloning channel and let $\mc V_{S}^{\mathrm{cl}}$ embed its $M$
polarization-qubit outputs as one photon on every rail in $S$, with vacuum on
all rails outside $S$. The embedded encoder is
\begin{equation}
    \mc E_{S,\pmb\beta}^{\mathrm{cl}}
    \coloneqq
    \mc V_{S}^{\mathrm{cl}}
    \circ
    \mc C_{\pmb\beta}^{(M)},
    \qquad
    \pmb\beta\in\mathfrak B_M .
    \label{eq:fixed-m-cloning-encoder}
\end{equation}
The admissible variable-$M$ family is
\begin{equation}
\begin{aligned}
    \mathfrak E_{\mathrm{cl}}^{\mathrm{var}}
    \coloneqq
    \bigcup_{M=1}^{N_{\mathrm t}}
    \ \bigcup_{\substack{S\subseteq\{1,\ldots,N_{\mathrm t}\}\\|S|=M}}
    \left\{
        \mc E_{S,\pmb\beta}^{\mathrm{cl}}:
        \pmb\beta\in\mathfrak B_M
    \right\}.
\end{aligned}
    \label{eq:variable-m-cloning-family}
\end{equation}
Each fixed-$M$ member emits exactly $M$ photons. A zero component of
$\pmb\beta$ changes the corresponding marginal clone quality but does not put
that rail in vacuum. In particular, the endpoint
$\pmb\beta=(1,0,\ldots,0)^{\mathsf T}$ still emits $M$ photons; it gives one
perfect marginal and $M-1$ maximally mixed marginals. Direct transmission is
instead the $M=1$ member
\begin{equation}
    \mc E_{\{i\},(1)}^{\mathrm{cl}}
    =
    \mc E_i^{\mathrm{dir}} .
    \label{eq:cloning-direct-member}
\end{equation}
Equal coefficients,
\begin{equation}
    \beta_m
    =
    \sqrt{\frac{2}{M(M+1)}},
    \qquad m=1,\ldots,M,
    \label{eq:symmetric-cloner-coefficients}
\end{equation}
give the symmetric $1\to M$ cloner; unequal feasible coefficients give the
asymmetric family. The marginal Haar-averaged state fidelity of output $m$ is
\begin{equation}
    F_m
    =
    \frac{1}{3}
    +
    \frac{1}{6}
    \left(
        \beta_m+\sum_{m'=1}^{M}\beta_{m'}
    \right)^2,
    \label{eq:universal-clone-fidelity}
\end{equation}
and the symmetric point gives
$F_m=(2M+1)/(3M)$~\cite{C:00:JMO,NPR:23:LMP}.
The fixed-$M$ cloning channel has Choi rank at most $M$.

The outputs are correlated and are not modeled as independent perfect copies.
The Rx applies one joint recovery channel or, for heralded operation, a
recovery instrument to the complete erasure-augmented receiver state. When an
instrument aborts, its trace deficit is assigned to the maximally mixed logical
state. See Appendix~\ref{app:heralded} for this completion. The cloning
coefficients require no spatial phase reference, although the end-to-end
objective can depend on field phases through multiphoton interference and
collision statistics. A phase-insensitive design may use power \ac{CSI};
optimization on the estimated composite channel uses field \ac{CSI} or a
stated model for phase uncertainty.

Increasing $M$ supplies more opportunities for at least one carrier to survive
loss, but it lowers the symmetric marginal fidelity and emits more photons.
The comparison therefore evaluates fidelity together with the emitted photon
number and the physical survival probability.

\subsubsection{Single-Photon Coherent Path Superposition}
\label{subsubsec:superposition-trajectories}

This strategy encodes the logical qubit in one photon distributed coherently
across $M$ selected spatial rails. At the Tx, injecting the polarization qubit
into one input of a lossless $M$-port interferometer prepares the normalized
complex rail amplitudes; the interferometer can be decomposed into two-port beam
splitters and phase shifters~\cite{RZBB:94:PRL,CHMKW:16:Optica} and is assumed
to act identically on both polarizations. Let
$\pmb t=(t_1,\ldots,t_M)^{\mathsf T}$ contain the selected rail indices in
increasing order,
$1\leq t_1<\cdots<t_M\leq N_{\mathrm t}$, and define the selection matrix
$\pmb T_{\pmb t}\coloneqq
[\pmb e_{t_1}\ \cdots\ \pmb e_{t_M}]$. The corresponding field and
internal-state submatrices are
$\pmb A_{\mathrm{eff}}^{(\pmb t)}
\coloneqq\pmb A_{\mathrm{eff}}\pmb T_{\pmb t}$ and
$\pmb G^{(\pmb t)}
\coloneqq\pmb T_{\pmb t}^{\dagger}\pmb G\pmb T_{\pmb t}$.
The normalized complex path weights are
\begin{equation}
    \pmb{w}
    =
    (w_1,\ldots,w_M)^{\mathsf{T}},
    \qquad
    \pmb{w}^{\dagger}\pmb{w}=1 .
    \label{eq:path-weights}
\end{equation}
Specifically, its isometry is
\begin{equation}
    \pmb{V}_{\pmb{w}}\ket{\psi}
    =
    \sum_{m=1}^{M}
    w_m
    \ket{\psi}_{t_m}
    \bigotimes_{\substack{i=1\\i\neq t_m}}^{N_{\mathrm{t}}}
    \ket{\mathrm{vac}}_{i},
    \label{eq:path-isometry}
\end{equation}
The encoded state contains exactly
one photon across the selected rails. We write
$\mc{E}_{\pmb{w}}^{\mathrm{coh}}(\pmb{\rho})
\coloneqq\pmb{V}_{\pmb{w}}\pmb{\rho}\pmb{V}_{\pmb{w}}^{\dagger}$ for the
corresponding encoder and $\mc{E}_{\mathrm{eq}}^{\mathrm{coh}}$ for its
equal-weight point $w_m=1/\sqrt{M}$. Normalization and removal of a global
phase leave $2M-2$ real degrees of freedom. Any normalized $\pmb{w}$ defines an
isometry that can be extended to a passive unitary over the rail modes; the
encoder therefore has Choi rank one.

Coherent superpositions of channel trajectories have been analyzed as a
resource in quantum Shannon theory, where the attributed advantage depends on
how each constituent channel is extended to act on the vacuum: distinct vacuum
extensions of the same channels yield distinct
performance~\cite{CK:19:PRSA,KCSEW:20:NJP}. Here
\eqref{eq:path-isometry} is a passive linear-optical transformation of the rail
modes, and the
physical propagation model fixes the vacuum extension. The same field-transfer
matrix $\pmb A(\omega)$ and Stinespring dilation \eqref{eq:loss-dilation} act
on every occupied and unoccupied rail. Thus the comparison concerns one
passive path encoder within a common propagation model, rather than alternative
causal or channel configurations. In the following design rules, a hat denotes the field,
depolarization, or internal-state estimate available to the optimizer. For a
field estimate, we define
$\widehat a_{\mathrm{eff},ji}
\coloneqq[\widehat{\pmb A}_{\mathrm{eff}}]_{ji}$ and
\begin{equation}
    \widehat{\pmb A}_{\mathrm{eff}}^{(\pmb t)}
    \coloneqq
    \widehat{\pmb A}_{\mathrm{eff}}\pmb T_{\pmb t},
    \qquad
    \widehat{\pmb G}^{(\pmb t)}
    \coloneqq
    \pmb T_{\pmb t}^{\dagger}\widehat{\pmb G}\pmb T_{\pmb t}.
    \label{eq:estimated-selected-submatrices}
\end{equation}
For a
chosen receive port $j$ and a coherent estimate
$\widehat{\pmb{A}}_{\mathrm{eff}}$, maximum-ratio
transmission maximizes the target-port power when the selected path components
have the same internal state:
\begin{equation}
    w_m
    =
    \frac{
        \widehat{a}_{\mathrm{eff},j t_m}^{*}
    }{
        \sqrt{
            \sum_{m'=1}^{M}
            \left|
                \widehat{a}_{\mathrm{eff},j t_{m'}}
            \right|^2
        }
    },
    \qquad m=1,\ldots,M ,
    \label{eq:maximum-ratio-path}
\end{equation}
provided that the selected portion of row $j$ has nonzero norm. Maximum-ratio
transmission is optimal for the stated objective, namely the power delivered to
one selected port, and is therefore the matched rule when the recovery reads a
single port. For a general selected-rail internal-state Gram matrix, we define
the
selected row
$\pmb b_j\coloneqq\pmb e_j^{\dagger}
\widehat{\pmb A}_{\mathrm{eff}}^{(\pmb t)}\in\mathbb C^{1\times M}$.
The corresponding target-port quadratic form is
$\widehat{\pmb G}^{(\pmb t)}\odot
(\pmb b_j^{\dagger}\pmb b_j)$, where $\odot$ denotes the Hadamard product;
its principal eigenvector is the optimal path-weight vector.

For an all-port deterministic recovery after the squashing map,
let
$\pmb{B}=\widehat{\pmb{A}}_{\mathrm{eff}}^{(\pmb{t})}$ and
$\pmb{Q}_{\mathrm{pol}}
=\operatorname{diag}(1-\widehat q_1,\ldots,1-\widehat q_{N_{\mathrm{r}}})$,
and let $\widehat{\pmb{G}}^{(\pmb{t})}$ be the Gram submatrix of the selected
transmit rails. Define
\begin{equation}
    \pmb{M}_{\mathrm{coh}}
    \coloneqq
    \widehat{\pmb{G}}^{(\pmb{t})}
    \odot
    \left(\pmb{B}^{\dagger}\pmb{Q}_{\mathrm{pol}}\pmb{B}\right).
    \label{eq:weighted-coherent-path-matrix}
\end{equation}
Assigning the state $\pmb I_2/2$ to a failed transmission gives the average
fidelity on the estimated channel
\begin{equation}
    F_{\mathrm{av}}^{\mathrm{des}}(\pmb{w})
    =
    \frac{1}{2}
    +
    \frac{1}{2}\pmb{w}^{\dagger}
    \pmb{M}_{\mathrm{coh}}\pmb{w}.
    \label{eq:weighted-coherent-path-fidelity}
\end{equation}
The optimal path-weight vector is therefore a principal eigenvector of
$\pmb{M}_{\mathrm{coh}}$.
When the selected internal states are identical,
$\widehat{\pmb{G}}^{(\pmb{t})}$ is the all-ones matrix and
$\pmb{M}_{\mathrm{coh}}$ reduces to
$\pmb{B}^{\dagger}\pmb{Q}_{\mathrm{pol}}\pmb{B}$, equivalently the Gram
matrix of $\pmb{Q}_{\mathrm{pol}}^{1/2}\pmb{B}$. Only when, in addition, all
$\widehat q_j=\widehat q$ does the solution
reduce to a dominant right singular vector of $\pmb{B}$ and
$F_{\mathrm{av}}^{\mathrm{des}}=1/2+
(1-\widehat q)\|\pmb{B}\pmb{w}\|^2/2$. Thus the unweighted
singular-vector rule is not generally optimal under unequal polarization
quality or rail-dependent internal states. The full $2M-2$ parameters also
remain available for prescribed-success or otherwise constrained designs.
Unlike cloning, coherent path design depends directly on the relative phases
of the field-transfer coefficients, so phase-estimation error reduces the
achieved concentration; it also requires the selected submatrix of the source
calibration $\pmb{G}$.

The strategies also differ in their \ac{CSI} requirements. Adaptive direct
transmission uses detector-plane column powers
$\{\lVert\pmb A_{\mathrm{eff}}\pmb e_i\rVert^2\}$ at the Tx for rail selection
and $\{|a_{\mathrm{eff},ji}|^2\}$ at the Rx for port selection. A fixed
\ac{QEC} placement uses no Tx \ac{CSI}, whereas adaptive placement uses power
\ac{CSI}. With Rx \ac{CSI}, the recovery is optimized on the receiver estimate;
otherwise one recovery is selected for the channel distribution. Cloning
adapts real coefficients from power \ac{CSI}, while optimization on the full
composite channel uses field \ac{CSI}. Coherent path superposition and the
alternating \ac{CPTP} design use field \ac{CSI}: the former needs the relative
phases of $\pmb A_{\mathrm{eff}}$ for its path weights, and the latter solves
each half-step on the estimated composite channel.

The role of $\pmb{G}$ depends on the encoder. Direct transmission launches
one photon in a definite transmit rail and therefore depends only on
$g_{ii}=1$; the off-diagonal overlaps are irrelevant. Coherent path superposition
uses amplitudes from several transmit rails, so the off-diagonal overlaps set
the path coherence that remains after the internal states are traced out.
Distinct internal states retain which-path information~\cite{E:96:PRL}.
Cloning and the stabilizer codes use multiple photons and can depend on
$\pmb{G}$ through interference between many-particle permutations. Thus,
$\pmb{G}$ specifies the overlap of the source internal states, but its effect
on the logical channel depends jointly on the encoder and the spatial
transformation $\pmb{W}(\omega)$. Its action cannot be modeled as a uniform
noise parameter determined only by photon number.

\begin{table*}[!t]
    \caption{Encoder resources, \ac{CSI} requirements, and receiver processing.
    Here $d_n=\binom{N_{\mathrm t}}{n}2^n$, and $M$ denotes the number of
    emitted clone outputs or selected coherent paths.}
    \label{tab:encoder-resource-comparison}
    \centering
    \footnotesize
    \setlength{\tabcolsep}{3pt}
    \renewcommand{\arraystretch}{1.16}
    \begin{tabular}{
        @{}>{\raggedright\arraybackslash}p{0.12\textwidth}
        >{\raggedright\arraybackslash}p{0.145\textwidth}
        >{\raggedright\arraybackslash}p{0.135\textwidth}
        >{\raggedright\arraybackslash}p{0.075\textwidth}
        >{\raggedright\arraybackslash}p{0.085\textwidth}
        >{\raggedright\arraybackslash}p{0.16\textwidth}
        >{\raggedright\arraybackslash}p{0.18\textwidth}@{}}
        \toprule
        \rowcolor{rowblue}
        Strategy
        & Encoder \ac{DoF}
        & Discrete design
        & \multicolumn{2}{c}{Physical resources}
        & \ac{CSI} for adaptation
        & Rx postprocessing
        \\
        \cmidrule(lr){4-5}
        \rowcolor{rowblue}
        & &
        & Choi rank
        & Photons
        & &
        \\
        \midrule

        Direct transmission
        & $0$
        & $N_{\mathrm t}$ transmit rails;
          $N_{\mathrm t}N_{\mathrm r}$ rail--port pairs
        & $1$
        & $1$
        & None for the fixed pair; Power \ac{CSI} for rail--port adaptation
        & Receive-port selection; failed trials map to $\pmb I_2/2$
        \\

        Fixed stabilizer \ac{QEC}
        & $0$
        & $\pi\in\mc Q_c$,
          $\lvert\mc Q_c\rvert=\binom{N_{\mathrm t}}{n_{\mathrm c}}$
        & $1$
        & $n_{\mathrm c}$
        & None for fixed placement; Power \ac{CSI} for placement adaptation
        & Channel-adapted recovery \ac{SDP}
        \\

        General \ac{CPTP} maps
        & $4D_{\mathrm{src}}^2-4$ with a shared phase reference;
          $4\sum_{n\in\mathsf N_{\mathrm{src}}}d_n^2-4$ without one
        & All \ac{CPTP} maps on $\mc H_{\mathrm{src}}$
        & $\leq 2D_{\mathrm{src}}$
        & Mean photon number over admitted sectors
        & Estimated composite channel; Field \ac{CSI} or a stated model for
          phase uncertainty
        & Alternating encoder and decoder \acp{SDP}; feasible lower bound
          without a global certificate
        \\

        Variable-$M$ universal asymmetric cloning
        & $M-1$ within a selected support
        & $2^{N_{\mathrm t}}-1$ nonempty supports, with $M=|S|$
        & $\leq M$
        & $M$
        & Power \ac{CSI} for phase-insensitive coefficients; Field \ac{CSI}
          for the estimated composite channel
        & One fixed all-port recovery under Tx \ac{CSI}; coefficient/support
          update and recovery \ac{SDP} under Full \ac{CSI}
        \\

        Single-photon coherent path superposition
        & $2M-2$
        & $\binom{N_{\mathrm t}}{M}$ selected-rail supports
        & $1$
        & $1$
        & Field \ac{CSI}; separate source calibration $\widehat{\pmb G}$ when
          internal states differ
        & Maximum-ratio transmission or principal-eigenvector design, followed
          by all-port recovery
        \\

        \bottomrule
    \end{tabular}
\end{table*}

The channel-estimation burden is distinct from the encoder design dimension.
The identifiable power- and field-\ac{CSI} coordinate counts in
\eqref{eq:csi-identifiable-counts} include the calibrated $\pmb q$ parameters
but exclude the separately calibrated $\pmb\tau$ and $\pmb G$.

\section{Channel State Information}
\label{sec:csi}

If the instantaneous realization is unavailable to both endpoints, the
relevant effective channel is the convex mixture
\begin{equation}
    \overline{\mc{N}}_{\mathrm{FSO}}
    =
    \mathbb{E}_{\omega}
    \left[
        \mc{N}_{\mathrm{FSO}}(\omega;\pmb G,\pmb\tau,\pmb q)
    \right].
    \label{eq:ensemble-channel}
\end{equation}
The channel is constructed for each realization before taking the expectation.
Replacing $\pmb{A}(\omega)$ by
$\mathbb{E}_{\omega}[\pmb{A}(\omega)]$ generally gives a different map because
field averaging can remove relative phases before the bosonic transformation
is formed. With realization \ac{CSI}, the encoder or decoder may instead
condition its admissible operation on the realized channel description.

\subsection{Estimation}
\label{subsec:csi-estimation}

\begin{figure}[!t]
    \centering
    \includegraphics[width=0.86\columnwidth]
        {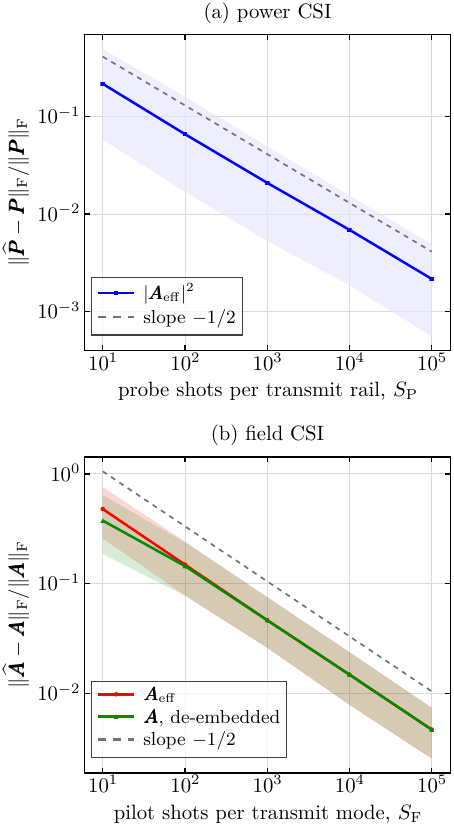}
    \caption{Normalized estimation error of the two \ac{CSI} tiers against the
    probe budget, on the $2\times2$ channel at $s_{\mathrm{rail}}/w(z)=2.5$, zero pointing and
    $\sigma_R^2=0.2$, over $256$ atmospheric realizations with eight estimator
    replicates each. Panel (a) is power \ac{CSI}, recovering
    $\lvert\pmb A_{\mathrm{eff}}\rvert^2$ from multinomial single-photon probe
    counts. Panel (b) is field \ac{CSI}, recovering the complex transfer matrix
    from a coherent heterodyne pilot of unit mean photon number, evaluated
    against the detector-plane matrix measured directly and the pre-receiver
    matrix after de-embedding $\pmb\tau^{\mathrm{rx}}$ and projection onto the set of
    contractions. Every field comparison uses a common row-and-column phase
    alignment, which makes the distance independent of the phase convention.
    Curves are medians and bands the $5$th to $95$th percentile range; the
    dashed line has slope $-1/2$.}
    \label{fig:csi-estimation}
\end{figure}

Channel-adapted recovery requires a channel description for the recovery
optimization~\cite{TKL:10:IEEE_J_IT}. Quantum process tomography can provide
this description, but unrestricted process reconstruction scales exponentially
with system size~\cite{MRL:08:PRA,TWACCEtAl:23:NC}. The structured channel
model permits a smaller set of physical quantities to be identified instead
of reconstructing an unrestricted process~\cite{BNWK:09:NJP,GRRWK:20:NC}.
For a fixed realization, the spatial estimands are
$\pmb{P}=\pmb{A}_{\mathrm{eff}}\odot\pmb{A}_{\mathrm{eff}}^{*}$ under power
\ac{CSI} and the complex $\pmb{A}_{\mathrm{eff}}$ under field \ac{CSI}, as
defined in \eqref{eq:effective-field-transfer}--\eqref{eq:power-csi}. Both refer
only to the selected transmit and receive mode subspaces; neither reconstructs
the full transverse-field operator $\mc K_{\mathrm{prop}}(\omega)$. The
receiver transmissivity $\pmb{\tau}$ is obtained from a local Rx calibration,
the effective per-port depolarization vector $\pmb q$ from known polarization
probes prepared at the Tx and measured at the Rx, and the source Gram matrix
$\pmb G$ from source characterization at the Tx. The vector $\pmb{\tau}$ is
needed only when the
pre-receiver matrix $\pmb{A}$ or receiver-side loss is to be specified
separately. For the fixed receiver and downstream processing considered here,
the spatial channel is determined by $\pmb{A}_{\mathrm{eff}}$. Separate
identification of $\pmb{A}$ and $\pmb{\tau}$ is required for physical
attribution, changes to the receiver, or mode processing applied before
receiver-side loss. Either form of \ac{CSI} may be available to the Tx, the Rx,
both, or neither.
Statistical knowledge of
$\sigma_R^2$ and receiver characteristics may be used to design a protocol for
an objective averaged over the channel-realization and estimation-error
distributions, but it does not specify an instantaneous realization. The internal-state Gram
matrix $\pmb{G}$ is treated as a source calibration and is updated on the
source-drift timescale.
Throughout this subsection, one realization $\omega$ is held fixed over the
estimation block and its explicit argument is suppressed.

\subsubsection{Power CSI}
\label{subsubsec:power-csi}

Power \ac{CSI} estimates the matrix $\pmb{P}(\omega)$ in
\eqref{eq:power-csi}. During the estimation of column $i$, one photon is
prepared in transmit mode $i$, and every other transmit mode is in vacuum. Let
$S_{\mathrm P}$ be the number of trials per input column,
$C_{ji}$ the number of detections in receive mode $j$, and $C_{0i}$
count the no-detection outcome. We define
$\pmb{c}_i\coloneqq(C_{0i},C_{1i},\ldots,C_{N_{\mathrm{r}}i})^{\mathsf{T}}$ and
$\pmb{p}_i^{\mathrm{out}}
\coloneqq(p_{0i},p_{1i},\ldots,p_{N_{\mathrm{r}}i})^{\mathsf{T}}$.
Then
\begin{equation}
\begin{aligned}
    \pmb{c}_i
    &\sim
    \operatorname{Multinomial}
    \left(S_{\mathrm{P}};\pmb{p}_i^{\mathrm{out}}\right),
    \\
    p_{0i}
    &=1-\sum_{j=1}^{N_{\mathrm{r}}}p_{ji}.
\end{aligned}
    \label{eq:power-csi-count-model}
\end{equation}
The complete power-\ac{CSI} estimation therefore uses
$B_{\mathrm P}\coloneqq N_{\mathrm t}S_{\mathrm P}$ single-photon probes.
For $j,k=1,\ldots,N_{\mathrm{r}}$, the maximum-likelihood estimators, their
plug-in marginal standard errors, and the exact within-column covariance are
\begin{equation}
\begin{aligned}
    \widehat p_{ji}
    &=\frac{C_{ji}}{S_{\mathrm{P}}},
    \\
    \widehat{\pmb{P}}
    &=\left[\widehat p_{ji}\right],
    \\
    \widehat p_{0i}
    &=\frac{C_{0i}}{S_{\mathrm{P}}}
    =1-\sum_j\widehat p_{ji},
    \\
    \widehat{\operatorname{se}}(\widehat p_{ji})
    &=
    \sqrt{
        \frac{\widehat p_{ji}(1-\widehat p_{ji})}{S_{\mathrm{P}}}
    },
    \\
    \operatorname{Cov}(\widehat p_{ji},\widehat p_{ki})
    &=
    \frac{\delta_{jk}p_{ji}-p_{ji}p_{ki}}{S_{\mathrm{P}}}.
    \label{eq:power-csi-estimator}
\end{aligned}
\end{equation}
The no-detection probability includes selected-mode leakage and receiver-side
loss. Equation~\eqref{eq:power-csi} shows that the end-to-end observation
identifies only $\tau_j^{\mathrm{rx}}|a_{ji}|^2$; it cannot separate
$\tau_j^{\mathrm{rx}}$ from $|a_{ji}|^2$ without an independent receiver calibration.
This non-identifiability does not alter the composite channel because its
spatial transformation depends on the product
$\pmb{A}_{\mathrm{eff}}=\pmb{D}_{\tau}\pmb{A}$. The separation becomes
operationally relevant when receiver-side loss and propagation are specified or
varied independently.

Power \ac{CSI} is supplemented by independently calibrated estimates
$\widehat{\pmb{\tau}}$ and $\widehat{\pmb{q}}$ when those mechanisms are
specified separately. The receiver-loss estimate comes from a separate local
calibration; its protocol, uncertainty, and resource cost must be declared when
$\pmb\tau$ is specified apart from $\pmb A_{\mathrm{eff}}$. To estimate
$\pmb q$, for port $j$ choose the lowest-index maximizer
$i_j\coloneqq\min\operatorname*{arg\,max}_{i}\widehat p_{ji}$.
In each basis $\mathsf b\in\{Z,X\}$, launch $S_{\mathrm Q}$ probes on
rail $i_j$, let $C^{\mathrm{det}}_{\mathsf b,j}$ be the number detected at port $j$,
and let $C^{\mathrm{err}}_{\mathsf b,j}$ be the detected polarization errors. Under the
isotropic depolarization model
in \eqref{eq:local-polarization-depolarization},
$C^{\mathrm{det}}_{\mathsf b,j}\sim
\operatorname{Binomial}(S_{\mathrm Q},p_{j i_j})$ and
$C^{\mathrm{err}}_{\mathsf b,j}\mid C^{\mathrm{det}}_{\mathsf b,j}\sim
\operatorname{Binomial}(C^{\mathrm{det}}_{\mathsf b,j},q_j/2)$. Using the Jeffreys
prior for this conditional Bernoulli error probability, we define the posterior
mean
$\widehat e_{\mathsf b,j}\coloneqq
(C^{\mathrm{err}}_{\mathsf b,j}+1/2)/(C^{\mathrm{det}}_{\mathsf b,j}+1)$, and let
$\Pi_{[0,1]}(x)\coloneqq\min\{1,\max\{0,x\}\}$ denote Euclidean projection
onto $[0,1]$. The physical estimator is
\begin{equation}
    \widetilde q_j
    \coloneqq
    \widehat e_{Z,j}+\widehat e_{X,j},
    \qquad
    \widehat q_j
    \coloneqq
    \Pi_{[0,1]}(\widetilde q_j).
    \label{eq:polarization-csi-estimator}
\end{equation}
This rule remains defined when a basis has no detection, in which case its
conditional error estimate is $1/2$. Agreement of the two error estimates checks
the assumed basis independence, but two bases do not establish full isotropy.
Calibrating all receive ports uses
$B_{\mathrm Q}=2N_{\mathrm r}S_{\mathrm Q}$ optical probes.

Power measurements discard the relative phases in
$\pmb{A}_{\mathrm{eff}}$. Consequently, $\widehat{\pmb{P}}$ supports
transmit-rail and receive-port selection and phase-insensitive clone allocation.
It specifies a family of compatible \ac{CPTP} maps rather than a unique map in
\eqref{eq:composite-fso-channel}; multiphoton interference can differ among
complex field matrices with the same power coefficients. Recovery design from
power \ac{CSI} therefore requires a specified statistical or worst-case model
for the compatible phases, whereas phase-sensitive optimization requires field
\ac{CSI}.

\subsubsection{Photon-Internal-State Calibration}
\label{subsubsec:internal-state-calibration}

Pairwise Hong--Ou--Mandel calibration uses a balanced beam splitter and
photon-number-resolving detection to estimate the magnitude of each
internal-state overlap before data transmission~\cite{HOM:87:PRL_2}. For
source rails $i\neq i'$, the probability that both photons leave through the
same beam-splitter output is
\begin{equation}
    p_{ii'}^{\mathrm{same}}
    =
    \frac{1+\left|g_{ii'}\right|^2}{2}
    =
    1-\frac{\zeta_{ii'}}{2}.
    \label{eq:hom-same-output-probability}
\end{equation}
Let $S_{\mathrm{HOM}}$ be the number of independent trials per
unordered rail pair. Calibrating every pair uses
$B_{\mathrm{HOM}}=\binom{N_{\mathrm t}}{2}S_{\mathrm{HOM}}$ two-photon trials,
or $2B_{\mathrm{HOM}}$ launched photons.
Inverting \eqref{eq:hom-same-output-probability} gives
$\left|g_{ii'}\right|^2=2p_{ii'}^{\mathrm{same}}-1$, so the same-output
frequency estimates the overlap directly. If $C_{ii'}$ same-output events are
observed in $S_{\mathrm{HOM}}$ independent trials, the count is binomially
distributed with success probability $p_{ii'}^{\mathrm{same}}$, and
$2C_{ii'}/S_{\mathrm{HOM}}-1$ is unbiased for
$\left|g_{ii'}\right|^2$. It can nonetheless fall below zero at small
$S_{\mathrm{HOM}}$, so it is projected back. The constrained estimator is
\begin{equation}
\begin{aligned}
    \widehat{|g_{ii'}|^2}
    &=
    \Pi_{[0,1]}
    \left(
        \frac{2C_{ii'}}{S_{\mathrm{HOM}}}-1
    \right),
    \\
    \widehat{\zeta}_{ii'}
    &=
    1-\widehat{|g_{ii'}|^2}.
    \label{eq:hom-distinguishability-estimator}
\end{aligned}
\end{equation}
Before enforcing the interval constraint, the standard error of
$\widehat{|g_{ii'}|^2}$ is
\begin{equation}
    \operatorname{se}
    \left(
        \widehat{|g_{ii'}|^2}
    \right)
    =
    2\sqrt{
        \frac{
            p_{ii'}^{\mathrm{same}}
            (1-p_{ii'}^{\mathrm{same}})
        }{
            S_{\mathrm{HOM}}
        }
    }.
    \label{eq:hom-overlap-standard-error}
\end{equation}
Since $p(1-p)\leq1/4$, this is bounded by $S_{\mathrm{HOM}}^{-1/2}$ uniformly
in the overlap, so $S_{\mathrm{HOM}}\geq\varepsilon^{-2}$ trials per rail pair
suffice for a standard error of at most $\varepsilon$. The bound is tight at
$|g_{ii'}|^2=0$ and loosens as the overlap approaches unity.

Equation \eqref{eq:hom-overlap-standard-error} describes the unconstrained
estimator. Enforcing the interval constraint in
\eqref{eq:hom-distinguishability-estimator} truncates the lower tail when the
true overlap lies near zero. The projected overlap estimate is then biased
upward and its dispersion is smaller than
\eqref{eq:hom-overlap-standard-error} indicates. The effect is negligible for
overlaps well inside the interval and is absent at unit overlap, where every
trial has a same-output event. It is largest for orthogonal internal states,
where $|g_{ii'}|^2=0$ places the true value on the lower boundary and the
constraint is active in approximately half of the realizations, so that
$\widehat{\zeta}_{ii'}$ is biased below unity at small
$S_{\mathrm{HOM}}$. Uncertainties quoted from
\eqref{eq:hom-overlap-standard-error} are therefore to be read as applying to
the unconstrained estimate.

Let $\widetilde{\pmb G}$ be the raw Hermitian overlap matrix supplied by the
calibration convention. Before it is used in the channel, enforce the
Gram-matrix constraints through the nearest correlation matrix
\begin{equation}
    \widehat{\pmb G}
    \coloneqq
    \underset{\substack{
        \pmb X\succeq\pmb 0,\;
        \operatorname{diag}(\pmb X)=\pmb 1_{N_{\mathrm t}}}}
    {\operatorname*{arg\,min}}
    \left\lVert\pmb X-\widetilde{\pmb G}\right\rVert_{\mathrm F}^{2}.
    \label{eq:gram-correlation-projection}
\end{equation}
Thus $\widehat{\pmb G}\succeq\pmb0$ and its diagonal is unity, as required by
the internal-state construction and the coherent-path Schur product.

These measurements determine overlap magnitudes but not overlap phases, which
limits what they specify as the number of rails grows. For
$N_{\mathrm{t}}=2$ the single magnitude $|g_{12}|$, equivalently $\zeta_{12}$,
fixes the Gram matrix up to a rephasing of the internal states, so pairwise
calibration is complete. For $N_{\mathrm{t}}\geq3$ the pairwise magnitudes are
still all recovered, but pairwise calibration leaves the closed-loop phases
defined in Section~\ref{subsubsec:multiphoton} undetermined. Recovering them
requires phase-sensitive characterization~\cite{T:15:PRA,S:15:PRA}. The
equicorrelated model in \eqref{eq:equicorrelated-gram-model} has real,
nonnegative overlaps and sets every closed-loop phase to zero. It is therefore
a restriction rather than a parameterization of the general case. This
limitation concerns the calibration of $\pmb{G}$ only and does not restrict
mode-resolved reception to two rails.

\subsubsection{Field CSI}
\label{subsubsec:field-csi}

Field \ac{CSI} estimates the complex detector-plane matrix
$\pmb{A}_{\mathrm{eff}}$ in \eqref{eq:effective-field-transfer}, including the
relative phases absent from $\pmb{P}$. We consider a phase-referenced coherent
probe of amplitude $\alpha_{\mathrm{p}}$ and mean photon number
$n_{\mathrm{p}}=|\alpha_{\mathrm{p}}|^2>0$, and let $S_{\mathrm F}$ be the
repetitions per matrix entry. We write
$\mc{CN}(0,v)$ for a circular complex Gaussian random variable $u$ satisfying
$\mathbb E[u]=0$ and $\mathbb E|u|^2=v$. For transmit mode $i$, the normalized
complex observation in receive mode $j$ on repetition $r$ is
\begin{equation}
\begin{aligned}
    y_{ji}^{(r)}
    &=
    \alpha_{\mathrm{p}}\,a_{\mathrm{eff},ji}
    +\nu_{ji}^{(r)},
    \\
    \nu_{ji}^{(r)}
    &\overset{\mathrm{iid}}{\sim}
    \mc{CN}(0,1),
    \qquad r=1,\ldots,S_{\mathrm{F}}.
\end{aligned}
    \label{eq:field-csi-observation-model}
\end{equation}
The sample-mean estimator is
\begin{equation}
\begin{aligned}
    \widetilde a_{\mathrm{eff},ji}
    &\coloneqq
    \frac{1}{S_{\mathrm{F}}\,\alpha_{\mathrm{p}}}
    \sum_{r=1}^{S_{\mathrm{F}}}\,y_{ji}^{(r)},
    \\
    \widetilde{\pmb{A}}_{\mathrm{eff}}
    &=
    \pmb{A}_{\mathrm{eff}}+\pmb{N}_{\mathrm F},
    \\
    \bar\nu_{ji}
    &=
    [\pmb{N}_{\mathrm F}]_{ji}
    \sim
    \mc{CN}(0,v_{\mathrm{eff},ji}),
    \\
    v_{\mathrm{eff},ji}
    &=
    \frac{1}{S_{\mathrm{F}}\,n_{\mathrm{p}}}.
\end{aligned}
    \label{eq:field-csi-sample-mean}
\end{equation}
Each real quadrature therefore has variance
$1/(2S_{\mathrm{F}}n_{\mathrm{p}})$. Coherent, phase-sensitive measurements
recover the moduli and relative phases of a finite optical transmission
matrix~\cite{PLCFBG:10:PRL,RBFFRW:13:OE}; the receiver and transmit-mode
controller must share a phase reference. Diagonal rephasing of the transmit and
receive bases changes the matrix representation without changing the physical
channel, provided the encoder and decoder use the same transformed bases.
Estimation errors are therefore compared after a common row and column phase
alignment, and transceiver design uses one fixed convention.
The estimation produces $N_{\mathrm t}N_{\mathrm r}S_{\mathrm F}$ scalar
complex observations. If all receive modes are observed in parallel, it uses
$B_{\mathrm F}^{\mathrm{par}}=N_{\mathrm t}S_{\mathrm F}$ optical probes;
serial observation uses
$B_{\mathrm F}^{\mathrm{ser}}=N_{\mathrm t}N_{\mathrm r}S_{\mathrm F}$.
The corresponding mean launched-photon budgets are
$n_{\mathrm p}B_{\mathrm F}^{\mathrm{par}}$ and
$n_{\mathrm p}B_{\mathrm F}^{\mathrm{ser}}$, respectively. Any finite-budget
comparison must declare which resource count is held fixed.

The sample mean $\widetilde{\pmb A}_{\mathrm{eff}}$ need not be passive. Let
$r_{\mathrm A}\coloneqq\min\{N_{\mathrm r},N_{\mathrm t}\}$. Using the
compact singular-value decomposition, let
$\pmb U_{\mathrm{eff}}\in\mathbb C^{N_{\mathrm r}\times r_{\mathrm A}}$ and
$\pmb V_{\mathrm{eff}}\in\mathbb C^{N_{\mathrm t}\times r_{\mathrm A}}$, and write
$\widetilde{\pmb A}_{\mathrm{eff}}
=\pmb U_{\mathrm{eff}}
\operatorname{diag}(\widetilde s_1,\ldots,\widetilde s_{r_{\mathrm A}})
\pmb V_{\mathrm{eff}}^{\dagger}$. The direct detector-plane estimate is
\begin{equation}
    \widehat{\pmb A}_{\mathrm{eff}}^{\mathrm{dir}}
    \coloneqq
    \pmb U_{\mathrm{eff}}
    \operatorname{diag}
    (\min\{\widetilde s_k,1\})_{k=1}^{r_{\mathrm A}}
    \pmb V_{\mathrm{eff}}^{\dagger}.
    \label{eq:detector-plane-passivity-projection}
\end{equation}
This is the Frobenius-nearest contraction and the Gaussian maximum-likelihood
estimator when the detector-plane entry errors are independent with equal
variance.

If a physical transmissivity estimate
$\widehat\tau_j^{\mathrm{rx}}\in[0,1]$ has also been calibrated for every port
and the propagation and receiver factors are to remain separate, let
$\widehat{\pmb{D}}_{\tau}
=\operatorname{diag}(\sqrt{\widehat\tau_1^{\mathrm{rx}}},\ldots,
\sqrt{\widehat\tau_{N_{\mathrm{r}}}^{\mathrm{rx}}})$. Treating
this calibration as fixed, the minimum-norm constrained Gaussian
maximum-likelihood estimator of the pre-receiver matrix is
obtained as follows. For a candidate matrix $\pmb X$, write
$x_{ji}\coloneqq[\pmb X]_{ji}$ and define
\begin{equation}
\begin{aligned}
    \mc J_{\lambda_{\mathrm{reg}}}(\pmb X)
    &\coloneqq
        \sum_{j=1}^{N_{\mathrm{r}}}
        \sum_{i=1}^{N_{\mathrm{t}}}
        \frac{
            |\widetilde a_{\mathrm{eff},ji}-\sqrt{\widehat\tau_j^{\mathrm{rx}}}\,x_{ji}|^2
        }{v_{\mathrm{eff},ji}}
        +\lambda_{\mathrm{reg}}\lVert\pmb X\rVert_{
        \mathrm F}^{2},
        \\
    \widehat{\pmb{A}}
    &\coloneqq
    \lim_{\lambda_{\mathrm{reg}}\downarrow0}
    \underset{\pmb{X}^{\dagger}\pmb{X}\preceq
    \pmb{I}_{N_{\mathrm{t}}}}
    {\operatorname*{arg\,min}}
    \mc J_{\lambda_{\mathrm{reg}}}(\pmb X).
    \label{eq:weighted-passive-field-estimator}
\end{aligned}
\end{equation}
Under the normalized observation model, all $v_{\mathrm{eff},ji}$ are equal, so
the data-fidelity term is proportional to
$\|\widehat{\pmb{D}}_{\tau}\pmb{X}
-\widetilde{\pmb{A}}_{\mathrm{eff}}\|_{\mathrm{F}}^2$; the vanishing
regularizer selects its minimum-norm constrained minimizer. A row with
$\widehat\tau_j^{\mathrm{rx}}=0$ contains no information about the corresponding row of
$\pmb{A}$; the minimum-norm representative sets that raw row to zero. When
$\widehat\tau_j^{\mathrm{rx}}$ is small, direct division by
$\sqrt{\widehat\tau_j^{\mathrm{rx}}}$ is ill
conditioned, whereas \eqref{eq:weighted-passive-field-estimator} remains in the
detector-plane coordinates. If receiver-calibration uncertainty is material,
it must be included in a joint likelihood rather than treated as fixed.

The calibrated estimate used for the spatial channel is
\begin{equation}
    \widehat{\pmb{A}}_{\mathrm{eff}}
    =
    \widehat{\pmb{D}}_{\tau}\widehat{\pmb{A}}.
    \label{eq:estimated-effective-field}
\end{equation}

The fixed receiver channel depends on $\pmb{A}$ and $\pmb{\tau}$ only
through $\pmb{A}_{\mathrm{eff}}$. The numbers of real detector-plane channel
coordinates in a fixed phase convention are therefore
\begin{equation}
\begin{aligned}
    N_{\mathrm{P}}
    &=
    N_{\mathrm{t}}N_{\mathrm{r}}+N_{\mathrm{r}},
    \\
    N_{\mathrm{F}}^{\mathrm{fix}}
    &=
    2N_{\mathrm{t}}N_{\mathrm{r}}+N_{\mathrm{r}},
    \\
    N_{\mathrm{F}}^{\mathrm{quot}}
    &=
    2N_{\mathrm{t}}N_{\mathrm{r}}-N_{\mathrm{t}}+1,
    \label{eq:csi-identifiable-counts}
\end{aligned}
\end{equation}
where the $N_{\mathrm{r}}$ terms account for depolarization. The fixed-convention
field count retains every real and imaginary part of $\pmb A_{\mathrm{eff}}$.
For a matrix whose nonzero support
connects all transmit and receive modes, quotienting the
$N_{\mathrm{t}}+N_{\mathrm{r}}-1$ redundant row and column phases gives the
second field count. Structural zeros or disconnected support reduce the gauge
orbit and must be counted separately. Thus a $2\times2$ link with connected
support uses six real power-\ac{CSI} coordinates, ten field coordinates in a fixed
phase convention, or seven physically distinct field coordinates modulo basis
rephasing. These counts include detector-plane \ac{CSI} and $\pmb{q}$, assume
polarization-independent spatial propagation, and exclude the separately
calibrated $\pmb{\tau}$ and $\pmb{G}$.

Substituting $\widehat{\pmb{A}}_{\mathrm{eff}}$ directly into the passive
dilation, together with $\widehat{\pmb{G}}$ and $\widehat{\pmb{q}}$ in the
subsequent maps, defines the estimated field-\ac{CSI} channel
$\widehat{\mc{N}}_{\mathrm{FSO}}$. The
calibrated factorization
$\widehat{\pmb{A}}_{\mathrm{eff}}
=\widehat{\pmb{D}}_{\tau}\widehat{\pmb{A}}$ gives the same fixed-receiver
channel. Under power \ac{CSI}, the set
$\{\widehat{\pmb{P}},\widehat{\pmb{q}},\widehat{\pmb{G}}\}$ lacks the relative
spatial phases and therefore describes a family of compatible channel maps
rather than a unique $\widehat{\mc{N}}_{\mathrm{FSO}}$. In either case the encoder and decoder
are selected using the estimated channel and evaluated on
\eqref{eq:composite-fso-channel}.

Figure~\ref{fig:csi-estimation} shows how far each estimate lies from the
exact channel as the probe budget grows. Both tiers are shot-noise limited, and
the fitted log-log slopes are $-0.498$ for power \ac{CSI} and $-0.501$ for the
field estimate of $\pmb A_{\mathrm{eff}}$, consistent with the $S^{-1/2}$ laws
predicted by the binomial standard error in~\eqref{eq:power-csi-estimator} and
the heterodyne quadrature variance. Reading the fitted power law at fixed
accuracy quantifies the cost of phase information: a $1\%$ normalized error
needs about $4.7\times10^{3}$ probe shots per transmit rail under power
\ac{CSI}, compared with $2.3\times10^{4}$ pilot shots per transmit mode under
field \ac{CSI}. The factor near five applies under the stated measurement and
resource conventions.

The de-embedded curve is not a pure $S^{-1/2}$ estimator, and the departure is
informative. Dividing out $\pmb\tau^{\mathrm{rx}}$ amplifies detector-plane
noise, yet the de-embedded error is \emph{lower} than the effective-target
error at small budgets, $0.376$ against $0.479$ at ten shots, and the two
coincide at $4.7\times10^{-3}$ by $10^{5}$ shots. Projection onto the set of
contractions explains this difference. At small budgets the unconstrained
estimate frequently violates $\pmb A^{\dagger}\pmb A\preceq\pmb I$, and the
projection reduces variance at the price of bias; once the noise is small the
constraint is inactive and the bias vanishes. Its fitted slope, $-0.479$, is
the furthest from $-1/2$ for this reason. The projection is therefore an
explicit estimation step with a measurable effect, not a formality.

\subsection{Availability}
\label{subsec:csi-availability}

The selected \ac{CSI} description may be available to neither endpoint, to the
Rx alone, to the Tx alone, or to both. Let $\widehat{\mc N}_{\omega}$ denote the
estimated composite channel for realization $\omega$ and let
$\overline{\mc N}_{\mathrm{FSO}}$ denote the ensemble channel
\eqref{eq:ensemble-channel}. For a fixed map at one endpoint, define the
one-sided optima
\begin{equation}
\begin{aligned}
    \mc D^{\star}(\mc E;\mc N)
    &\in\argmax_{\mc D\in\mathfrak D_{\mathsf a}}
    F_{\mathrm{av}}(\mc D\circ\mc N\circ\mc E),
    \\
    \mc E^{\star}(\mc D;\mc N)
    &\in\argmax_{\mc E\in\mathfrak E_{\mathsf a}}
    F_{\mathrm{av}}(\mc D\circ\mc N\circ\mc E),
\end{aligned}
\label{eq:availability-one-sided-optima}
\end{equation}
where $\mathfrak E_{\mathsf a}$ and $\mathfrak D_{\mathsf a}$ are the admissible
encoder and decoder classes of strategy $\mathsf a$. In either one-sided case,
the map at the uninformed endpoint is independent of $\omega$. It may differ
from the separately optimized No-\ac{CSI} map because its offline optimization
includes the adaptive policy used at the informed endpoint. During operation,
the fixed map operates without $\widehat{\mc N}_{\omega}$ or any unobserved
realization-dependent control. Equation~\eqref{eq:availability-one-sided-optima}
gives the informed endpoint's update when the opposite-side map is fixed. The
joint offline optimization of that fixed map and the adaptive policy is
described explicitly for Tx \ac{CSI} below.

The availability comparison carries the fixed \ac{SISO} baseline above
unchanged across all four cases. Hence the availability-dependent maps below
belong only to asymmetric cloning and coherent path superposition.

\paragraph{No \ac{CSI}.}
Neither $\mc E$ nor $\mc D$ of a multirail strategy depends on $\omega$, and
long-term channel statistics may be used once to choose its fixed
encoder--decoder pair.
\begin{itemize}
    \item \emph{Asymmetric cloning.} The clone count $M_0$, support $S_0$, and
    coefficients $\pmb\beta_0\in\mathfrak B_{M_0}$ are fixed on
    $\overline{\mc N}_{\mathrm{FSO}}$. Equal coefficients recover the symmetric
    $1\to M_0$ cloner in \eqref{eq:symmetric-cloner-coefficients}. The Rx uses
    one fixed joint all-port recovery for the resulting ensemble channel.

    \item \emph{Coherent path superposition.} The ordered support $\pmb t_0$
    and normalized path-weight vector $\pmb w_0$ are fixed on
    $\overline{\mc N}_{\mathrm{FSO}}$, together with one joint all-port
    recovery.
\end{itemize}

\paragraph{Rx \ac{CSI}.}
Only the receiver map depends on $\widehat{\mc N}_{\omega}$.
\begin{itemize}
    \item \emph{Asymmetric cloning.} The Tx retains the fixed
    $(M_0,S_0,\pmb\beta_0)$ encoder, while the Rx applies
    $\mc D^{\star}(\mc E_{S_0,\pmb\beta_0}^{\mathrm{cl}};
    \widehat{\mc N}_{\omega})$. Thus the realized channel changes the joint
    recovery while the emitted photon number, support, and clone coefficients
    remain fixed.

    \item \emph{Coherent path superposition.} The Tx retains
    $(\pmb t_0,\pmb w_0)$, while the Rx applies
    $\mc D^{\star}(\mc E_{\pmb w_0}^{\mathrm{coh}};
    \widehat{\mc N}_{\omega})$.
\end{itemize}

\paragraph{Tx \ac{CSI}.}
Only the encoder depends on $\widehat{\mc N}_{\omega}$. The Rx uses a single
decoder $\mc D_{\mathsf a,\mathrm{Tx},0}$ optimized offline for the agreed
adaptive transmitter policy; it receives neither the realized channel estimate
nor the encoder choice for an individual channel use. This decoder generally
differs from the No-\ac{CSI} design. Let
$\Omega_{\mathrm{des}}$ contain the design realizations used in this offline
optimization. Starting from an admissible fixed decoder, iteration $k+1$
computes
\begin{subequations}
\label{eq:tx-csi-offline-design}
\begin{align}
    \mc E_{\mathsf a,k+1}(\widehat{\mc N}_{\omega})
    &\in
    \argmax_{\mc E\in\mathfrak E_{\mathsf a}}\;
    F_{\mathrm{av}}\!\left(
        \mc D_{\mathsf a,\mathrm{Tx},0}^{(k)}
        \circ\widehat{\mc N}_{\omega}\circ\mc E
    \right),
    \nonumber
    \\[-2pt]
    &\hphantom{{}\in{}}
    \qquad\qquad\qquad\qquad\quad
    \omega\in\Omega_{\mathrm{des}},
    \label{eq:tx-csi-encoder-policy-update}
    \\[4pt]
    \mc D_{\mathsf a,\mathrm{Tx},0}^{(k+1)}
    &\in
    \argmax_{\mc D\in\mathfrak D_{\mathsf a}}\;
    \frac{1}{|\Omega_{\mathrm{des}}|}
    \sum_{\omega\in\Omega_{\mathrm{des}}}
    \nonumber
    \\[-2pt]
    &\hphantom{{}\in{}}
    \quad\times
    F_{\mathrm{av}}\!\left(
        \mc D\circ\widehat{\mc N}_{\omega}
        \circ\mc E_{\mathsf a,k+1}(\widehat{\mc N}_{\omega})
    \right).
    \label{eq:tx-csi-common-decoder-update}
\end{align}
\end{subequations}
The same decoder variable appears in every summand of
\eqref{eq:tx-csi-common-decoder-update}. It is fixed before evaluation and
operates independently of $\widehat{\mc N}_{\omega}$. A preset convergence
tolerance and iteration cap determine the stopping iterate, and the
initialization with the largest design objective is retained. The updates are
block-coordinate steps; they do not certify the global optimum of the joint
policy-decoder problem~\cite{RW:05:PRL_2,TKL:10:IEEE_J_IT}.

\begin{itemize}
    \item \emph{Variable-$M$ asymmetric cloning.} The Tx chooses
    \begin{equation}
    \begin{aligned}
        &\bigl(M^{\star},S^{\star},\pmb\beta^{\star}\bigr)(\omega)
        \in
        \\
        &\qquad
        \argmax_{\substack{
            1\leq M\leq N_{\mathrm t},\ |S|=M,\\[1pt]
            \pmb\beta\in\mathfrak B_M}}\;
        F_{\mathrm{av}}\!\left(
            \mc D_{\mathrm{cl},\mathrm{Tx},0}
            \circ\widehat{\mc N}_{\omega}
            \circ\mc E_{S,\pmb\beta}^{\mathrm{cl}}
        \right).
    \end{aligned}
        \label{eq:availability-tx-csi-variable-cloning}
    \end{equation}
    The receiver uses one fixed joint all-port decoder for every $M$, $S$, and
    $\pmb\beta$. For a fixed decoder, each support gives a small quadratic
    coefficient problem; enumerating all nonempty supports gives the complete
    encoder half-step. The family contains the $M=1$ single-photon member and
    every symmetric $1\to M$ encoder.

    \item \emph{Coherent path superposition.} The Tx optimizes the normalized
    field-weight vector $\pmb w^{\star}(\omega)$ against one fixed all-port
    decoder. Under the deterministic recovery in
    \eqref{eq:weighted-coherent-path-fidelity}, the weight update is the
    principal-eigenvector rule stated there.
\end{itemize}

\paragraph{Full \ac{CSI}.}
Both endpoints use $\widehat{\mc N}_{\omega}$, and both maps adapt to the
realization. They use the same estimate and deterministic tie-breaking, so
both endpoints derive the same selected mode without a separate mode label.
\begin{itemize}
    \item \emph{Variable-$M$ asymmetric cloning.} With the decoder fixed, the
    Tx enumerates every nonempty support and solves the exact coefficient
    half-step over $\mathfrak B_M$. Before alternating, every nonempty support
    is assigned its equal-$\beta$ encoder and an independently optimized
    deterministic all-port recovery. The best of these initial designs
    initializes the alternating encoder--decoder updates. A
    coefficient boundary remains in its $M$-photon sector. The
    block-coordinate sequence is monotone after physical rescoring, but it does
    not certify the joint global optimum.

    \item \emph{Coherent path superposition.} The Tx adapts
    $\pmb w^{\star}(\omega)$ through \eqref{eq:weighted-coherent-path-matrix},
    and the Rx adapts the all-port recovery.
\end{itemize}

Let $s$ index the channel operating point and $\mathsf a$ the strategy, and let
an overline denote the mean over the common realizations. The common
reliability target is the largest deterministic Full \ac{CSI} mean over the
strategies,
\begin{equation}
    F_{\mathrm{rel}}(s)
    \coloneqq
    \max_{\mathsf a}\overline F_{\mathsf a,\mathrm{Full}}^{\mathrm{det}}(s).
    \label{eq:availability-reliability-target}
\end{equation}
For availability case $\mathsf v$, let
$\overline p_{\mathrm{s},\mathsf a,\mathsf v}(\mc E,\mc D_{\mathrm s};s)$ and
$\overline f_{\mathrm{s},\mathsf a,\mathsf v}(\mc E,\mc D_{\mathrm s};s)$ be
the ensemble success probability and success-weighted fidelity. For a feasible
target, the optimum success probability solves
\begin{equation}
\begin{aligned}
    p_{\mathrm{s},\mathsf a,\mathsf v}^{\star}(s)
    &\coloneqq
    \max_{(\mc E,\mc D_{\mathrm{s}})
    \in\mathfrak H_{\mathsf a,\mathsf v}}
    \overline p_{\mathrm{s},\mathsf a,\mathsf v}(\mc E,\mc D_{\mathrm s};s),
    \\
    \operatorname{subject\ to}\quad
    &\overline f_{\mathrm{s},\mathsf a,\mathsf v}(\mc E,\mc D_{\mathrm s};s)
    \geq
    \bigl[F_{\mathrm{rel}}(s)-\epsilon_F\bigr]
    \overline p_{\mathrm{s},\mathsf a,\mathsf v}(\mc E,\mc D_{\mathrm s};s),
    \\
    &\mc D_{\mathrm{s}}\ \text{completely positive},
    \\
    &\mc D_{\mathrm{s}}\ \text{trace nonincreasing}.
\end{aligned}
\label{eq:availability-reliability-success-probability}
\end{equation}
Here $\mathfrak H_{\mathsf a,\mathsf v}$ is the set of pairs admitted by
strategy $\mathsf a$ under availability case $\mathsf v$, including the fixed
opposite-side map of each one-sided case, and $\epsilon_F$ is the fidelity
tolerance. The linear fidelity constraint preserves the decoder problem as an
\ac{SDP}. Let $(\mc E^{\star},\mc D_{\mathrm s}^{\star})$ be a maximizing pair
for a feasible target and define
\begin{equation}
    G_{\mathsf a,\mathsf v}^{\star}(s)
    \coloneqq
    \overline f_{\mathrm{s},\mathsf a,\mathsf v}
    (\mc E^{\star},\mc D_{\mathrm s}^{\star};s).
    \label{eq:availability-reliability-weighted-fidelity}
\end{equation}
For an infeasible target, set
$p_{\mathrm{s},\mathsf a,\mathsf v}^{\star}
=G_{\mathsf a,\mathsf v}^{\star}=0$. The common target compares the
success-weighted fidelity of the strategies at a common conditional-fidelity
threshold. Here $p_{\mathrm{s},\mathsf a,\mathsf v}^{\star}$ is the heralding
success probability of the recovery instrument.

\section{Numerical Results}
\label{sec:numerical-results}

\subsection{Simulation Setup}
\label{subsec:simulation-setup}

For the numerical study, $\omega$ indexes a transfer-matrix realization drawn
from the distribution induced by the link geometry and turbulence strength
$\sigma_R^2$. Each operating point uses $R_{\omega}$ independent atmospheric
realizations. Within each comparison, every communication strategy is evaluated
on the same realizations.

The model includes only the photon-number sectors in
$\mathsf N_{\mathrm{src}}$ and assumes no background photons or dark counts. A
collision produces an erasure under the squashing receiver $\mc{S}$ but may be
resolved by the mode-resolved decoder $\mc{D}_{\mathrm{mr},\pmb{q}}^{(\varpi)}$.
Each receive port carries its own label, so a photon arriving at a port other
than the one addressed by its transmit rail counts as crosstalk, and it
contributes to the logical output only when the recovery map reads that port.

\begin{table}[!t]
    \caption{Channel-model parameters.}
    \label{tab:numerical-realization-parameters}
    \centering
    \small
    \renewcommand{\arraystretch}{1.15}
    \begin{tabular}{@{}>{\raggedright\arraybackslash}p{0.52\columnwidth}
                      >{\raggedright\arraybackslash}p{0.36\columnwidth}@{}}
        \toprule
        \rowcolor{rowblue}
        Parameter (Symbol) & Value
        \\
        \midrule
        Spatial rails (\mbox{$N_{\mathrm{t}}=N_{\mathrm{r}}$})
        & $2$ to $5$
        \\
        Optical wavelength ($\lambda_0$)
        & $809\,\mathrm{nm}$
        \\
        Propagation distance ($z$)
        & $1000\,\mathrm{m}$
        \\
        Gaussian beam waist ($w_0$)
        & $20\,\mathrm{mm}$
        \\
        Geometry pairs
        $(s_{\mathrm{rail}}/w(z),x_{\mathrm p}/w(z))$
        & $(1.5,0)$, $(2.5,0)$, $(3.5,0)$, $(2.5,0.5)$, $(2.5,1)$
        \\
        Transverse sampling (\mbox{$N_{\mathrm{grid}},L_{\mathrm{win}}$})
        & $512$ points per axis; $0.30\,\mathrm{m}$ square window
        \\
        Phase screens ($N_{\mathrm{scr}}$)
        & $34$
        \\
        Multiscale-screen oversampling
        & $2$
        \\
        Power-extinction coefficient ($\alpha_{\mathrm{ext}}$)
        & $0$
        \\
        Receive aperture ($\mc A_{\mathrm{rx}}$)
        & $\mathbb R^2$ (no finite aperture mask)
        \\
        Outer and inner scales
        (\mbox{$\ell_{\mathrm{out}},\ell_{\mathrm{in}}$})
        & $80\,\mathrm{m}$, $1\,\mathrm{mm}$
        \\
        Rytov variance ($\sigma_R^2$)
        & $0.02$, $0.05$, $0.10$, $0.20$, $0.50$,
          \newline $1$, $3$, $10$, $20$, $50$
        \\
        Receiver transmissivity ($\tau_j^{\mathrm{rx}}$)
        & $0.92$ for every $j$
        \\
        Depolarization probability ($q_j$)
        & $0.01$ for every $j$
        \\
        Realizations per point ($R_{\omega}$)
        & $256$ for the principal sweeps; otherwise stated with each comparison
        \\
        Baseline internal states ($\pmb G$)
        & $\pmb 1_{N_{\mathrm t}}\pmb 1_{N_{\mathrm t}}^{\dagger}$
        ($\zeta=0$)
        \\
        Transmit and receive mode basis
        & Displaced Gaussians, symmetrically orthonormalized
        \\
        \bottomrule
    \end{tabular}
\end{table}

See Appendix~\ref{app:phase-screens} for the stochastic phase-screen
construction, the screen-count criterion, the finite-grid assumptions used in
the simulation, and the transverse-grid validation of the descriptors shown
below.

The channel realizations are generated with HCIPy version
0.7.0~\cite{PHRDKB:18:SPIE}. We use HCIPy because its sampled transverse grids,
multiscale phase-screen generator, and Fresnel propagator implement the
stochastic paraxial model within a common wave-optics framework. For each
realization, the propagated field realizes $\mc K_{\mathrm{prop}}(\omega)$.
Projection onto the transmit and receive modes gives $\pmb A(\omega)$ through
\eqref{eq:field-transfer}; applying the receiver attenuation gives
$\pmb A_{\mathrm{eff}}(\omega)$ and $\pmb P(\omega)$ through
\eqref{eq:effective-field-transfer} and \eqref{eq:power-csi}. The fixed
receiver-side attenuation and effective per-port polarization noise are
composed after the sampled atmospheric spatial transformation through
$\tau_j^{\mathrm{rx}}$ and $q_j$, respectively, so every receiver setting uses
the same atmospheric realization.

We use a common simulation seed across rail counts, geometry settings, and
Rytov values, so comparisons across operating points are paired; the $R_{\omega}$
realizations within each operating point remain independent.
Channel-descriptor statistics use the exact sampled $\pmb A_{\mathrm{eff}}$,
equivalently $\pmb P$. Full strategy-channel evaluations additionally use the
exact $\pmb q$ and $\pmb G$; receiver loss is already contained in
$\pmb A_{\mathrm{eff}}$ and is separated as $\pmb\tau$ only for a
factorized ablation. Finite-shot \ac{CSI} evaluations state the probe counts
$S_{\mathrm P}$, $S_{\mathrm F}$, $n_{\mathrm p}$, $S_{\mathrm Q}$, and
$S_{\mathrm{HOM}}$, and whether field estimation is serial or parallel. The
baseline values of
$\pmb\tau$, $\pmb q$, and $\pmb G$ are treated as exact calibrations; their
calibration cost is excluded unless stated otherwise.
Finite field-CSI results that operate directly at the detector plane use
\eqref{eq:detector-plane-passivity-projection}; results that separate
propagation from receiver loss use
\eqref{eq:weighted-passive-field-estimator} and
\eqref{eq:estimated-effective-field}.

The computational grid has $N_{\mathrm{grid}}$ points on each axis and a
square side length $L_{\mathrm{win}}$. The transmit and receive modes
are displaced Gaussians with the same
$1/e$ field-amplitude radius $w_0$ and zero phase curvature at their respective
planes. For $N$ rails, both endpoint arrays lie on the $x$ axis at
\begin{equation}
    (x_i,y_i)
    =
    \left(\left[i-\frac{N+1}{2}\right]s_{\mathrm{rail}},0\right),
    \qquad i=1,\ldots,N .
    \label{eq:rail-positions}
\end{equation}
The deterministic pointing setting $x_{\mathrm p}$ is applied along $+x$ and
corresponds to $\pmb\theta=(x_{\mathrm p}/z,0)^{\mathsf T}$; no random pointing
jitter is added. Each mode set is orthonormalized separately. Thus, $\pmb{A}$ includes
mismatch to the receive projection basis as well as atmospheric modal
scattering. The rail separation is comparable to the propagated spot size, so
the raw beam sets are not orthogonal.
For these parameters $z_R=1.553\,\mathrm{km}$ and $z/z_R=0.644$, so the link is
not in the far field and $w(z)=23.79\,\mathrm{mm}$ rather than the
$12.88\,\mathrm{mm}$ far-field value. Rail separations are therefore quoted as
multiples of $w(z)$ rather than in absolute units.
The numerical study uses weighted symmetric (L\"owdin) orthonormalization with
minimum-eigenvalue tolerance $10^{-12}$ on the normalized sampled Gram matrix,
which returns
the orthonormal set closest to the original in the least-squares sense and
commutes with a permutation of the input
modes~\cite{L:50:JCP,CB:02:JMC}. Sequential Gram-Schmidt orthogonalization is
avoided because it leaves the first mode unchanged and distorts the remainder,
imposing an ordering on rails that the geometry does not distinguish.

Split-step phase-screen
propagation reproduces the turbulence spectrum only over the spatial
frequencies supported by the transverse window, so the lowest-frequency
content, which contributes most of the beam wander, is attenuated relative to
the continuous spectrum. The accuracy of a screen realization consequently
depends on the window, the sampling, and the treatment of subharmonic
components~\cite{KS:23:PRA}. Absolute collected powers obtained under this
discretization may therefore be optimistic. Grid and window refinement is
required before assigning quantitative outer-scale effects or claiming a
discretization-independent strategy ranking.

\begin{figure}[!t]
    \centering
    \includegraphics[width=0.92\columnwidth]
        {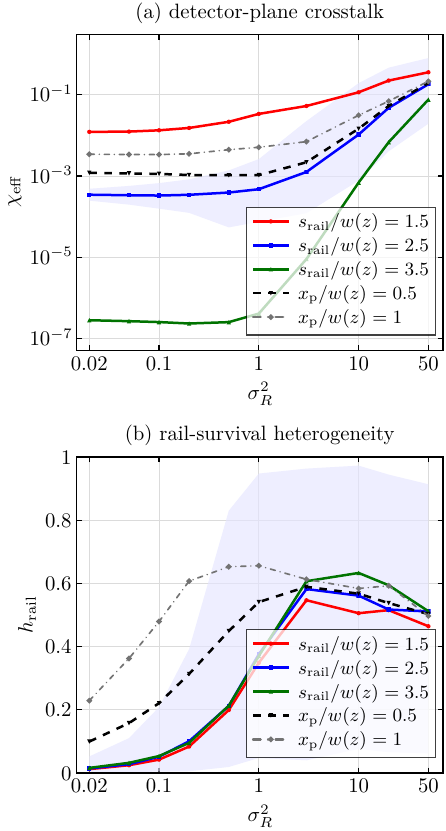}
    \caption{Detector-plane crosstalk $\chi_{\mathrm{eff}}$ (a) and normalized
    rail-survival heterogeneity $h_{\mathrm{rail}}$ (b) of the $2\times2$
    channel against turbulence strength, for rail spacings $s_{\mathrm{rail}}/w(z)=1.5$, $2.5$
    and $3.5$ at zero pointing and pointing offsets
    $x_{\mathrm p}/w(z)=0.5$ and $1$ at spacing $2.5$. Curves are medians over
    the $256$ atmospheric realizations at each operating point; the shaded band is
    $5$th to $95$th percentile range at spacing $2.5$. Panel (a) has
    logarithmic axes.}
    \label{fig:n2-descriptors}
\end{figure}

\begin{figure}[!t]
    \centering
    \includegraphics[width=0.86\columnwidth]
        {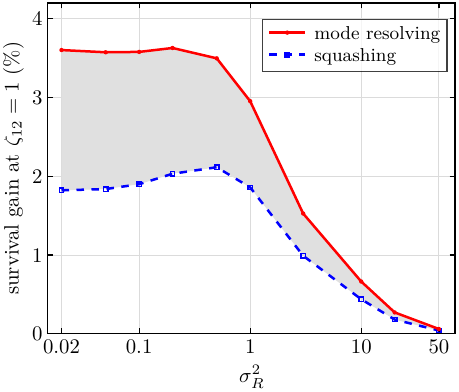}
    \caption{Relative gain in the mean two-photon survival probability at
    distinguishability $\zeta_{12}=1$ over its value at $\zeta_{12}=0$, against turbulence
    strength, under the squashing receiver and the mode-resolving receiver.
    A symmetric $1\to2$ cloner occupies both transmit rails of the $2\times2$
    channel at spacing $s_{\mathrm{rail}}/w(z)=1.5$ and zero pointing, with $256$ atmospheric
    realizations. The shaded region is the difference between the two
    receivers.}
    \label{fig:n2-survival-pair}
\end{figure}

\begin{figure}[!t]
    \centering
    \includegraphics[width=0.92\columnwidth]
        {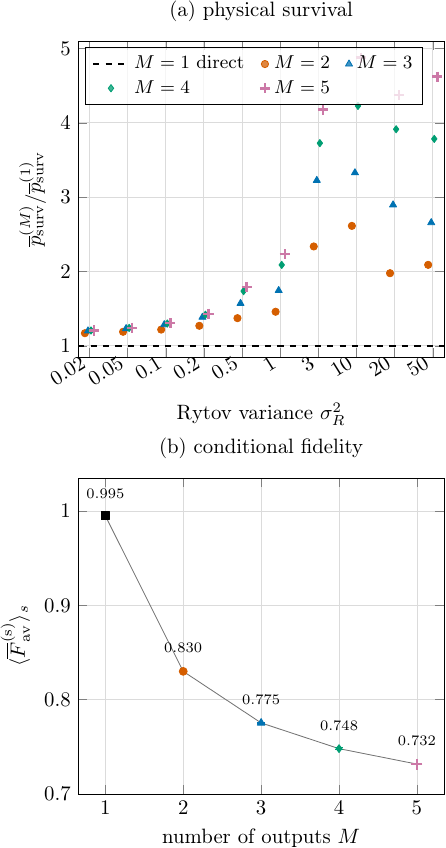}
    \caption{Symmetric-cloning dilution under Rx \ac{CSI} on the five-rail
    channel. All $256$ realizations select the rail placements; plotted means
    use $16$ of them, making this an in-sample comparison. The $M=1$ baseline is direct
    transmission; $M=2,\ldots,5$ use a symmetric $1\to M$ cloner and
    realization-adapted all-port recovery. Panel (a) shows the mean probability
    that at least one port is non-erased relative to direct transmission.
    Panel (b) gives mean survival-conditioned fidelity across the ten Rytov points.}
    \label{fig:cloning-dilution}
\end{figure}

Each realization therefore yields the detector-plane field matrix
$\pmb{A}_{\mathrm{eff}}(\omega)$ and the power matrix $\pmb{P}(\omega)$ defined
in \eqref{eq:power-csi}. Direct matrix comparisons become difficult as the
number of rails grows. We use normalized descriptors for mean single-photon
survival, detector-plane crosstalk, and rail-survival heterogeneity. These
quantities summarize the channel features that govern the strategy comparisons;
they do not uniquely determine the quantum channel.

\begin{figure*}[!t]
    \centering
    \includegraphics[width=0.96\textwidth]
        {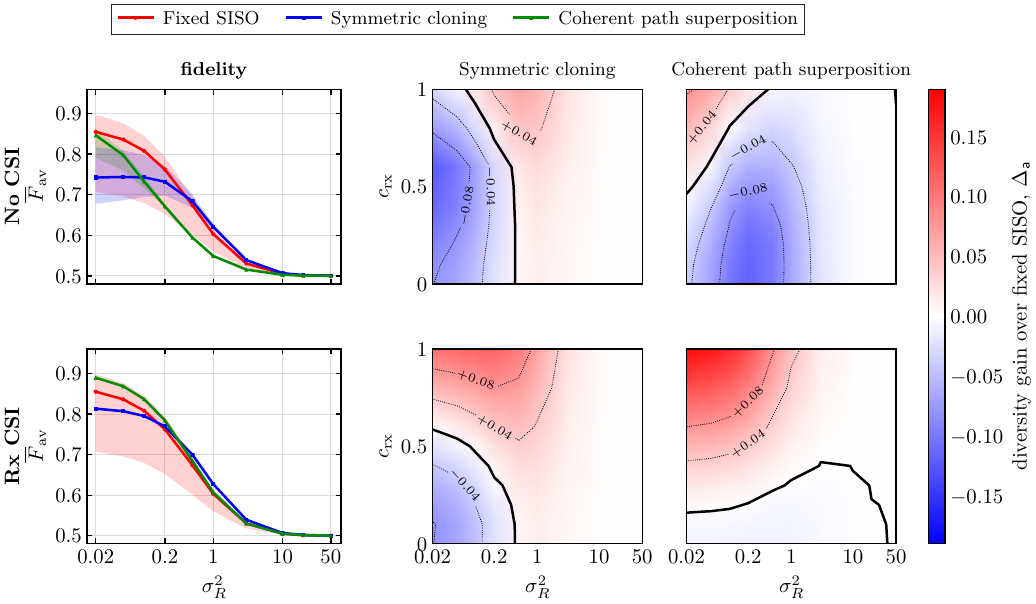}
    \caption{Held-out mean fidelity and diversity gain over the fixed \ac{SISO}
    baseline on the $2\times2$ channel without receiver \ac{CSI} (top row) and
    with receiver \ac{CSI} (bottom row), for symmetric $1\to2$ cloning and
    equal-weight coherent path superposition. The left panel in each row shows the
    median, over $c_{\mathrm{rx}}$, of the held-out mean fidelity; its band spans the full
    sampled range and therefore measures sensitivity to receive-mode
    scrambling rather than statistical uncertainty. The maps show the
    per-setting diversity gain $\Delta_{\mathsf a}$. Red denotes an advantage
    over the baseline, the bold contour marks $\Delta_{\mathsf a}=0$, and the
    dotted contours mark $\Delta_{\mathsf a}=\pm0.04$ and $\pm0.08$. Colour is
    interpolated between the
    $10\times9$ evaluated settings.}
    \label{fig:controlled-rx-csi-mesh}
\end{figure*}

For a photon launched in transmit rail $i$, the detector-plane survival and
loss probabilities are
\begin{align}
    \eta_i^{\mathrm{eff}}(\omega)
    &\coloneqq
    \sum_{j=1}^{N_{\mathrm{r}}}p_{ji}(\omega)
    =
    \sum_{j=1}^{N_{\mathrm{r}}}
    \tau_j^{\mathrm{rx}}\eta_{ji}(\omega)
    \\[2pt]
    &\phantom{{}\coloneqq{}}
    =
    \left\lVert
        \pmb{A}_{\mathrm{eff}}(\omega)\pmb{e}_i
    \right\rVert^2 ,
    \\[6pt]
    p_i^{\mathrm{loss}}(\omega)
    &\coloneqq
    1-\eta_i^{\mathrm{eff}}(\omega)
    \\[2pt]
    &\phantom{{}\coloneqq{}}
    =
    p_i^{\mathrm{leak}}(\omega)
    +
    \sum_{j=1}^{N_{\mathrm{r}}}
    (1-\tau_j^{\mathrm{rx}})\eta_{ji}(\omega).
\end{align}
Consequently, $p_i^{\mathrm{loss}}(\omega)\geq
p_i^{\mathrm{leak}}(\omega)$.
The total loss $p_i^{\mathrm{loss}}$ includes selected-mode leakage and
receiver-side loss, but not multiphoton collision erasure. Averaging over the
transmit rails gives
\begin{equation}
    \bar{\eta}_{\mathrm{eff}}(\omega)
    \coloneqq
    \frac{1}{N_{\mathrm{t}}}
    \sum_{i=1}^{N_{\mathrm{t}}}\eta_i^{\mathrm{eff}}(\omega)
    =
    \frac{\left\lVert\pmb{A}_{\mathrm{eff}}(\omega)\right\rVert_{\mathrm{F}}^2}
         {N_{\mathrm{t}}} .
    \label{eq:mean-detector-plane-survival}
\end{equation}
Thus $\bar{\eta}_{\mathrm{eff}}\in[0,1]$ measures the mean detector-plane
single-photon survival. It is not the arithmetic mean of the receiver-side
transmissivities $\tau_j^{\mathrm{rx}}$ because it also includes the loss present in
$\pmb{A}(\omega)$. For a completely dark realization,
$\pmb{A}_{\mathrm{eff}}=\pmb{0}$ and
$\bar{\eta}_{\mathrm{eff}}=0$. The remaining normalized descriptors are
undefined in this boundary case because their denominators vanish.

For paired transmit and receive bases with
$N_{\mathrm{t}}=N_{\mathrm{r}}=N$, the detector-plane crosstalk fraction is
\begin{equation}
    \chi_{\mathrm{eff}}(\omega)
    \coloneqq
    \frac{
        \displaystyle
        \sum_{\substack{i,j=1\\i\neq j}}^{N}p_{ji}(\omega)
    }{
        \displaystyle
        \sum_{i,j=1}^{N}p_{ji}(\omega)
    } .
    \label{eq:detector-plane-crosstalk}
\end{equation}
For nonzero detector-plane power, $\chi_{\mathrm{eff}}$ is the fraction that
arrives at noncorresponding receive ports. Under a uniform input prior, this
global ratio weights each input by its survival probability and therefore
differs from an equal-input average of conditional per-column crosstalk.
Coherent port mixing can in principle be processed before detection, whereas
lost amplitude cannot be deterministically
recovered~\cite{RZBB:94:PRL,CHMKW:16:Optica,TRS:18:PRX}. All crosstalk statistics shown use $\chi_{\mathrm{eff}}$.

The normalized rail-survival heterogeneity is
\begin{equation}
    h_{\mathrm{rail}}(\omega)
    \coloneqq
    \sqrt{
        \frac{
            \sum_{i=1}^{N_{\mathrm{t}}}
            \left[\eta_i^{\mathrm{eff}}(\omega)
            -\bar{\eta}_{\mathrm{eff}}(\omega)\right]^2
        }{
            N_{\mathrm{t}}\left(N_{\mathrm{t}}-1\right)
            \bar{\eta}_{\mathrm{eff}}^{\,2}(\omega)
        }
    } .
    \label{eq:rail-heterogeneity}
\end{equation}
For $N_{\mathrm{t}}\geq2$ and $\bar{\eta}_{\mathrm{eff}}>0$,
$h_{\mathrm{rail}}\in[0,1]$. Equal survival on all rails gives
$h_{\mathrm{rail}}=0$, while survival confined to one rail gives
$h_{\mathrm{rail}}=1$. The descriptor is invariant under common attenuation
$\eta_i^{\mathrm{eff}}\mapsto c\eta_i^{\mathrm{eff}}$.

Mean survival, crosstalk, and rail heterogeneity depend only on $\pmb{P}$ and are
available from power \ac{CSI}. Turbulence changes both the
off-diagonal coefficients $p_{ji}$ and the per-input survival probabilities
$\eta_i^{\mathrm{eff}}$. It can therefore produce realization-dependent
crosstalk and unequal rail survival. A fixed transmit rail, receive port, or
coherent path-weight vector cannot track those changes, which motivates the
\ac{CSI}-adaptive methods evaluated below. The descriptors are analysis
coordinates rather than substitutes for the full power or field matrix used by
the transceiver. Because turbulence, pointing, rail geometry, and receiver
attenuation can change several descriptors simultaneously, comparisons at fixed
survival, crosstalk, or heterogeneity require matched realizations,
stratification, or a channel family constructed so that the remaining
descriptors are held fixed.

Each descriptor is characterized over the ensemble by its sample mean, median, and
$5$th and $95$th percentiles, computed from the realization-level values. Strategy
performance is summarized with the same statistics, and its sample mean estimates
the expectation in \eqref{eq:design-objective}.

On the $2\times2$ channel, rail spacing primarily controls crosstalk, whereas
pointing primarily controls rail imbalance under weak turbulence in the
power-\ac{CSI} descriptors of Fig.~\ref{fig:n2-descriptors}. Neither descriptor
is strictly monotone across every geometry, but both follow a clear overall
turbulence trend. At
$s_{\mathrm{rail}}/w(z)=3.5$, the weak-turbulence crosstalk median is
$2.8\times10^{-7}$ and reaches $7.6\times10^{-2}$ at $\sigma_R^2=50$; the
tightest spacing begins about $4.6$ decades higher. Pointing also raises
crosstalk at fixed spacing. The three zero-pointing heterogeneity curves remain
close under weak turbulence, while pointing raises the imbalance to
$h_{\mathrm{rail}}=0.23$ at $\sigma_R^2=0.02$, compared with $0.02$ at zero
pointing. Geometry differences narrow for $\sigma_R^2\gtrsim3$, and the
medians lie near $1/2$ at $\sigma_R^2=50$. Because this value equals
$\mathbb E[h_{\mathrm{rail}}]$ for two independent exponentially distributed
rail powers, approaching it is consistent with strong rail decorrelation
rather than a normalization ceiling. The two descriptors therefore separate
geometric coupling from rail imbalance. Crosstalk is more sensitive to spacing
than heterogeneity, so both descriptors are needed to compare geometries and
explain differences among strategies.

Internal-state distinguishability provides a few-percent survival benefit on
this link, with a shallow maximum under weak-to-moderate turbulence, as shown
in Fig.~\ref{fig:n2-survival-pair}. Let
$\kappa_{12}\coloneqq\sum_j\lvert a_{j1}\rvert^2\lvert a_{j2}\rvert^2$ denote the
same-port overlap of the two transmit columns. Bosonic statistics give the
collision probability $(2-\zeta_{12})\kappa_{12}$, so the survival gain is
linear in distinguishability and proportional to this overlap. For balanced,
approximately symmetric channels with low crosstalk,
$\kappa_{12}\simeq2\chi_{\mathrm{eff}}\bar\eta_{\mathrm{eff}}^2$; hence,
distinguishability affects survival only when spatial mixing and photon
survival are both appreciable. Increasing turbulence strengthens the collision
contribution through modal mixing but also removes photon pairs through loss.
This competition produces the shallow maximum, followed by a decay beyond
$\sigma_R^2\approx1$. At $\sigma_R^2=50$, both receiver gains are below
$0.06\%$. The common baseline at $\zeta_{12}=0$ falls from $0.930$ to
$0.0012$, while changing $\zeta_{12}$ modifies survival by only a few percent.
The mode-resolving receiver recovers collisions whose photons occupy distinct
internal modes. Its maximum gain is $3.63\%$ at $\sigma_R^2=0.2$, compared
with $2.11\%$ at $\sigma_R^2=0.5$ for the squashing receiver, and the ratio
between their gains falls from $1.98$ at the weakest turbulence to $1.48$ at
the strongest.

The descriptor trends identify three operating regimes for the strategy
comparisons. When several rails provide distinct survival opportunities,
phase-insensitive cloning trades clone fidelity for higher survival without
requiring field phase information. With stable coherent coupling and accurate
field \ac{CSI}, single-photon coherent path weights concentrate amplitude at
the receiver without a cloning penalty. When one rail and receive port
dominate, direct transmission avoids both the cloning penalty and the
additional path parameters. The comparisons below test these regimes without
assuming a fixed ordering among the strategies.

\begin{figure*}[!t]
    \centering
    \includegraphics[width=0.98\textwidth]
        {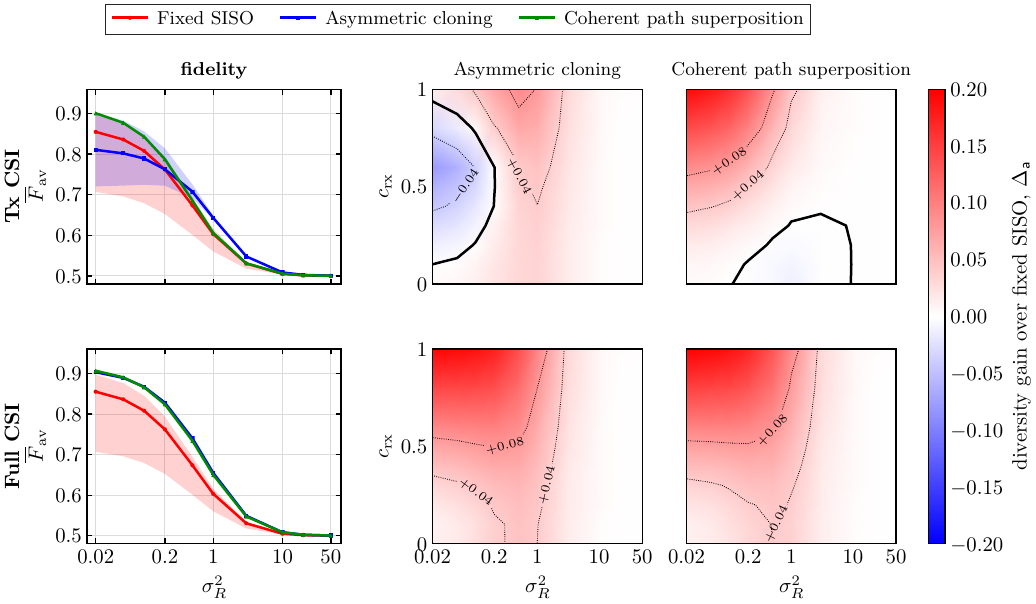}
    \caption{Fidelity and diversity gain over the fixed \ac{SISO} baseline on the
    $2\times2$ channel under Tx \ac{CSI} (top row) and Full \ac{CSI} (bottom
    row), for variable-$M$ asymmetric cloning and optimized coherent path
    superposition. Tx \ac{CSI} adapts only the encoder; Full \ac{CSI} also
    adapts all-port recovery. Curves show median held-out fidelity over
    $c_{\mathrm{rx}}$; bands give its range. Interpolation is
    display-only: red and blue indicate the gain sign; bold contours show zero; dotted
    levels are $\pm0.04$ and $\pm0.08$.}
    \label{fig:controlled-tx-full-csi-mesh}
\end{figure*}

\subsection{Strategy Metrics and Plot Conventions}
\label{subsec:strategy-plot-conventions}

All comparisons use the Haar-averaged fidelity defined in
Section~\ref{subsec:system-model-strategies}. The fixed \ac{SISO} baseline is
the non-adaptive configuration $(i_0,j_0)=(1,1)$: it launches transmit mode~1,
reads receive mode~1, and uses no rail selection, beamforming, receiver
rotation, or realization-dependent decoder update. The same realization-paired
$\overline F_{\mathrm{SISO}}$ is used when a strategy is compared across
\ac{CSI}-availability cases. By contrast, direct transmission under Full
\ac{CSI} enumerates the $N_{\mathrm t}N_{\mathrm r}$ rail--port pairs and
selects the best pair for the realized channel. This distinction separates the
benefit of multirail processing and \ac{CSI} adaptation from the fixed
single-mode baseline.

For the cloning-dilution study, let
$\pmb\rho_{\mathrm{o}}^{(M)}(\omega;s)$ be the output of the squashing map for
an $M$-output encoder at operating point $s$, before logical recovery.
Using the erasure projector already defined in
Section~\ref{subsubsec:receiver-maps}, the physical survival probability is
\begin{equation}
    p_{\mathrm{surv}}^{(M)}(\omega;s)
    \coloneqq
    1-
    \Tr\!\left[
        \pmb\Pi_{\mathrm e}^{\otimes N_{\mathrm r}}
        \pmb\rho_{\mathrm{o}}^{(M)}(\omega;s)
    \right].
    \label{eq:cloning-survival-probability}
\end{equation}
It is the probability that at least one receive port carries a non-erasure
polarization output after loss and receiver squashing. Its ensemble mean is
\begin{equation}
    \overline p_{\mathrm{surv}}^{(M)}(s)
    \coloneqq
    \mathbb E_{\omega}\!\left[p_{\mathrm{surv}}^{(M)}(\omega;s)\right].
    \label{eq:mean-cloning-survival-probability}
\end{equation}
The survival gain of an $M$-output encoder over direct transmission is the ratio
$\overline p_{\mathrm{surv}}^{(M)}(s)/\overline p_{\mathrm{surv}}^{(1)}(s)$, which
equals one at $M=1$.
The quantity $p_{\mathrm{surv}}^{(M)}$ is also distinct from the decoder success
probability $\overline p_{\mathrm s}$: the deterministic recovery used here is
completed to a \ac{CPTP} map and therefore has unit success trace. See
Appendix~\ref{app:heralded} for the decoder success definition. Conditioning on
the physical survival event gives the survival-conditioned fidelity, denoted by
$F_{\mathrm{av}}^{(\mathrm{surv})}$.

Cloning trades survival-conditioned fidelity for higher survival at every
turbulence strength sampled in Fig.~\ref{fig:cloning-dilution}. Increasing $M$
raises the survival gain but lowers the fidelity of each surviving
transmission. At $\sigma_R^2\leq0.2$, the $M=5$ gain lies between $1.21$ and
$1.43$, while the survival-conditioned fidelity falls from $0.995$ for direct
transmission to about $0.732$. At $\sigma_R^2=50$, the survival gain reaches
$4.62$, and the fidelity cost persists.

The mode-resolving receiver has a physical selection event, distinct from
\eqref{eq:cloning-survival-probability}: it accepts every configuration the
squashing receiver accepts, and in addition those spatial collisions whose
photons occupy distinct internal modes. Its selection probability therefore
dominates $p_{\mathrm{surv}}^{(M)}$, with equality when all photons share one
internal mode, and the difference between the two is what
Figure~\ref{fig:n2-survival-pair} measures. See
Appendix~\ref{app:receiver-map-realizations} for the resolvable set and its
projector.

\subsection{CSI-Adaptive Diversity on the $2\times2$ Channel}
\label{subsec:controlled-rx-csi-gain}

We next isolate how \ac{CSI} converts the two spatial paths into diversity. We use
\eqref{eq:controlled-composite-fso-channel} with
$N_{\mathrm t}=N_{\mathrm r}=2$,
$s_{\mathrm{rail}}/w(z)=2.5$, and zero pointing. The turbulence strength spans
the ten Rytov variances in
Table~\ref{tab:numerical-realization-parameters}, and the receive-mode
scrambling parameter takes
$c_{\mathrm{rx}}\in\{0,0.1,0.2,0.3,0.4,0.5,0.6,0.8,1\}$. Hence each
availability case contains $10\times9=90$ operating points. The receiver
transmissivity is $\tau_j^{\mathrm{rx}}=0.92$, the depolarization probability
is $q_j=0.01$, and the baseline internal states are indistinguishable
($\zeta=0$). At each setting, $64$ realizations characterize the channel
ensemble and provide the statistical \ac{CSI} used to determine any encoder or
decoder that does not depend on instantaneous \ac{CSI}. Performance is
evaluated over a disjoint set of $192$ realizations, each representing an
independent instantaneous channel state. For each evaluation realization, an
endpoint with \ac{CSI} adapts its map to the realized channel, whereas an
endpoint without \ac{CSI} uses the map determined from the channel ensemble. Common random numbers pair all strategy comparisons.

The No- and Rx-\ac{CSI} comparisons use two realization-independent transmit
encoders: symmetric $1\to2$ cloning and the fixed equal-weight coherent path
superposition $\pmb w=(1,1)^{\mathsf T}/\sqrt{2}$. At the Rx, both strategies
are restricted to the same $114$-element codebook of passive two-mode
transformations followed by the same downstream recovery map. In the No-\ac{CSI}
case, for each strategy and operating point, the codebook element with the
largest mean fidelity over the $64$ ensemble realizations is used for all $192$
evaluation realizations. The receiver therefore uses long-term channel
statistics but no instantaneous \ac{CSI}. In the Rx-\ac{CSI} case, the receiver
uses the instantaneous \ac{CSI} of each evaluation realization to select the
codebook element with the largest fidelity for that realized channel. Using the
same codebook and downstream recovery in both cases isolates the benefit of
realization-dependent receiver selection.

The Tx- and Full-\ac{CSI} comparisons use variable-$M$ asymmetric cloning and
optimized coherent path superposition. Under Tx \ac{CSI}, the encoder is chosen
from the instantaneous \ac{CSI} of each realized channel, while one all-port
recovery determined from the $64$ ensemble realizations is used for all $192$
evaluation realizations. Under Full \ac{CSI}, the encoder and the recovery are
both adapted to each instantaneous channel state. In every case, the fixed
\ac{SISO} baseline launches mode~1, reads receive mode~1, and uses no
realization-dependent adaptation.

Throughout, $s$ is the channel operating point and $\mathsf a$ the strategy, as in
Subsection~\ref{subsec:csi-availability}. We define the fidelity benefit
obtained from the two spatial paths as
\begin{equation}
    \Delta_{\mathsf a}(s)
    \coloneqq
    \overline F_{\mathsf a}(s)-\overline F_{\mathrm{SISO}}(s).
    \label{eq:strategy-margin}
\end{equation}
We refer to $\Delta_{\mathsf a}$ as the diversity gain over the fixed single-mode
baseline. A positive value means that spatial diversity and the available \ac{CSI}
more than compensate for the encoder's physical cost; a negative value means
that the fixed single-mode transmission remains preferable at that channel
setting.

Rx \ac{CSI} gives its largest fidelity improvement under maximum scrambling
and weak turbulence. In Fig.~\ref{fig:controlled-rx-csi-mesh}, it increases the
symmetric-cloning fidelity by as much as $13.48\pm0.88$ percentage points at
$(\sigma_R^2,c_{\mathrm{rx}})=(0.02,1)$, from $0.678$ to $0.812$. The
equal-weight coherent path superposition encoder gains as much as
$12.07\pm1.33$ percentage points at $(0.1,1)$, from $0.716$ to $0.837$. The
left panels show the median fidelity over $c_{\mathrm{rx}}$, with the band
spanning its full range, while the maps show $\Delta_{\mathsf a}$ in each of
the $90$ settings. The uncertainties are paired $95\%$ confidence-interval
half-widths over the $192$ evaluation realizations.

Across the full grid, receiver adaptation expands the region in which both two-path
strategies outperform the fixed \ac{SISO} baseline. Without \ac{CSI}, symmetric
cloning and equal-weight coherent path superposition exceed the baseline in
only $57$ and $9$ of the $90$ settings, respectively; Rx \ac{CSI} raises these
counts to $69$ and $62$, with maximum diversity gains of $11.68$ and $18.25$
percentage points. The improvement is concentrated at large
$c_{\mathrm{rx}}$, where a fixed receiver is poorly aligned with the realized
receive-mode basis. Under strong turbulence, the improvement vanishes: at
$(\sigma_R^2,c_{\mathrm{rx}})=(50,1)$, receiver adaptation changes either
two-path strategy by at most $0.012$ percentage points. Rx \ac{CSI} therefore
corrects modal misalignment within the retained receive space but cannot
restore power scattered outside that space. This distinction explains why the
scrambling bands contract under Rx \ac{CSI}, while every strategy approaches
the $1/2$ erasure limit as loss becomes dominant.

Transmitter adaptation makes the diversity gain positive across most of the
grid, while joint transmitter and receiver adaptation makes it positive
throughout the grid in
Fig.~\ref{fig:controlled-tx-full-csi-mesh}. With Tx \ac{CSI}, asymmetric cloning
and coherent path superposition yield positive gain in $72$ and $77$ of the
$90$ settings, with maxima of $9.10$ and $19.22$ percentage points. The
remaining negative regions occur under weak turbulence, with worst deficits of
$7.90$ and $1.12$ percentage points. Full \ac{CSI} removes these regions: both
encoders exceed the fixed \ac{SISO} baseline in all $90$ settings and reach
maximum diversity gains of $19.56$ and $19.96$ percentage points at
$(\sigma_R^2,c_{\mathrm{rx}})=(0.02,1)$. Adding receiver adaptation to Tx
\ac{CSI} improves asymmetric cloning by as much as $18.20\pm1.26$ percentage
points and coherent path superposition by $5.01\pm0.75$ percentage points.
Transmitter adaptation selects how the two paths are used, whereas receiver
adaptation removes the realization-dependent output-basis mismatch. Both help
when scrambling is strong, but neither overcomes the strong-loss regime.

The two adaptive encoders impose different \ac{CSI} requirements. A
phase-insensitive cloning policy can select its support and real allocation
coefficients from power \ac{CSI}, whereas coherent path superposition requires
field \ac{CSI} because its complex weights depend on both the amplitudes and
relative phases of $\pmb{A}_{\mathrm{eff}}$. For this $2\times2$ channel,
Section~\ref{sec:csi} gives six real coordinates for power \ac{CSI}, compared
with ten for field \ac{CSI} in a fixed phase convention or seven after
removing basis-phase redundancy. In Fig.~\ref{fig:csi-estimation}, a $1\%$
normalized error requires approximately $4.7\times10^3$ probe shots per
transmit rail for power \ac{CSI} and $2.3\times10^4$ pilot shots per transmit
mode for field \ac{CSI}, a factor of about five. This lower requirement applies
to phase-insensitive cloning adaptation. Cloning optimized against the complete
composite channel, as in the Full \ac{CSI} row, also uses field \ac{CSI};
coherent path superposition requires it intrinsically to set the relative
phases.

With sequential probing at pulse rate $R_{\mathrm p}$, the total estimation
times are $T_{\mathrm P}=N_{\mathrm t}S_{\mathrm P}/R_{\mathrm p}$ and
$T_{\mathrm F}=N_{\mathrm t}S_{\mathrm F}/R_{\mathrm p}$. A 10~MHz rate has
been demonstrated on a microsatellite link, while Micius operated at 100~MHz
and was later upgraded to 200~MHz~\cite{TCFKST:17:NPh,SBMPO:23:CP}. For the
$2\times2$ channel at $R_{\mathrm p}=100\,\mathrm{MHz}$, the $1\%$ error points
correspond to $94\,\mu\mathrm{s}$ for power \ac{CSI} and $0.46\,\mathrm{ms}$
for field \ac{CSI}. A representative satellite-to-Earth model gives an
atmospheric coherence time of $2.29\,\mathrm{ms}$, so both estimates fit within
that interval under the assumed rate~\cite{WZMAGR:24:NJP}. This timing
comparison supports rapid Rx estimation when $\omega$ remains effectively
constant during the probing block. Tx and Full \ac{CSI} additionally require
signaling, processing, and transceiver reconfiguration within the same
interval and are therefore treated as ideal availability benchmarks.

\begin{figure}[!t]
    \centering
    \includegraphics[width=0.92\columnwidth]
        {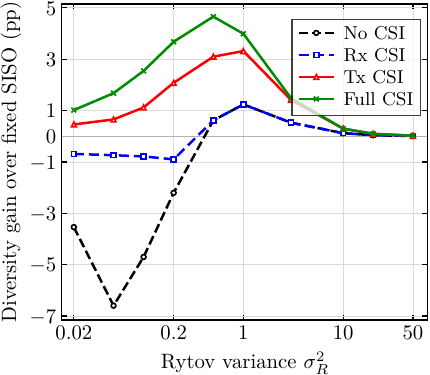}
    \caption{Largest observed two-path diversity gain over the fixed
    \ac{SISO} baseline on the $2\times2$ channel. At each turbulence strength
    and \ac{CSI}-availability case, the ordinate is the largest held-out mean among
    the two encoders over the sampled $c_{\mathrm{rx}}$ values, minus the fixed
    \ac{SISO} mean at $c_{\mathrm{rx}}=0$, expressed in percentage points. No
    \ac{CSI} and Rx \ac{CSI} use symmetric cloning and equal-weight coherent
    path superposition; Tx \ac{CSI} and Full \ac{CSI} use variable-$M$ asymmetric
    cloning and optimized coherent path superposition. Negative values mean that
    both two-path encoders remain below the single-mode baseline.}
    \label{fig:gain-over-fixed-siso-by-csi}
\end{figure}

Full \ac{CSI} gives the largest best-observed diversity gain, while Tx \ac{CSI}
removes the weak-turbulence penalty for the corresponding adaptive encoders in
Fig.~\ref{fig:gain-over-fixed-siso-by-csi}. At each Rytov variance, the plotted
value is the largest held-out mean over the two encoders and sampled
$c_{\mathrm{rx}}$ values, relative to the fixed \ac{SISO} baseline at
$c_{\mathrm{rx}}=0$. Each point is therefore the best observed configuration
on the sampled grid, not an average over receiver scrambling; the two mesh
figures retain the variation across all channel settings. No \ac{CSI} and Rx
\ac{CSI} remain below the baseline through $\sigma_R^2=0.2$, and their largest
gain is only $1.24$ percentage points. Tx \ac{CSI} has positive gain at every
evaluated turbulence strength and reaches $3.32$ percentage points at
$\sigma_R^2=1$. Full \ac{CSI} increases the maximum to $4.66$ percentage
points at $\sigma_R^2=0.5$. Under strong turbulence, all four curves approach
zero because propagation loss removes the benefit of spatial adaptation and
drives every strategy toward the common erasure-limited fidelity of $1/2$.

\begin{figure}[!t]
    \centering
    \includegraphics[width=0.92\columnwidth]
        {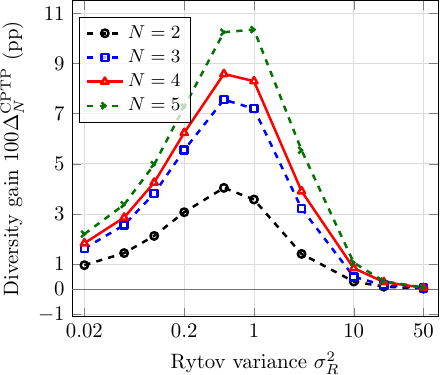}
    \caption{Attained Full \ac{CSI} general-\ac{CPTP} diversity gain
    $100\Delta_N^{\mathrm{CPTP}}$ over the fixed \ac{SISO} baseline for
    $N\times N$ channels, $N\in\{2,3,4,5\}$. Each point is a held-out mean over the same $192$
    atmospheric realizations at $s_{\mathrm{rail}}/w(z)=2.5$ and zero pointing.
    Each curve is the largest value attained across the finite
    initializations of the alternating optimization and does not certify a
    globally optimal encoder--recovery pair.}
    \label{fig:cptp-mimo-scaling}
\end{figure}

\begin{figure}[!t]
    \centering
    \begin{subfigure}[t]{\linewidth}
        \centering
        \includegraphics[width=0.94\linewidth]
            {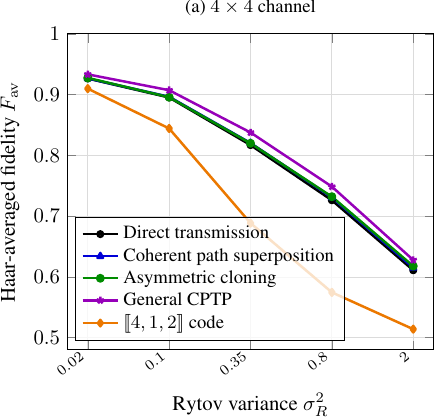}
    \end{subfigure}
    \par\medskip
    \begin{subfigure}[t]{\linewidth}
        \centering
        \includegraphics[width=0.94\linewidth]
            {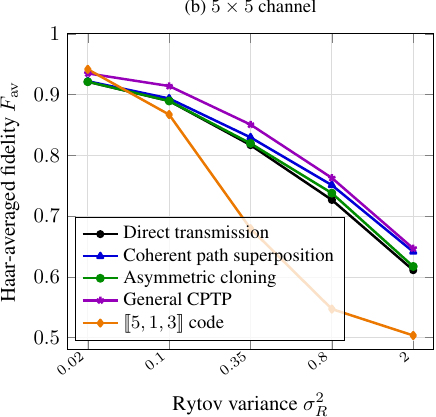}
    \end{subfigure}
    \caption{Haar-averaged state fidelity $F_{\mathrm{av}}$ of five strategies
    on the (a) $4\times4$ and (b) $5\times5$ channels under Full \ac{CSI}, at
    $\sigma_R^2\in\{0.02,0.1,0.35,0.8,2\}$. Every point in both panels is an
    ensemble mean over $128$ atmospheric realizations. Direct transmission,
    coherent path superposition, asymmetric cloning, and the attained
    general-\ac{CPTP} design use realization-dependent information at both
    endpoints. The
    $\left[\!\left[4,1,2\right]\!\right]$ and
    $\left[\!\left[5,1,3\right]\!\right]$ curves use fixed code isometries in
    their respective panels with channel-adapted recoveries.}
    \label{fig:four-five-rail-strategies}
\end{figure}

\begin{figure}[!t]
    \centering
    \includegraphics[width=0.92\columnwidth]
        {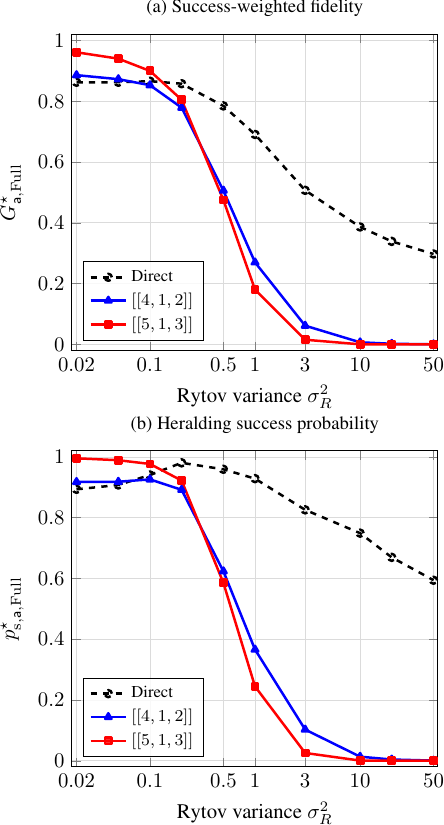}
    \caption{Reliability-constrained Full-\ac{CSI} comparison for Direct
    transmission and the $\left[\!\left[4,1,2\right]\!\right]$ and
    $\left[\!\left[5,1,3\right]\!\right]$ codes on the five-rail channel:
    (a) success-weighted fidelity $G_{\mathsf a,\mathrm{Full}}^{\star}$ and
    (b) heralding success probability
    $p_{\mathrm{s},\mathsf a,\mathrm{Full}}^{\star}$. Each point averages the
    same $16$ realizations. The common reliability target is conditional
    fidelity at least $F_{\mathrm{rel}}(s)-2\times10^{-5}$, where
    $F_{\mathrm{rel}}(s)$ is the deterministic Full-\ac{CSI} reference
    fidelity.}
    \label{fig:full-csi-reliability-yield}
\end{figure}

\subsection{Diversity Scaling with Rail Count under Full CSI}
\label{subsec:cptp-mimo-scaling}

We next examine how the attained general-\ac{CPTP} design changes with the
number of spatial rails. The study uses square channels with
$N_{\mathrm t}=N_{\mathrm r}=N\in\{2,3,4,5\}$,
$s_{\mathrm{rail}}/w(z)=2.5$, zero pointing, and
$c_{\mathrm{rx}}=1$. The ten Rytov variances are those in
Table~\ref{tab:numerical-realization-parameters}. Each turbulence point is a
held-out mean over the same $192$ atmospheric realizations. For each realized
field matrix, Full \ac{CSI} is available at both endpoints and the encoder and
deterministic recovery are adapted by the alternating optimization in
Subsubsection~\ref{subsubsec:iterative-cptp-optimization}. The source space
contains every nonvacuum sector,
$\mathsf N_{\mathrm{src}}^{(N)}=\{1,\ldots,N\}$. The fixed \ac{SISO} baseline
launches transmit mode~1, reads receive mode~1, and uses
$c_{\mathrm{rx}}=0$ without realization-dependent adaptation.

For rail count $N$, the diversity gain of the attained general-\ac{CPTP}
design is
\begin{equation}
\begin{aligned}
    \Delta_N^{\mathrm{CPTP}}(\sigma_R^2)
    &\coloneqq
    \overline F_{\mathrm{av},\mathrm{CPTP}}^{(N)}
    (\sigma_R^2;c_{\mathrm{rx}}=1)
    \\
    &\quad-
    \overline F_{\mathrm{av},\mathrm{SISO}}
    (\sigma_R^2;c_{\mathrm{rx}}=0).
\end{aligned}
    \label{eq:cptp-mimo-scaling-margin}
\end{equation}
Figure~\ref{fig:cptp-mimo-scaling} shows
$100\Delta_N^{\mathrm{CPTP}}$ in percentage points. The same
$c_{\mathrm{rx}}$ and Full \ac{CSI} rule are used for all four multirail
curves, so their ordering isolates the attained change with $N$ within this
design class. The absolute margin over the fixed \ac{SISO} baseline also includes the value
of endpoint adaptation. Moreover, because the admitted photon-number sectors
expand from $\{1,2\}$ to $\{1,2,3,4,5\}$, this experiment increases spatial
dimension and the available photon-number resource together; it is not a
photon-matched comparison of rail count alone.

Additional rails increase the mean attained Full \ac{CSI} diversity gain at all
ten turbulence strengths, with the ordering
$\Delta_5^{\mathrm{CPTP}}>\Delta_4^{\mathrm{CPTP}}>
\Delta_3^{\mathrm{CPTP}}>\Delta_2^{\mathrm{CPTP}}>0$. The benefit is largest
under moderate turbulence. At $\sigma_R^2=0.5$, the gains are
$4.041\pm0.951$, $7.560\pm1.265$, $8.584\pm1.151$, and $10.248\pm1.325$
percentage points for $N=2$, $3$, $4$, and $5$, respectively; the
uncertainties are paired $95\%$ confidence-interval half-widths relative to the
fixed \ac{SISO} samples. In particular, the paired $N=5$ minus $N=4$
increment is $1.664\pm0.923$ percentage points at this turbulence strength.
At $\sigma_R^2=50$, the corresponding gains fall to $0.0226$, $0.0375$,
$0.0558$, and $0.0686$ percentage points as every design approaches the $1/2$
erasure limit. The rail-count increments therefore become negligible under
strong turbulence because loss suppresses the benefit of additional spatial
rails. Although the $N=5$ sample mean remains above $N=4$ at every sampled
turbulence strength, its paired difference at $\sigma_R^2=20$ is not resolved
from zero at the $95\%$ confidence level.

\subsection{Strategy Comparison at Four and Five Rails under Full CSI}
\label{subsec:strategy-rail-count}

We next compare the structured strategies on four- and five-rail channels.
Figure~\ref{fig:four-five-rail-strategies} shows their performance for the
$4\times4$ and $5\times5$ configurations over the five-point turbulence grid
$\sigma_R^2\in\{0.02,0.1,0.35,0.8,2\}$. Both panels use the channel
construction in Table~\ref{tab:numerical-realization-parameters}, with
$\tau_j^{\mathrm{rx}}=0.92$ and $q_j=0.01$. Each point is an ensemble mean over
$128$ atmospheric realizations.

The adaptive strategies use Full \ac{CSI}. Direct transmission selects its
rail--port pair for the realized channel. Coherent path superposition optimizes
the complex path weights and all-port recovery, while asymmetric cloning adapts
within the cloning family. The general-\ac{CPTP} curve is the value
attained by the alternating encoder--recovery design. The stabilizer encoders
are fixed isometries: the
$\left[\!\left[4,1,2\right]\!\right]$ code occupies all four rails in panel
(a), and the $\left[\!\left[5,1,3\right]\!\right]$ code occupies all five rails
in panel (b). They introduce no realization-adaptive encoder variables; their
recoveries are optimized for the encoded channel as defined in
Subsubsection~\ref{subsubsec:quantum-error-correction}.

On the four-rail channel, the attained general-\ac{CPTP} design has the largest
mean at all five turbulence strengths and exceeds the best structured adaptive
strategy by as much as $1.73$ percentage points at $\sigma_R^2=0.35$. Direct
transmission, coherent path superposition, and asymmetric cloning remain close,
although their spread increases from $0.097$ percentage points at
$\sigma_R^2=0.02$ to $0.682$ percentage points at $\sigma_R^2=2$. By contrast,
the fixed $\left[\!\left[4,1,2\right]\!\right]$ encoder remains below direct
transmission at all five points, with its largest deficit of $15.16$ percentage
points at $\sigma_R^2=0.8$.

The five-rail channel shows larger differences among the structured strategies. From
$\sigma_R^2=0.1$ onward, the attained general-\ac{CPTP} curve has the largest
mean. Coherent path superposition is the strongest structured adaptive
strategy at every evaluated turbulence strength, and its advantage over direct
transmission increases from $0.089$ percentage points at
$\sigma_R^2=0.02$ to $3.03$ percentage points at $\sigma_R^2=2$. From
$\sigma_R^2=0.1$ onward, the fixed $\left[\!\left[5,1,3\right]\!\right]$ code
falls below direct transmission, with its largest deficit of $17.96$ percentage
points at $\sigma_R^2=0.8$. The general-\ac{CPTP} formulation includes every
admissible encoder and recovery map on the declared spaces and therefore
defines a resource-matched upper bound for the structured strategies. We
approximate this bound by alternating optimization. Because the joint problem
is biconvex, the plotted curve is the best attained general-\ac{CPTP} value,
not a certified upper bound. A structured value above it therefore indicates
an optimization gap rather than a violation of the general-\ac{CPTP} bound.

The channel descriptors explain this ordering. The
$\left[\!\left[4,1,2\right]\!\right]$ code corrects one erasure at a known
position, and the $\left[\!\left[5,1,3\right]\!\right]$ code corrects two, but
spatial crosstalk and multiphoton collisions do not reveal independent erased
code positions at the decoder. A fixed code also distributes its physical
qubits across rails whose survival probabilities become increasingly unequal.
In contrast, a Full \ac{CSI} strategy can concentrate the logical state on the
strongest realized spatial directions. Rail heterogeneity
$h_{\mathrm{rail}}$ therefore explains the adaptive-strategy ordering, while
crosstalk $\chi_{\mathrm{eff}}$ marks the failure of the known-position erasure
model. The weak-turbulence five-rail point identifies the regime in which code
distance remains useful; the subsequent ordering shows why channel-adaptive
diversity becomes preferable as rail heterogeneity develops.

Figures~\ref{fig:cptp-mimo-scaling} and
\ref{fig:four-five-rail-strategies} separate resource scaling from strategy
selection. Increasing the spatial and photon-number resources raises the
attained general-\ac{CPTP} gain, but the structured comparison shows that the
practical benefit depends on how the encoder uses the realized channel. This
comparison does not establish a universal ordering between stabilizer codes
and channel-adaptive encoders; it contrasts fixed code isometries with adaptive
continuous strategies on this turbulence ensemble.

The heralded comparison favors the
$\left[\!\left[5,1,3\right]\!\right]$ code under the weakest turbulence but
direct transmission as rail heterogeneity develops. At $\sigma_R^2=0.02$, the
code has the largest sample mean,
$G_{\mathsf a,\mathrm{Full}}^{\star}=0.961$, with
$p_{\mathrm{s},\mathsf a,\mathrm{Full}}^{\star}=0.996$. Direct transmission
reaches the largest sample mean at higher turbulence, with
$G^{\star}=0.858$ at $\sigma_R^2=0.2$ and $0.387$ at $\sigma_R^2=10$. In
Fig.~\ref{fig:full-csi-reliability-yield}, $G^{\star}$ is the success-weighted
fidelity corresponding to the maximum heralding success probability under the
common conditional-fidelity target in
\eqref{eq:availability-reliability-success-probability}. Full \ac{CSI} adapts
the recovery but does not convert turbulence-induced crosstalk and collisions
into erasures at known code positions for fixed stabilizer encoders. These
orderings apply to the sample means shown; the figure provides no confidence
intervals for their differences.

\section{Conclusion}

We represent a discrete-variable \ac{FSO} \ac{QuMIMO} link as a single
\ac{CPTP} map from an unknown polarization qubit to an erasure-augmented
receiver state. The map composes the passive complex field-transfer matrix, the
Fock-space Stinespring dilation of the associated bosonic pure-loss channel,
the internal-state Gram matrix for partial photon distinguishability, the
receiver map to erasure-augmented polarization qubits, and effective per-port
polarization noise. This representation separates spatial crosstalk, photon
loss, multiphoton-collision handling, and polarization noise. Within the same
model and Haar-averaged fidelity metric, we compare five strategies with
different photon budgets and \ac{CSI} requirements.

\Ac{CSI} availability determines whether spatial diversity raises mean
fidelity. Although no-cloning forbids transmitting identical copies of an
unknown state, multiple rails provide distinct propagation paths. Approximate
cloning distributes several photons across those paths, whereas coherent path
superposition distributes one photon's amplitude. Relative to the unscrambled
fixed \ac{SISO} baseline, the best No- and Rx-\ac{CSI} configurations remain
below the baseline through the four weakest evaluated turbulence strengths.
Rx \ac{CSI} nevertheless reduces sensitivity to receive-mode scrambling across
matched scrambling settings. With Full \ac{CSI}, both adaptive encoders exceed
the matched fixed \ac{SISO} baseline at all operating points tested in this
work. Their largest diversity gains are $19.56$ percentage points for
asymmetric cloning and $19.96$ percentage points for coherent path
superposition.

Asymmetric cloning and coherent path superposition use different resources
to obtain spatial diversity. An $M$-clone encoder emits $M$ photons and raises
survival at the cost of survival-conditioned fidelity. Across the ten Rytov
values, the latter falls from $0.995$ for direct transmission to about $0.732$
at five clones, while the weak-turbulence survival gain peaks near $1.43$.
Cloning can be favorable when the survival increase outweighs this fidelity
reduction. Coherent path superposition retains a single-photon budget and uses
complex amplitude weights, but it requires phase-sensitive field \ac{CSI} and
cross-rail internal-state overlap to preserve path coherence. Partial photon
distinguishability therefore affects the strategies differently. In the
evaluated $2\times2$ channel, it suppresses multiphoton bunching collisions
and raises cloning survival by a few percent, but degrades the coherence used
by single-photon path superposition. Direct transmission launches one photon
into one rail and is unaffected by photon distinguishability.

The evaluated fixed stabilizer encoders perform poorly when rail survival is
unequal and the channel errors are not erasures at known positions. The
$\left[\!\left[4,1,2\right]\!\right]$ encoder remains below direct
transmission at all five evaluated turbulence strengths, with a maximum deficit
of $15.16$ percentage points at $\sigma_R^2=0.8$. The
$\left[\!\left[5,1,3\right]\!\right]$ encoder falls below direct transmission
from $\sigma_R^2=0.1$ onward and reaches a maximum deficit of $17.96$
percentage points at $\sigma_R^2=0.8$. These results are consistent with fixed
codes distributing physical qubits across rails whose survival probabilities
become unequal, while crosstalk and multiphoton collisions differ from erasures
at known code positions. The recoveries are channel adapted, but the code
isometries remain fixed, whereas the other adaptive encoders change with each
realization. The comparison therefore does not establish a universal ordering
between fixed stabilizer codes and channel-adaptive encoders.

The general design searches over pairs of \ac{CPTP} maps, one encoder and one
recovery. Its feasible set contains every resource-matched structured strategy
on the same source space. The unknown global optimum therefore upper-bounds
those strategies, while the
alternating optimization provides an attained lower bound without a global
optimality certificate. This Full-\ac{CSI} benchmark has an encoder parameter
count that scales with the squares of the source-space or photon-number-sector
dimensions, so it serves as a numerical benchmark rather than a practical
transceiver design. The attained diversity gain increases with rail count at
each sampled turbulence strength. At $\sigma_R^2=0.5$, the gains over the fixed
\ac{SISO} baseline are $4.041$, $7.560$, $8.584$, and $10.248$ percentage
points for two, three, four, and five rails, respectively. This comparison
includes every available nonvacuum photon-number
sector and therefore increases spatial dimension and the allowed total photon
numbers together. Separating these resources, evaluating finite-shot \ac{CSI}
adaptation, and extending the framework to multi-qubit transmission remain
open.

\appendices

\section{Phase-Screen Channel Realization}
\label{app:phase-screens}

Restricting the continuous atmospheric propagation operator to the selected
transmit and receive bases gives $\pmb A(\omega)$ in
\eqref{eq:field-transfer}. This appendix specifies that operator, its
finite-grid realization, and the longitudinal-discretization check.

\begin{figure*}[!t]
    \centering
    \includegraphics[width=0.92\textwidth]
        {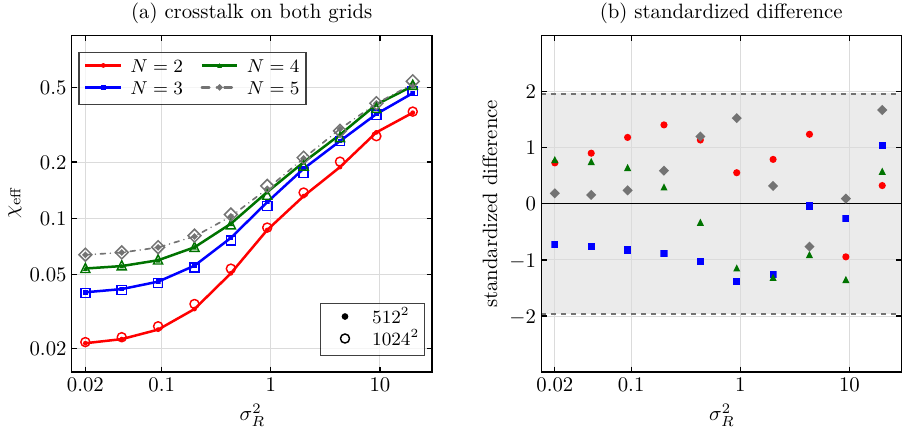}
    \caption{Transverse-grid validation of the phase-screen channel at the
    reference geometry. Both grids use matched seeds for $512$ realizations at
    rail counts $N=2,\ldots,5$ and ten values of $\sigma_R^2$. Panel (a) shows
    detector-plane crosstalk at both resolutions, with filled markers and lines at
    $512^2$ and open markers at $1024^2$. Panel (b) shows the difference
    $(\chi_{1024}-\chi_{512})/\sqrt{\mathrm{sem}_{1024}^2+\mathrm{sem}_{512}^2}$,
    standardized by the quadrature sum of the two sample-mean standard errors, at
    the $40$ operating points; the shaded band marks magnitudes below $1.96$.
    The two resolutions use the same screen count at each point, with counts
    $10$, $14$, and $21$ across this convergence sweep. The fixed
    $0.30\,\mathrm{m}$ window isolates transverse sampling within this
    configuration.}
    \label{fig:grid-convergence}
\end{figure*}

\begin{figure*}[!t]
    \centering
    \includegraphics[width=\textwidth]
        {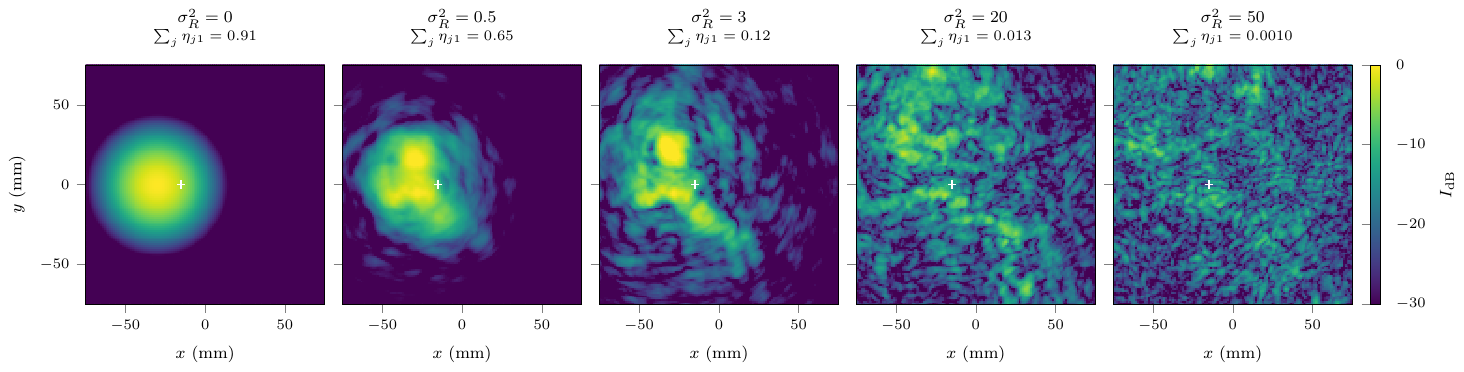}
    \caption{Effect of turbulence strength alone on the propagated field of one
    transmit rail. A two-rail link is simulated at $\lambda_0=809\,\mathrm{nm}$
    over $z=1000\,\mathrm{m}$ with waist $w_0=20\,\mathrm{mm}$, rail separation
    $s_{\mathrm{rail}}=2.5\,w(z)$, thirty-four phase screens drawn from a von
    K\'arm\'an spectrum with outer and inner scales of $80\,\mathrm{m}$ and
    $1\,\mathrm{mm}$, and a $512\times512$ transverse grid over
    $0.30\,\mathrm{m}\times0.30\,\mathrm{m}$. The central $256\times256$
    samples are displayed after $2\times2$ binning. All five panels use the
    same phase-screen realization and launched rail, so they differ only in
    the plane-wave Rytov variance
    $\sigma_R^2$. Colour shows intensity in decibels on a shared scale referred
    to the peak of the turbulence-free beam. Above each panel,
    $\sum_j\eta_{j1}$ is the fraction of launched power collected into the two
    modelled receive modes for that draw.}
    \label{fig:phase-screen-realizations}
\end{figure*}

\subsection{Continuous Atmospheric Model}

Let $\lambda_0>0$ be the optical wavelength, $z>0$ the propagation distance,
$\pmb{x}_{\perp}=(x,y)^{\mathsf{T}}\in\mathbb{R}^2$ the transverse coordinate,
and $\xi\in[0,z]$ the longitudinal coordinate. We define the optical wavenumber
$\kappa_0\coloneqq2\pi/\lambda_0$ and the transverse Laplacian
$\nabla_{\perp}^{2}\coloneqq\partial_x^2+\partial_y^2$. For a realization
$\omega$ that is constant over one channel use, let $\delta n_{\omega}$ denote
the real-valued refractive-index fluctuation. The scalar quasi-monochromatic
field envelope $\Psi_{\omega}(\pmb{x}_{\perp},\xi)$ satisfies
\begin{equation}
    2\mathrm{i}\kappa_0
    \frac{\partial\Psi_{\omega}}{\partial\xi}
    +
    \nabla_{\perp}^{2}\Psi_{\omega}
    +
    2\kappa_0^{2}\delta n_{\omega}(\pmb{x}_{\perp},\xi)
    \Psi_{\omega}
    =0.
    \label{eq:stochastic-paraxial-equation}
\end{equation}
Let $c_n^2$ be the constant refractive-index structure parameter and
$\ell_{\mathrm{in}},\ell_{\mathrm{out}}>0$ the turbulence inner and outer
scales. For a homogeneous path, $\delta n_{\omega}$ is modeled as a zero-mean,
statistically homogeneous and isotropic Gaussian random field with modified
von K\'arm\'an spectrum. Let
$\pmb{\kappa}_3=(\pmb{\kappa}_{\perp},\kappa_{\xi})\in\mathbb{R}^{3}$ be the
angular spatial-frequency vector, where $\pmb{\kappa}_{\perp}$ and
$\kappa_{\xi}$ are its transverse and longitudinal components. We define
$\kappa_{\mathrm{in}}\coloneqq5.92/\ell_{\mathrm{in}}$ and
$\kappa_{\mathrm{out}}\coloneqq2\pi/\ell_{\mathrm{out}}$. The
three-dimensional refractive-index spectrum is
\begin{equation}
    \Phi_n(\pmb{\kappa}_3)
    =
    0.033\,c_n^2
    \exp\!\left[-\frac{\|\pmb{\kappa}_3\|^2}{\kappa_{\mathrm{in}}^2}\right]
    \left(\|\pmb{\kappa}_3\|^2+\kappa_{\mathrm{out}}^2\right)^{-11/6}.
    \label{eq:refractive-index-spectrum}
\end{equation}

\subsection{Thin-Screen Statistics}

The Markov thin-screen approximation divides the path into
$N_{\mathrm{scr}}$ slabs of length
$\Delta z=z/N_{\mathrm{scr}}$ and replaces each slab by an independent,
zero-mean, real Gaussian phase field $\phi_m$, with
$m=1,\ldots,N_{\mathrm{scr}}$. Its Fried parameter and transverse phase
spectrum are
\begin{equation}
\begin{aligned}
    r_{0,m}
    &\coloneqq
    \left(0.423\,\kappa_0^2 c_n^2\Delta z\right)^{-3/5},
    \\
    \Phi_{\phi,m}(\pmb{\kappa}_{\perp})
    &\coloneqq
    0.0229\,r_{0,m}^{-5/3}
    \left(
        \frac{
            \|\pmb{\kappa}_{\perp}\|^2+\kappa_{\mathrm{out}}^2
        }{(2\pi)^2}
    \right)^{-11/6}
    \\
    &\quad\times
    \exp\left[
        -\frac{\|\pmb{\kappa}_{\perp}\|^2}{\kappa_{\mathrm{in}}^2}
    \right].
    \label{eq:phase-screen-spectrum}
\end{aligned}
\end{equation}
Let
$\widetilde\phi_m(\pmb{\kappa}_{\perp})
\coloneqq\int_{\mathbb{R}^2}\phi_m(\pmb{x}_{\perp})
e^{-\mathrm{i}\pmb{\kappa}_{\perp}^{\mathsf{T}}\pmb{x}_{\perp}}
\,\mathrm{d}^2\pmb{x}_{\perp}$.
For $m,m'\in\{1,\ldots,N_{\mathrm{scr}}\}$ and transverse wavevectors
$\pmb{\kappa}_{\perp},\pmb{\kappa}'_{\perp}\in\mathbb{R}^2$, the following
covariance fixes the Fourier convention. Here $\delta_{mm'}$ denotes the
Kronecker delta and $\delta^{(2)}$ denotes the two-dimensional Dirac delta:
\begin{equation}
\begin{aligned}
    \mathbb{E}\!\left[
        \widetilde\phi_m(\pmb{\kappa}_{\perp})
        \widetilde\phi_{m'}^{*}(\pmb{\kappa}'_{\perp})
    \right]
    &=(2\pi)^2\delta_{mm'}
    \\
    &\quad\times
    \delta^{(2)}(\pmb{\kappa}_{\perp}-\pmb{\kappa}'_{\perp})
    \\
    &\quad\times
    \Phi_{\phi,m}(\pmb{\kappa}_{\perp}).
\end{aligned}
    \label{eq:phase-screen-covariance}
\end{equation}
The sampled spectrum sets the zero-frequency coefficient to zero and thereby
fixes the common phase of each screen. This convention changes neither the
modal powers nor the relative phases of the transfer matrix. These screen
statistics are the thin-screen
counterpart of
$\phi_m(\pmb{x}_{\perp})
=\kappa_0\int_{(m-1)\Delta z}^{m\Delta z}
\delta n_{\omega}(\pmb{x}_{\perp},\xi)\,\mathrm{d}\xi$
under the Markov approximation~\cite{BGL:20:JOSA,KS:23:PRA,C:20:JOSA}.

\subsection{Split-Step Propagation and Physical Scales}

For $d\geq0$, $f\in L^2(\mathbb{R}^2)$, and pointing angle
$\pmb{\theta}(\omega)\in\mathbb{R}^2$, define the vacuum Fresnel propagator
$\mc{U}_{\mathrm{Fr}}(d)\coloneqq
\exp[\mathrm{i}d\nabla_{\perp}^{2}/(2\kappa_0)]$, the phase multiplier
$(\mc{M}_{\phi_m}f)(\pmb{x}_{\perp})
=\exp[\mathrm{i}\phi_m(\pmb{x}_{\perp};\omega)]f(\pmb{x}_{\perp})$, and the
pointing multiplier
$(\mc{M}_{\pmb{\theta}}f)(\pmb{x}_{\perp})
=\exp[\mathrm{i}\kappa_0\pmb{\theta}^{\mathsf{T}}\pmb{x}_{\perp}]
f(\pmb{x}_{\perp})$. Let the receive aperture be the measurable set
$\mc{A}_{\mathrm{rx}}$. Using $\mathbbm{1}_{S}$ for the indicator function of
a set $S$, define
$(\mc{M}_{\mathrm{ap}}f)(\pmb{x}_{\perp})
=\mathbbm{1}_{\mc{A}_{\mathrm{rx}}}(\pmb{x}_{\perp})
f(\pmb{x}_{\perp})$. Let $\alpha_{\mathrm{ext}}\geq0$ be the
power-extinction coefficient. With later slabs ordered to the left, the
single-photon transverse-field propagation operator is
\begin{equation}
\begin{aligned}
    \mc K_{\mathrm{prop}}(\omega)
    &\coloneqq
    e^{-\alpha_{\mathrm{ext}}z/2}
    \mc{M}_{\mathrm{ap}}
    \\
    &\quad\times
    \overleftarrow{\prod_{m=1}^{N_{\mathrm{scr}}}}
    \left(
        \begin{gathered}
            \mc{U}_{\mathrm{Fr}}(\Delta z/2)
            \mc{M}_{\phi_m(\omega)}
            \\
            {}\times\mc{U}_{\mathrm{Fr}}(\Delta z/2)
        \end{gathered}
    \right)
    \mc{M}_{\pmb{\theta}(\omega)} .
\end{aligned}
    \label{eq:transverse-field-propagation-operator}
\end{equation}
This is the symmetric split-step representation of
\eqref{eq:stochastic-paraxial-equation}; its accuracy depends
on the longitudinal step, phase-screen statistics, and transverse spectral
support~\cite{KS:23:PRA,C:20:JOSA}. For a fixed screen realization, the
Fresnel, phase, and pointing factors are unitary on $L^2(\mathbb R^2)$, while
aperture truncation and extinction are contractions. Writing
$\pmb I_{L^2(\mathbb R^2)}$ for the identity on $L^2(\mathbb R^2)$, we obtain
$\mc K_{\mathrm{prop}}^{\dagger}(\omega)\mc K_{\mathrm{prop}}(\omega)
\preceq\pmb I_{L^2(\mathbb R^2)}$.

For a homogeneous path, the plane-wave Rytov variance is
\begin{equation}
    \sigma_R^2
    =
    1.23\,c_n^2 \kappa_0^{7/6}z^{11/6}.
    \label{eq:plane-wave-rytov-variance}
\end{equation}
At fixed $\lambda_0$ and $z$, $\sigma_R^2$ selects $c_n^2$ and hence the
atmospheric ensemble, not an instantaneous $\pmb A(\omega)$. For finite-waist
modes, it need not equal the scintillation variance of an individual propagated
mode. The standard regime labels use $\sigma_R^2\lesssim1$ for weak
fluctuations and $\sigma_R^2\gtrsim1$ for strong fluctuations, where the
scintillation index saturates rather than following $\sigma_R^2$
~\cite{KS:23:PRA,VSV:16:PRL}.

For a Gaussian waist $w_0$ at the transmitter, define the Rayleigh range and
propagated spot size by~\cite{KL:66:AO}
\begin{equation}
    z_R=\frac{\pi w_0^2}{\lambda_0},
    \qquad
    w(z)=w_0\sqrt{1+\left(\frac{z}{z_R}\right)^2}.
    \label{eq:gaussian-spot-size}
\end{equation}
The far-field approximation $w(z)\simeq\lambda_0z/(\pi w_0)$ requires
$z\gg z_R$. Four dimensionless quantities separate the main physical and
geometric effects: $\sigma_R^2$ sets the turbulence strength;
$N_{\mathrm{Fr}}=z_R/z$ controls diffraction and, with the aperture and receive
basis, the collected fraction; $s_{\mathrm{rail}}/w(z)$ measures rail separation
at the receiver; and $s_{\mathrm{rail}}/w_0$ controls the raw overlap of
equal-waist displaced basis functions. Loss and crosstalk can therefore vary
together, and neither is determined by $\sigma_R^2$ alone.

\subsection{Finite-Grid Realization and Screen Count}

Each rail is a Gaussian of
waist $w_0=20\,\mathrm{mm}$, taken as the $1/e$ field-amplitude radius.
For the reference geometry, \eqref{eq:gaussian-spot-size} gives a Rayleigh range
$z_R=1.55\,\mathrm{km}$ and propagated spot size $w(z)=23.8\,\mathrm{mm}$, so
$z/z_R=0.64$ and $s_{\mathrm{rail}}/w(z)=2.5$. The numerical sweep also varies
the normalized rail separation as specified in
Table~\ref{tab:numerical-realization-parameters}. The exact spot size exceeds
the far-field value $\lambda_0z/(\pi w_0)=12.9\,\mathrm{mm}$ by a factor of
$1.8$, so the far-field form of \eqref{eq:gaussian-spot-size} is not applicable
at this reference geometry. The transmit and receive bases are orthonormalized
before use.

The finite-grid realization uses $N_{\mathrm{scr}}=34$ equal slabs, with each
screen at its slab midplane as in
\eqref{eq:transverse-field-propagation-operator}. The phase spectrum in
\eqref{eq:phase-screen-spectrum} uses
$\ell_{\mathrm{out}}=80\,\mathrm{m}$ and
$\ell_{\mathrm{in}}=1\,\mathrm{mm}$; its exponential factor supplies the
inner-scale cutoff. All transmit modes in one realization traverse the same
screens, so inter-rail coupling reflects one common atmosphere rather than
independent per-rail draws.

The Rytov variance and screen count play different roles. The physical
parameter $\sigma_R^2$ selects $c_n^2$ through
\eqref{eq:plane-wave-rytov-variance} and fixes the atmospheric ensemble. The
numerical parameter $N_{\mathrm{scr}}$ divides the same turbulence into shorter
slabs at fixed $\sigma_R^2$ without changing the physical link strength. Because $\sigma_R^2$ scales as
$z^{11/6}$, splitting the path into $N_{\mathrm{scr}}$ equal slabs gives the
per-slab plane-wave Rytov variance
\begin{equation}
    \sigma_{\mathrm{slab}}^{2}
    =
    \frac{\sigma_R^{2}}{N_{\mathrm{scr}}^{11/6}} .
    \label{eq:slab-rytov-variance}
\end{equation}
The split-step construction treats each screen as a weak perturbation, so the
discretization is resolved only while $\sigma_{\mathrm{slab}}^{2}$ remains in
the weak-fluctuation range. We therefore fix the count from the strongest
operating point, $\sigma_{R,\max}^{2}\coloneqq50$, by requiring
$\sigma_{\mathrm{slab}}^{2}\leq0.08$ there. This gives
\begin{equation}
    N_{\mathrm{scr}}
    =
    \left\lceil
        \left(
            \frac{\sigma_{R,\max}^{2}}{0.08}
        \right)^{6/11}
    \right\rceil
    = 34,
    \label{eq:screen-count}
\end{equation}
and an achieved $\sigma_{\mathrm{slab}}^{2}=0.0779$. Every weaker operating
point is resolved
by a larger margin, since \eqref{eq:slab-rytov-variance} is linear in
$\sigma_R^{2}$ at fixed $N_{\mathrm{scr}}$. This is a longitudinal criterion
only. The treatment of the lowest spatial frequencies, whose attenuation is
noted in the main text, is known to differ among screen-generation
methods~\cite{C:20:JOSA,KS:23:PRA}. The
transverse grid is a separate convergence axis from
\eqref{eq:slab-rytov-variance} and must be refined independently.

The transverse-grid check in Fig.~\ref{fig:grid-convergence} places all
$40$ standardized differences below $1.96$, with a maximum of $1.67$. The two
resolutions use matched seeds, equal realization counts, and the same screen
count at each operating point; \eqref{eq:slab-rytov-variance} gives counts of
$10$, $14$, and $21$ across this separate sweep. The largest absolute changes
are $0.018$ in $\chi_{\mathrm{eff}}$ and $0.008$ in mean survival
$\bar{\eta}_{\mathrm{eff}}$. For the tested reference geometry, the check
resolves no change in the descriptor trends of
Section~\ref{sec:numerical-results}. It addresses transverse sampling only and
does not bound the window-size or subharmonic effects discussed in
Section~\ref{subsec:simulation-setup}; those effects involve the lowest spatial
frequencies that refinement within a fixed window cannot restore.

Figure~\ref{fig:phase-screen-realizations} isolates the physical parameter. One
atmospheric realization is evaluated at five values of $\sigma_R^2$. The
thirty-four screen positions and the launched rail are held fixed, so the
panels differ only in turbulence strength. Intensities are shown in decibels
relative to the peak of the turbulence-free beam. Let
$I_{\sigma_R^2}(x,y)$ be the intensity at turbulence strength $\sigma_R^2$ and
$I_0(x,y)$ the turbulence-free intensity for the same launched rail. The
displayed intensity is
\begin{equation}
    I_{\mathrm{dB}}(x,y)
    =
    10\log_{10}
    \left[
        \frac{I_{\sigma_R^{2}}(x,y)}
             {\max_{x,y}I_{0}(x,y)}
    \right],
    \label{eq:displayed-intensity}
\end{equation}
on one shared colour scale. A common reference preserves both the change in
beam shape and the reduction in peak intensity; normalizing each panel
independently would display the former while concealing the latter. The screen
count is a fixed numerical discretization parameter and is not varied with
$\sigma_R^2$. The collected fraction $\sum_j\eta_{j1}$ shown above each
panel falls from $0.91$ to $0.0010$ across the sweep, and the five values span
the weak, transition, and strong-fluctuation regimes. Increasing $\sigma_R^2$
moves power out of the intended mode and out of the collected subspace, which
enter $\pmb A(\omega)$ as off-diagonal coupling and as a column-power deficit,
respectively.

\section{Bosonic Fock Space}
\label{app:fock}

For any single-photon Hilbert space $\mathfrak h$, let
$\operatorname{Sym}^{n}(\mathfrak h)$ denote the symmetric subspace of
$\mathfrak h^{\otimes n}$, with
$\operatorname{Sym}^{0}(\mathfrak h)\coloneqq
\operatorname{span}\{\ket{\mathrm{vac}}\}$, where $\ket{\mathrm{vac}}$ is the
vacuum. Symmetrization imposes the exchange statistics of photons. For the
spaces defined in
Section~\ref{subsubsec:multiphoton},
$\dim\mathfrak h_{\mathrm{in}}=2N_{\mathrm t}$ and
$\dim\mathfrak h_{\mathrm{out}}=2(N_{\mathrm r}+N_{\mathrm t})$. An
orthonormal basis of $\operatorname{Sym}^{n}(\mathfrak h_{\mathrm{in}})$ is
given by occupation-number vectors $\ket{\pmb n}$ with
$\pmb n=(n_1,\ldots,n_{2N_{\mathrm t}})$ and $\sum_k n_k=n$, where $n_k$
counts photons in input mode $k$, indexed by rail and polarization. In this basis the
second quantization $\Gamma(\pmb W(\omega)\otimes\pmb I_2)$ of
Section~\ref{subsubsec:multiphoton} acts blockwise, one photon-number sector at
a time, as used in \eqref{eq:fock-lift}.

A common direct-sum description is required because the strategies occupy
different photon-number sectors. The universal asymmetric cloner of
Section~\ref{subsubsec:asymmetric-cloning-purification} maps one logical qubit
onto $M$ approximate copies in the $M$-photon sector, whereas rail selection
and coherent path superposition occupy the single-photon sector.

The passive Stinespring dilation of \eqref{eq:loss-dilation} preserves total
photon number across receiver and environment modes. After the environment is
traced out, the accessible receiver state from an $n$-photon input has support
on sectors $0,1,\ldots,n$. Photon loss contributes erasure outcomes in the
reduced description, while local multiphoton occupancy provides a second
erasure mechanism.

Multiphoton interference requires a bosonic Fock-space description. The
transition amplitude between occupation-number vectors is a matrix permanent,
and partial distinguishability modifies that structure directly: reducing the
internal-state overlaps in $\pmb{G}$ reweights the interfering permutations and
alters the Hong--Ou--Mandel statistics within the same fixed-$n$ sector. Its
effect is therefore not reducible to independent single-photon channels.

The bosonic lift is defined on the full input Fock space, but the physical
channel is restricted to the admissible source subspace
$\mc H_{\mathrm{src}}
\subset\mc F_{\mathrm s}(\mathfrak h_{\mathrm{in}})$ in
\eqref{eq:source-space}. With
$n_{\max}$ as defined in Section~\ref{subsubsec:multiphoton}, the accessible
receiver state, including its internal modes, lies after the environment trace
in
$\bigoplus_{n=0}^{n_{\max}}
\operatorname{Sym}^{n}(\mathfrak h_{\mathrm{rec}}
\otimes\mathfrak h_{\mathrm{int}})$.
The squashing map sends this finite receiver Fock space to the
finite-dimensional receiver space $\mc H_{\mathrm{rx}}$ in
\eqref{eq:receiver-space}. Its erasure flag $\ket{\mathrm e}$ identifies photon
loss or unresolved local multiphoton occupancy.

\section{Internal-State Assignment}
\label{app:internal-state-assignment}

The internal-state assignment realizes $\pmb G$ as an isometry on each
admissible source sector. For the Gram matrix defined in
\eqref{eq:internal-state-distinguishability} and the states
$\{\ket{\varphi_i}\}$ chosen in Section~\ref{subsubsec:multiphoton}, let
$r_{\mathrm G}\coloneqq\operatorname{rank}(\pmb G)$, set
$\mathfrak h_{\mathrm{int}}\coloneqq\operatorname{span}\{\ket{\varphi_i}\}$,
and choose an orthonormal basis
$\{\ket{\gamma}\}_{\gamma=1}^{r_{\mathrm G}}$ for that space. Let
$\widehat a_{i,\sigma}^{\dagger}$ create a photon in transmit rail $i$ with
polarization $\sigma$, and let $\widehat a_{i,\sigma,\gamma}^{\dagger}$
create one carrying in addition the internal mode $\gamma$. We define
$c_{i\gamma}\coloneqq\langle\gamma|\varphi_i\rangle$ and
$\widehat a_{i,\sigma}^{\dagger}(\varphi_i)
\coloneqq\sum_{\gamma=1}^{r_{\mathrm G}}
c_{i\gamma}\widehat a_{i,\sigma,\gamma}^{\dagger}$. For each
$n\in\mathsf N_{\mathrm{src}}$ with
$n\geq1$, the isometry $\mc V_{\pmb G}^{(n)}$ assigns
$\ket{\varphi_i}$ to the photon supplied by every occupied transmit rail $i$.
For $\sigma\in\{\mathrm H,\mathrm V\}$ and
occupation numbers
$n_{i,\sigma}\in\{0,1\}$ satisfying
$\sum_{\sigma}n_{i,\sigma}\leq1$ and
$\sum_{i,\sigma}n_{i,\sigma}=n$, write
\begin{equation}
\begin{aligned}
    \ket{\pmb n}
    &\coloneqq
    \prod_{i,\sigma}
    (\widehat a_{i,\sigma}^{\dagger})^{n_{i,\sigma}}
    \ket{\mathrm{vac}},
    \\
    \mc V_{\pmb G}^{(n)}\ket{\pmb n}
    &=
    \prod_{i,\sigma}
    [\widehat a_{i,\sigma}^{\dagger}(\varphi_i)]^{n_{i,\sigma}}
    \ket{\mathrm{vac}}.
\end{aligned}
    \label{eq:fock-basis-state}
\end{equation}
Orthogonality of the spatial rails makes this assignment
isometric on every admissible input sector. If
$0\in\mathsf N_{\mathrm{src}}$, the vacuum has no
internal label and
$\mc V_{\pmb G}^{(0)}\ket{\mathrm{vac}}=\ket{\mathrm{vac}}$. Thus
$\mc V_{\pmb G}=\bigoplus_{n\in\mathsf N_{\mathrm{src}}}
\mc V_{\pmb G}^{(n)}$ is an isometry on $\mc H_{\mathrm{src}}$.

\section{Receiver-Map Realizations}
\label{app:receiver-map-realizations}

The single-photon direct-sum decomposition separates the receiver and
environment Fock spaces through the canonical bosonic factorization
\begin{equation}
\begin{aligned}
    \mc F_{\mathrm s}
    ((\mathfrak h_{\mathrm{rec}}\oplus\mathfrak h_{\mathrm E})
    \otimes\mathfrak h_{\mathrm{int}})
    \cong{}&
    \mc F_{\mathrm s}(\mathfrak h_{\mathrm{rec}}
    \otimes\mathfrak h_{\mathrm{int}})
    \\
    &\otimes
    \mc F_{\mathrm s}(\mathfrak h_{\mathrm E}
    \otimes\mathfrak h_{\mathrm{int}}).
\end{aligned}
    \label{eq:bosonic-factorization}
\end{equation}
Because the source has support only on photon numbers not exceeding
$n_{\max}$, tracing the second factor maps the propagated source support into
$\mc F_{\mathrm{rec}}$ in \eqref{eq:receiver-fock-space}.

The per-port polarization map in
\eqref{eq:local-polarization-depolarization} has the following Kraus
realization on all of $\mathrm L(\mc H_{\mathrm{rx},j})$. Let
$\sigma_x,\sigma_y,\sigma_z$ be the Pauli operators on the
$\mathrm H/\mathrm V$ block, and let $0_{\mathrm e}$ be the zero operator on
the erasure subspace:
\begin{equation}
\begin{aligned}
    \pmb K^{\mathrm{pol}}_{j,0}
    &=
    \sqrt{1-\tfrac{3q_j}{4}}\,\pmb\Pi_{\mathrm{pol}}
    +\pmb\Pi_{\mathrm e},
    \\
    \pmb K^{\mathrm{pol}}_{j,\alpha}
    &=
    \sqrt{\tfrac{q_j}{4}}\,\widetilde\sigma_\alpha,
    \qquad \alpha\in\{x,y,z\},
    \\
    \widetilde\sigma_\alpha
    &=\sigma_\alpha\oplus0_{\mathrm e}.
\end{aligned}
    \label{eq:polarization-kraus}
\end{equation}
The completeness relation
$\pmb K_{j,0}^{\mathrm{pol}\dagger}\pmb K_{j,0}^{\mathrm{pol}}
+\sum_{\alpha\in\{x,y,z\}}
\pmb K_{j,\alpha}^{\mathrm{pol}\dagger}
\pmb K_{j,\alpha}^{\mathrm{pol}}
=\pmb I_{\mc H_{\mathrm{rx},j}}$ proves that the map is
trace preserving. On the block-diagonal image of $\mc S_j$, this realization
reduces to \eqref{eq:local-polarization-depolarization}.

The mode-resolved decoder $\mc{D}_{\mathrm{mr},\pmb{q}}^{(\varpi)}$ of
Section~\ref{subsec:system-model-strategies} accepts a physical selection event
that differs from both heralded decoder success and the survival event in
\eqref{eq:cloning-survival-probability}. Let
$\pmb{n}=(n_{j,\mathrm{P},\gamma})$ index the receiver Fock occupation basis of
$\mc F_{\mathrm{rec}}$, where $j$ runs over receive ports,
$\mathrm{P}\in\{\mathrm H,\mathrm V\}$ over polarization, and $\gamma$ over the
calibrated orthonormal internal modes. A bin is a pair $(j,\gamma)$. The
resolvable set is
\begin{equation}
    \mathsf{R}
    \coloneqq
    \left\{
        \pmb{n}
        \;:\;
        \exists\,(j,\gamma),\;
        n_{j,\mathrm{H},\gamma}+n_{j,\mathrm{V},\gamma}=1
    \right\},
    \label{eq:resolvable-set}
\end{equation}
with projector
$\pmb\Pi_{\mathrm{mr}}\coloneqq\sum_{\pmb{n}\in\mathsf{R}}
\ketbra{\pmb{n}}{\pmb{n}}$. Writing
$\pmb\rho_{\mathrm{rec}}^{(M)}(\omega;s)$ for the receiver Fock state before
$\mc S$, the mode-resolved selection probability is
\begin{equation}
\resizebox{\columnwidth}{!}{$\displaystyle
    p_{\mathrm{sel}}^{(M)}(\omega;s)
    \coloneqq
    \Tr\!\left[
        \pmb\Pi_{\mathrm{mr}}
        \pmb\rho_{\mathrm{rec}}^{(M)}(\omega;s)
    \right],
    \qquad
    \overline p_{\mathrm{sel}}^{(M)}(s)
    \coloneqq
    \mathbb E_{\omega}\!\left[p_{\mathrm{sel}}^{(M)}(\omega;s)\right].
$}
    \label{eq:mode-resolved-selection-probability}
\end{equation}
The port priority $\varpi$ determines which occupied bin is read and leaves
$p_{\mathrm{sel}}^{(M)}$ unchanged. Because the squashing
receiver erases a port whose \emph{total} occupation differs from one, whereas
$\pmb\Pi_{\mathrm{mr}}$ tests each internal mode separately, every configuration
accepted by $\mc S$ is accepted here, giving
$p_{\mathrm{sel}}^{(M)}\geq p_{\mathrm{surv}}^{(M)}$. At $\zeta=0$ there is a
single internal mode, the bins coincide with the ports, and the two
probabilities are equal. Their gap is the probability of configurations rejected
by the squashing receiver but accepted by the mode-resolved receiver: no spatial
port is singly occupied, but at least one internal-mode bin is.

\section{Choi Convention}
\label{app:choi}

The encoder, propagation, and decoder maps use the input-first, unnormalized
Choi convention. For a channel
$\mc{N}:\mathrm{L}(\mc H_X)\rightarrow\mathrm{L}(\mc H_Y)$ with
$\mc H_X\cong\mathbb{C}^{D}$, choose an orthonormal input
basis $\{\ket{k}_X\}_{k=0}^{D-1}$ and an isomorphic reference system $X'$ with
$\mc H_{X'}\cong\mc H_X$ and corresponding basis
$\{\ket{k}_{X'}\}_{k=0}^{D-1}$. The maximally entangled state is
\begin{equation}
    \ket{\Phi^{+}}
    =
    \frac{1}{\sqrt{D}}
    \sum_{k=0}^{D-1}
    \ket{k}_{X'}\otimes\ket{k}_{X}.
    \label{eq:maximally-entangled-state}
\end{equation}
The Choi operator is
\begin{equation}
\begin{aligned}
    \pmb{J}^{\mc{N}}_{X\rightarrow Y}
    &\coloneqq
    D
    \left(
        \operatorname{id}_{\mc H_{X'}}\otimes\mc{N}
    \right)
    \left(
        \ketbra{\Phi^{+}}{\Phi^{+}}
    \right)
    \\
    &\in
    \mathrm{L}(\mc H_{X'}\otimes\mc H_Y).
\end{aligned}
    \label{eq:choi-definition}
\end{equation}
Although $\ket{\Phi^+}$ is normalized, the factor $D$ in
\eqref{eq:choi-definition} makes $\pmb{J}^{\mc{N}}_{X\rightarrow Y}$ unnormalized.
Identifying the reference basis with the input basis, the channel action is
recovered through
\begin{equation}
    \mc{N}(\pmb{\rho})
    =
    \Tr_{X'}
    \left[
        \left(
            \pmb{\rho}^{\mathsf{T}}_{X'}
            \otimes
            \pmb{I}_{\mc H_Y}
        \right)
        \pmb{J}^{\mc{N}}_{X\rightarrow Y}
    \right].
    \label{eq:choi-action}
\end{equation}
This convention expresses channel composition as tensor contractions and is
used for every encoder, propagation channel, and decoder in this manuscript.

\section{Haar-Averaged State Fidelity}
\label{app:haar-average}

The main fidelity metric follows by Haar averaging the input-output state
fidelity. For a pure logical input $\pmb{\psi}$, this fidelity is
$\Tr[\pmb{\psi}\Lambda(\pmb{\psi})]$; under the normalized measure
$\int\mathrm{d}\psi=1$, its average is
\begin{equation}
\begin{aligned}
    F_{\mathrm{av}}(\Lambda)
    &=
    \int
    \Tr\!\left[
        \pmb{\psi}\,
        \Lambda(\pmb{\psi})
    \right]
    \mathrm{d}\psi                                                     \\
    &=
    \Tr\!\left[
        \pmb{J}^{\Lambda}
        \int
        \pmb{\psi}^{\mathsf{T}}\otimes\pmb{\psi}\,
        \mathrm{d}\psi
    \right].
    \label{eq:haar-average-start}
\end{aligned}
\end{equation}
Let $\pmb I_{D^2}$ denote the identity on the two-system space of dimension
$D^2$. The Haar second moment is
\begin{equation}
    \int
    \pmb{\psi}^{\mathsf{T}}\otimes\pmb{\psi}\,
    \mathrm{d}\psi
    =
    \frac{
        \pmb{I}_{D^2}+D\pmb{\Phi}_D
    }{
        D(D+1)
    }.
    \label{eq:haar-second-moment}
\end{equation}
Substituting \eqref{eq:haar-second-moment} into
\eqref{eq:haar-average-start} gives
\begin{equation}
\begin{aligned}
    F_{\mathrm{av}}(\Lambda)
    &=
    \frac{
        \Tr\pmb{J}^{\Lambda}
        +
        D\bra{\Phi^{+}}
        \pmb{J}^{\Lambda}
        \ket{\Phi^{+}}
    }{
        D(D+1)
    }                                                        \\
    &=
    \frac{DF_{\mathrm{e}}(\Lambda)+1}{D+1}
    =
    \frac{2F_{\mathrm{e}}(\Lambda)+1}{3},
    \qquad D=2 .
    \label{eq:haar-average-derived}
\end{aligned}
\end{equation}
This recovers the main-text relation
\eqref{eq:average-entanglement-fidelity-relation}.

\section{Heralded Recovery}
\label{app:heralded}

A heralded decoder is specified by a trace-nonincreasing success branch
$\Lambda_{\mathrm{s}}$ with Choi operator $\pmb{J}^{\mathrm{s}}$. For input
$\pmb{\psi}$, its recovery success probability is
$p_{\mathrm{s}}(\pmb{\psi})
\coloneqq\Tr[\Lambda_{\mathrm{s}}(\pmb{\psi})]$.
The success probability averaged over all pure inputs is
\begin{equation}
    \overline{p}_{\mathrm{s}}
    \coloneqq
    \int p_{\mathrm{s}}(\pmb{\psi})\,\mathrm{d}\psi
    =
    \frac{\Tr\pmb{J}^{\mathrm{s}}}{D}.
    \label{eq:average-success-probability}
\end{equation}
The contribution of successful trials to the Haar-averaged state fidelity is
\begin{equation}
\begin{aligned}
    \overline{f}_{\mathrm{s}}
    &\coloneqq
    \int
    \Tr\!\left[
        \pmb{\psi}\,
        \Lambda_{\mathrm{s}}(\pmb{\psi})
    \right]
    \mathrm{d}\psi
    \\
    &=
    \frac{
        \Tr\pmb{J}^{\mathrm{s}}
        +
        D\bra{\Phi^{+}}\pmb{J}^{\mathrm{s}}\ket{\Phi^{+}}
    }{
        D(D+1)
    } .
    \label{eq:success-weighted-average-fidelity}
\end{aligned}
\end{equation}
For $\overline{p}_{\mathrm{s}}>0$, the success-conditioned fidelity is
\begin{equation}
    F_{\mathrm{av}}^{(\mathrm{s})}
    \coloneqq
    \frac{\overline{f}_{\mathrm{s}}}{\overline{p}_{\mathrm{s}}}
    =
    \frac{
        \Tr\pmb{J}^{\mathrm{s}}
        +
        D\bra{\Phi^{+}}\pmb{J}^{\mathrm{s}}\ket{\Phi^{+}}
    }{
        (D+1)\Tr\pmb{J}^{\mathrm{s}}
    }.
    \label{eq:accepted-ensemble-fidelity}
\end{equation}
Inputs with a larger value of $p_{\mathrm{s}}(\pmb{\psi})$ contribute more
strongly to the accepted ensemble. If the success probability is independent
of the input, \eqref{eq:accepted-ensemble-fidelity} equals the uniform Haar
average of the accepted-output fidelity.

To include failed trials, each failure is assigned the maximally mixed logical
state $\pmb\rho_{\mathrm{mix}}$. Let $\Lambda_{\mathrm{s}}^{\dagger}$ denote
the Hilbert--Schmidt adjoint of the success branch. The resulting
trace-preserving logical channel is
\begin{equation}
    \Lambda_{\mathrm{all}}(\pmb{\rho})
    =
    \Lambda_{\mathrm{s}}(\pmb{\rho})
    +
    \Tr\!\left[
        \left(
            \pmb{I}_{\mc H_{\mathrm L}}
            -
            \Lambda_{\mathrm{s}}^{\dagger}
            (\pmb{I}_{\mc H_{\mathrm L}})
        \right)
        \pmb{\rho}
    \right]\pmb\rho_{\mathrm{mix}}.
    \label{eq:abort-to-mixed-completion}
\end{equation}
Since
$\Tr[\pmb{\psi}\pmb\rho_{\mathrm{mix}}]=1/D$ for every pure logical input, the
unconditional average fidelity of this channel is
\begin{equation}
    F_{\mathrm{av}}(\Lambda_{\mathrm{all}})
    =
    \overline{f}_{\mathrm{s}}
    +
    \frac{1-\overline{p}_{\mathrm{s}}}{D}.
    \label{eq:unconditional-average-fidelity}
\end{equation}

We fix an encoder and a receiver information value
$\widehat{\mc I}_{\mathrm r}$. Let
$\mc N_{\mathrm{pre}}^{(\widehat{\mc I}_{\mathrm r})}:
\mathrm L(\mc H_{\mathrm L})
\rightarrow\mathrm L(\mc H_{\mathrm{rx}})$ be the conditional
channel
before logical recovery,
\begin{equation}
    \mc N_{\mathrm{pre}}^{(\widehat{\mc I}_{\mathrm r})}
    \coloneqq
    \mathbb E_{\omega,\widehat{\mc I}_{\mathrm t}\mid
    \widehat{\mc I}_{\mathrm r}}
    \left[
        \mc N_{\mathrm{FSO}}(\omega;\pmb G,\pmb\tau,\pmb q)
        \circ\mc E(\widehat{\mc I}_{\mathrm t})
    \right].
    \label{eq:conditional-pre-recovery-channel}
\end{equation}
Thus every channel variable unavailable to the receiver is averaged before
recovery. We define
$\pmb\rho_{\mathrm R}(\psi\mid\widehat{\mc I}_{\mathrm r})
\coloneqq\mc N_{\mathrm{pre}}^{(\widehat{\mc I}_{\mathrm r})}
(\ketbra{\psi}{\psi})$. The Haar-averaged correlation and normalization
operators are
\begin{equation}
\begin{aligned}
    \pmb C_{\mathrm{Haar}}
    &=
    \int
    \pmb\rho_{\mathrm R}(\psi\mid\widehat{\mc I}_{\mathrm r})
    \otimes\ketbra{\psi}{\psi}\,
    \mathrm{d}\psi,
    \\
    \pmb N_{\mathrm{Haar}}
    &=
    \int
    \pmb\rho_{\mathrm R}(\psi\mid\widehat{\mc I}_{\mathrm r})
    \otimes\pmb I_{\mc H_{\mathrm L}}\,
    \mathrm{d}\psi .
    \label{eq:purification-operators}
\end{aligned}
\end{equation}
They represent, respectively, the averaged correlation between the receiver
state and target and the linear functional giving the mean success probability.
Thus, if the Rx has realization information, the receiver state is conditioned
on its available estimate; otherwise it is averaged over the hidden channel
variables. The recovery is fixed for that conditional ensemble. After Haar
integration, both operators are independent of the transmitted state
~\cite{N:02:PL}.

A semidefinite program determines the success branch at each operating point. Let
$\pmb{J}^{\mc{D}_{\mathrm{s}}}$ be the Choi operator of the success branch from
$\mc H_{\mathrm{rx}}$ to $\mc H_{\mathrm L}$. For each fixed
$\widehat{\mc I}_{\mathrm r}$ and target success probability
$p_{\mathrm{tar}}\in(0,1]$, the recovery that maximizes the
success-conditioned fidelity is
\begin{equation}
\begin{aligned}
    \underset{\pmb{J}^{\mc{D}_{\mathrm{s}}}}{\operatorname{maximize}}
    \quad&
    \frac{1}{p_{\mathrm{tar}}}
    \Tr\!\left[
        \pmb{J}^{\mc{D}_{\mathrm{s}}}
        \pmb C_{\mathrm{Haar}}^{
        \mathsf{T}_{\mc H_{\mathrm{rx}}}}
    \right]
    \\
    \operatorname{subject\ to}
    \quad&
    \pmb{J}^{\mc{D}_{\mathrm{s}}}\succeq\pmb{0},
    \qquad
    \Tr\!\left[
        \pmb{J}^{\mc{D}_{\mathrm{s}}}
        \pmb N_{\mathrm{Haar}}^{
        \mathsf{T}_{\mc H_{\mathrm{rx}}}}
    \right]
    =
    p_{\mathrm{tar}},
    \\
    &
    \Tr_{\mc H_{\mathrm L}}\pmb{J}^{\mc{D}_{\mathrm{s}}}
    \preceq
    \pmb{I}_{\mc H_{\mathrm{rx}}}.
    \label{eq:purification-sdp}
\end{aligned}
\end{equation}
Replacing the equality of \eqref{eq:decoder-sdp} by
$\Tr_{\mc H_{\mathrm L}}\pmb{J}^{\mc{D}_{\mathrm{s}}}
\preceq\pmb{I}_{\mc H_{\mathrm{rx}}}$ admits a
completely positive, trace-nonincreasing map, and the linear constraint on
$p_{\mathrm{tar}}$
fixes the operating point in the tradeoff between success probability and
fidelity.

Lowering the target success probability permits stronger postselection, which
can raise the fidelity of the retained output above the deterministic value.
A heralded recovery can therefore attain a higher success-conditioned fidelity
than the deterministic decoder, whereas
\eqref{eq:unconditional-average-fidelity} accounts for rejected trials through
their fixed logical completion.
The states reaching the Rx are correlated through the shared channel rather
than independent noisy copies of a common target with the product structure
assumed in standard purification analyses~\cite{yao_protocols_2025}. The
encoders considered here produce correlated outputs, so
\eqref{eq:purification-sdp} is solved on the joint receiver space rather than
on a marginal.

\section{Multistart Bound for the Alternating Design}
\label{app:multistart}

Alternating optimization produces a nondecreasing objective sequence for each
initialization. Using the notation of
Section~\ref{subsubsec:iterative-cptp-optimization}, fix an initialization $u$
and let $\mc{D}_u^\star$ be the decoder obtained from
\eqref{eq:decoder-sdp} for the encoder $\mc{E}_u$. Each exact half-step
optimizes one map with the other held fixed over a feasible set containing the
current iterate and therefore cannot decrease the objective. Because fidelity
is bounded above, the sequence converges to a value at least
$F_{\mathrm{av}}(\mc{D}_u^\star\circ\mc{N}_{\mathrm{FSO}}\circ\mc{E}_u)$.
Let $F_u^{\mathrm{alt}}$ denote the terminal fidelity from initialization $u$
and define
$F_{\mathrm{best}}\coloneqq\max_{1\leq u\leq N_{\mathrm{init}}}
F_u^{\mathrm{alt}}$. Then
\begin{equation}
    F_{\mathrm{best}}
    \geq
    \max_{1\leq u\leq N_{\mathrm{init}}}
    F_{\mathrm{av}}
    \left(
        \mc{D}_u^\star
        \circ
        \mc{N}_{\mathrm{FSO}}
        \circ
        \mc{E}_u
    \right),
    \label{eq:multistart-cptp-bound}
\end{equation}
for exact half-steps. The attained numerical values satisfy this inequality
within solver tolerance.

Because $F_{\mathrm{best}}$ is attained by an encoder and decoder pair feasible
within solver tolerance, it is a tolerance-scoped lower bound on the globally
optimal fidelity. Equation~\eqref{eq:multistart-cptp-bound} shows that this
bound is at least as tight as decoder-only optimization from the best available
initialization. The argument does not certify local or global optimality because
the joint problem is biconvex rather than jointly
convex~\cite{RW:05:PRL_2,KL:09:QIP_2}. If exact half-steps reach a fixed point,
they establish only coordinatewise optimality.

\balance
\bibliographystyle{IEEEtran}
\bibliography{TQE-QMIMO}

@IEEEtranBSTCTL{ResearchBrain:BSTcontrol,
  CTLuse_forced_etal       = "yes",
  CTLmax_names_forced_etal = "5",
  CTLnames_show_etal       = "5"
}

@STRING{AO           = "Appl. Opt."}

@STRING{CP           = "Commun. Physics"}

@STRING{EPJ          = "EPJ Quantum Technol."}

@STRING{IJSAC        = "{IEEE} J. Select. Areas in Commun."}

@STRING{IN           = "{IEEE} Network"}

@STRING{JLT          = "J. of Lightwave Technol."}

@STRING{JMO          = "J. of Modern Opt."}

@STRING{JOSA         = "J. Opt. Soc. of America A"}

@STRING{LMP          = "Lett. in Math. Physics"}

@STRING{NC           = "Nat. Commun."}

@STRING{NJP          = "New J. Phys."}

@STRING{NQI          = "npj Quantum Inform."}

@STRING{Nat          = "Nature"}

@STRING{PLA          = "Phys. Lett. A"}

@STRING{PQ           = "PRX Quantum"}

@STRING{PRA          = "Phys. Rev. A"}

@STRING{PRL          = "Phys. Rev. Lett."}

@STRING{PRR          = "Physical Rev. Res."}

@STRING{QIC          = "Quantum Inform. and Computation"}

@STRING{QIP          = "Quantum Inf. Process."}

@STRING{QST          = "Quantum Sci. Technol."}

@STRING{Qu           = "Quantum"}

@STRING{IEEE_J_COML  = "{IEEE} Commun. Lett."}

@STRING{IEEE_J_IT    = "{IEEE} Trans. Inf. Theory"}

@article{ANV:08:AO,
  author    = {Jaime A. Anguita and Mark A. Neifeld and Bane V. Vasic},
  title     = {{Turbulence-induced channel crosstalk in an orbital angular momentum-multiplexed free-space optical link}},
  journal   = AO,
  volume    = {47},
  number    = {13},
  pages     = {2414},
  publisher = {Optica Publishing Group},
  month     = apr,
  year      = {2008},
  doi       = {10.1364/ao.47.002414},
  url       = {https://doi.org/10.1364/ao.47.002414}
}

@article{BDS:97:PRL_2,
  author    = {Charles H. Bennett and David P. DiVincenzo and John A. Smolin},
  title     = {{Capacities of Quantum Erasure Channels}},
  journal   = PRL,
  volume    = {78},
  number    = {16},
  pages     = {3217-3220},
  publisher = {American Physical Society (APS)},
  month     = apr,
  year      = {1997},
  doi       = {10.1103/physrevlett.78.3217},
  url       = {http://dx.doi.org/10.1103/physrevlett.78.3217}
}

@article{BGL:20:JOSA,
  author    = {Liliana Borcea and Josselin Garnier and Knut Sølna},
  title     = {{Multimode communication through the turbulent atmosphere}},
  journal   = JOSA,
  volume    = {37},
  number    = {5},
  pages     = {720},
  publisher = {Optica Publishing Group},
  year      = {2020},
  doi       = {10.1364/JOSAA.384007},
  url       = {https://doi.org/10.1364/josaa.384007}
}

@article{BMT:08:PRL_2,
  author    = {Normand J. Beaudry and Tobias Moroder and Norbert Lütkenhaus},
  title     = {{Squashing Models for Optical Measurements in Quantum Communication}},
  journal   = PRL,
  volume    = {101},
  number    = {9},
  pages     = {093601},
  publisher = {American Physical Society},
  month     = aug,
  year      = {2008},
  doi       = {10.1103/PhysRevLett.101.093601},
  url       = {https://link.aps.org/doi/10.1103/PhysRevLett.101.093601}
}

@article{C:00:JMO,
  author    = {Nicolas J. Cerf},
  title     = {{Asymmetric quantum cloning in any dimension}},
  journal   = JMO,
  volume    = {47},
  number    = {2-3},
  pages     = {187-209},
  publisher = {Informa UK Limited},
  month     = feb,
  year      = {2000},
  doi       = {10.1080/09500340008244036},
  url       = {http://dx.doi.org/10.1080/09500340008244036}
}

@article{CC:97:PRA,
  author    = {Nicolas J. Cerf and Richard Cleve},
  title     = {{Information-theoretic interpretation of quantum error-correcting codes}},
  journal   = PRA,
  volume    = {56},
  number    = {3},
  pages     = {1721--1732},
  publisher = {American Physical Society},
  month     = sep,
  year      = {1997},
  doi       = {10.1103/PhysRevA.56.1721},
  url       = {https://link.aps.org/doi/10.1103/PhysRevA.56.1721}
}

@article{FSW:07:PR,
  author    = {Andrew S. Fletcher and Peter W. Shor and Moe Z. Win},
  title     = {{Optimum quantum error recovery using semidefinite programming}},
  journal   = PRA,
  volume    = {75},
  number    = {1},
  pages     = {012338},
  publisher = {American Physical Society},
  month     = jan,
  year      = {2007},
  doi       = {10.1103/PhysRevA.75.012338},
  url       = {https://link.aps.org/doi/10.1103/PhysRevA.75.012338}
}

@article{GA:06:JLT,
  author    = {M. Gabay and S. Arnon},
  title     = {{Quantum key distribution by a free-space MIMO system}},
  journal   = JLT,
  volume    = {24},
  number    = {8},
  pages     = {3114-3120},
  publisher = {Institute of Electrical and Electronics Engineers (IEEE)},
  month     = aug,
  year      = {2006},
  doi       = {10.1109/jlt.2006.878043},
  url       = {http://dx.doi.org/10.1109/jlt.2006.878043}
}

@article{GBP:97:PR,
  author    = {M. Grassl and Th. Beth and T. Pellizzari},
  title     = {{Codes for the quantum erasure channel}},
  journal   = PRA,
  volume    = {56},
  number    = {1},
  pages     = {33-38},
  publisher = {American Physical Society (APS)},
  month     = jul,
  year      = {1997},
  doi       = {10.1103/physreva.56.33},
  url       = {http://dx.doi.org/10.1103/physreva.56.33}
}

@article{HOM:87:PRL_2,
  author    = {C. K. Hong and Z. Y. Ou and L. Mandel},
  title     = {{Measurement of subpicosecond time intervals between two photons by interference}},
  journal   = PRL,
  volume    = {59},
  number    = {18},
  pages     = {2044-2046},
  publisher = {American Physical Society},
  month     = nov,
  year      = {1987},
  doi       = {10.1103/PhysRevLett.59.2044},
  url       = {https://link.aps.org/doi/10.1103/PhysRevLett.59.2044}
}

@article{LWHRK:04:PRL,
  author    = {Thomas Legero and Tatjana Wilk and Markus Hennrich and Gerhard Rempe and Axel Kuhn},
  title     = {{Quantum Beat of Two Single Photons}},
  journal   = PRL,
  volume    = {93},
  number    = {7},
  pages     = {070503},
  publisher = {American Physical Society},
  month     = aug,
  year      = {2004},
  doi       = {10.1103/PhysRevLett.93.070503},
  url       = {https://link.aps.org/doi/10.1103/PhysRevLett.93.070503}
}

@article{SGSREBS:23:PQ,
  author    = {Laura Serino and Jano Gil-Lopez and Michael Stefszky and Raimund Ricken and Christof Eigner and Benjamin Brecht and Christine Silberhorn},
  title     = {{Realization of a Multi-Output Quantum Pulse Gate for Decoding High-Dimensional Temporal Modes of Single-Photon States}},
  journal   = PQ,
  volume    = {4},
  number    = {2},
  pages     = {020306},
  publisher = {American Physical Society},
  month     = apr,
  year      = {2023},
  doi       = {10.1103/PRXQuantum.4.020306},
  url       = {https://link.aps.org/doi/10.1103/PRXQuantum.4.020306}
}

@article{Scheel:08:APS,
  author        = {Stefan Scheel},
  title         = {{Permanents in linear optical networks}},
  journal       = {Acta Physica Slovaca},
  volume        = {58},
  pages         = {675--680},
  year          = {2008},
  eprint        = {quant-ph/0406127},
  archivePrefix = {arXiv},
  url           = {https://arxiv.org/abs/quant-ph/0406127}
}

@article{ISS:11:PRA,
  author    = {J. Solomon Ivan and Krishna Kumar Sabapathy and R. Simon},
  title     = {{Operator-sum representation for bosonic Gaussian channels}},
  journal   = PRA,
  volume    = {84},
  number    = {4},
  publisher = {American Physical Society (APS)},
  month     = oct,
  year      = {2011},
  doi       = {10.1103/PhysRevA.84.042311},
  url       = {https://doi.org/10.1103/physreva.84.042311}
}

@article{KS:23:PRA,
  author    = {M. Klen and A. A. Semenov},
  title     = {{Numerical simulations of atmospheric quantum channels}},
  journal   = PRA,
  volume    = {108},
  number    = {3},
  pages     = {033718},
  publisher = {American Physical Society},
  month     = sep,
  year      = {2023},
  doi       = {10.1103/PhysRevA.108.033718},
  url       = {https://link.aps.org/doi/10.1103/PhysRevA.108.033718}
}

@article{LMPZ:96:PRL,
  author    = {Raymond Laflamme and Cesar Miquel and Juan Pablo Paz and Wojciech Hubert Zurek},
  title     = {{Perfect Quantum Error Correcting Code}},
  journal   = PRL,
  volume    = {77},
  number    = {1},
  pages     = {198-201},
  publisher = {American Physical Society},
  month     = jul,
  year      = {1996},
  doi       = {10.1103/PhysRevLett.77.198},
  url       = {https://link.aps.org/doi/10.1103/PhysRevLett.77.198}
}

@article{N:02:PL,
  author    = {Michael A Nielsen},
  title     = {{A simple formula for the average gate fidelity of a quantum dynamical operation}},
  journal   = PLA,
  volume    = {303},
  number    = {4},
  pages     = {249-252},
  publisher = {Elsevier BV},
  month     = oct,
  year      = {2002},
  doi       = {10.1016/s0375-9601(02)01272-0},
  url       = {https://doi.org/10.1016/s0375-9601(02)01272-0}
}

@book{NC:10:NoVenue,
  author    = {Michael A. Nielsen and Isaac L. Chuang},
  title     = {{Quantum computation and quantum information}},
  publisher = {Cambridge University Press},
  address   = {Cambridge ; New York},
  edition   = {10th anniversary ed},
  year      = {2010}
}

@misc{NPR:23:LMP,
  author    = {Ion Nechita and Clément Pellegrini and Denis Rochette},
  title     = {{The Asymmetric Quantum Cloning Region}},
  journal   = LMP,
  volume    = {113},
  number    = {3},
  pages     = {74},
  month     = jun,
  year      = {2023},
  doi       = {10.1007/s11005-023-01694-8},
  url       = {http://arxiv.org/abs/2209.11999}
}

@article{RW:05:PRL_2,
  author    = {M. Reimpell and R. F. Werner},
  title     = {{Iterative Optimization of Quantum Error Correcting Codes}},
  journal   = PRL,
  volume    = {94},
  number    = {8},
  publisher = {American Physical Society (APS)},
  month     = mar,
  year      = {2005},
  doi       = {10.1103/physrevlett.94.080501},
  url       = {https://doi.org/10.1103/physrevlett.94.080501}
}

@article{S:15:PRA,
  author    = {V. S. Shchesnovich},
  title     = {{Partial indistinguishability theory for multiphoton experiments in multiport devices}},
  journal   = PRA,
  volume    = {91},
  number    = {1},
  publisher = {American Physical Society (APS)},
  year      = {2015},
  doi       = {10.1103/PhysRevA.91.013844},
  url       = {https://doi.org/10.1103/physreva.91.013844}
}

@article{T:15:PRA,
  author    = {Malte C. Tichy},
  title     = {{Sampling of partially distinguishable bosons and the relation to the multidimensional permanent}},
  journal   = PRA,
  volume    = {91},
  number    = {2},
  publisher = {American Physical Society (APS)},
  year      = {2015},
  doi       = {10.1103/PhysRevA.91.022316},
  url       = {https://doi.org/10.1103/physreva.91.022316}
}

@article{Peng2026_3AU6T4M9,
  author    = {Heyang Peng and Seid Koudia and Symeon Chatzinotas},
  title     = {{Security Analysis of MDI-QKD in Turbulent Free-Space Polarization Channels---A Composite Channel Framework}},
  journal   = IJSAC,
  volume    = {44},
  pages     = {5190--5205},
  year      = {2026},
  doi       = {10.1109/JSAC.2026.3706034},
  url       = {https://ieeexplore.ieee.org/document/11574653/}
}

@article{Koudia2025_JJ3WBISQ,
  author    = {Seid Koudia and Symeon Chatzinotas},
  title     = {{Crosstalk-resilient quantum MIMO for scalable quantum communications}},
  journal   = NQI,
  volume    = {11},
  number    = {1},
  pages     = {162},
  publisher = {Springer Science and Business Media LLC},
  year      = {2025},
  doi       = {10.1038/s41534-025-01113-x},
  url       = {https://doi.org/10.1038/s41534-025-01113-x}
}

@misc{Koudia2024_ACUR6M78,
  author    = {Seid Koudia and Leonardo Oleynik and Mert Bayraktar and Junaid ur Rehman and Symeon Chatzinotas},
  title     = {{Physical Layer Aspects of Quantum Communications: A Survey}},
  month     = jul,
  year      = {2024},
  doi       = {10.48550/arXiv.2407.09244},
  url       = {https://arxiv.org/abs/2407.09244}
}

@article{TCFKST:17:NPh,
  author  = {Hideki Takenaka and Alberto Carrasco-Casado and Mikio Fujiwara and Mitsuo Kitamura and Masahide Sasaki and Morio Toyoshima},
  title   = {{Satellite-to-ground quantum-limited communication using a 50-kg-class microsatellite}},
  journal = {Nature Photonics},
  volume  = {11},
  pages   = {502--508},
  year    = {2017},
  doi     = {10.1038/nphoton.2017.107},
  url     = {https://doi.org/10.1038/nphoton.2017.107}
}

@article{SBMPO:23:CP,
  author    = {Jasminder S. Sidhu and Thomas Brougham and Duncan McArthur and Roberto G. Pousa and Daniel K. L. Oi},
  title     = {{Finite key performance of satellite quantum key distribution under practical constraints}},
  journal   = CP,
  volume    = {6},
  number    = {1},
  pages     = {210},
  publisher = {Springer Nature},
  month     = aug,
  year      = {2023},
  doi       = {10.1038/s42005-023-01299-6},
  url       = {https://doi.org/10.1038/s42005-023-01299-6}
}

@article{WZMAGR:24:NJP,
  author    = {Matthew S. Winnel and Ziqing Wang and Robert Malaney and Ryan Aguinaldo and Jonathan Green and Timothy C. Ralph},
  title     = {{Classical-quantum dual encoding for laser communications in space}},
  journal   = NJP,
  volume    = {26},
  number    = {3},
  pages     = {033012},
  publisher = {IOP Publishing},
  month     = mar,
  year      = {2024},
  doi       = {10.1088/1367-2630/ad295a},
  url       = {https://doi.org/10.1088/1367-2630/ad295a}
}

@article{NMRFHFW:13:NPh,
  author  = {Sebastian Nauerth and Florian Moll and Markus Rau and Christian Fuchs and Joachim Horwath and Stefan Frick and Harald Weinfurter},
  title   = {{Air-to-ground quantum communication}},
  journal = {Nature Photonics},
  volume  = {7},
  pages   = {382--386},
  month   = may,
  year    = {2013},
  doi     = {10.1038/nphoton.2013.46},
  url     = {https://doi.org/10.1038/nphoton.2013.46}
}

@article{Tariq2026_878BK5NE,
  author    = {Shehbaz Tariq and Symeon Chatzinotas},
  title     = {{Design and optimization of adaptive diversity schemes in quantum MIMO channels}},
  journal   = NQI,
  publisher = {Springer Science and Business Media LLC},
  month     = jun,
  year      = {2026},
  doi       = {10.1038/s41534-026-01308-w},
  url       = {https://doi.org/10.1038/s41534-026-01308-w}
}

@article{urRehman2025_39RDN6DM,
  author    = {Junaid ur Rehman and Leonardo Oleynik and Seid Koudia and Mert Bayraktar and Symeon Chatzinotas},
  title     = {{Diversity and multiplexing in quantum MIMO channels}},
  journal   = EPJ,
  volume    = {12},
  number    = {1},
  pages     = {18},
  month     = feb,
  year      = {2025},
  doi       = {10.1140/epjqt/s40507-025-00324-7},
  url       = {https://doi.org/10.1140/epjqt/s40507-025-00324-7}
}

@article{Rehman2025_HGQ2Z98S,
  author    = {Junaid ur Rehman and Syed Muhammad Abuzar Rizvi and Seid Koudia and Symeon Chatzinotas and Hyundong Shin},
  title     = {{MIMO Quantum Communication in Correlated Pure-Loss Channels}},
  journal   = IEEE_J_COML,
  volume    = {29},
  number    = {7},
  pages     = {1544--1548},
  month     = jul,
  year      = {2025},
  doi       = {10.1109/LCOMM.2025.3567246},
  url       = {https://ieeexplore.ieee.org/document/10988590}
}

@article{VSV:16:PRL,
  author    = {D. Vasylyev and A. A. Semenov and W. Vogel},
  title     = {{Atmospheric Quantum Channels with Weak and Strong Turbulence}},
  journal   = PRL,
  volume    = {117},
  number    = {9},
  pages     = {090501},
  publisher = {American Physical Society},
  month     = aug,
  year      = {2016},
  doi       = {10.1103/PhysRevLett.117.090501},
  url       = {https://link.aps.org/doi/10.1103/PhysRevLett.117.090501}
}

@article{WZ:82:Nat,
  author    = {W. K. Wootters and W. H. Zurek},
  title     = {{A single quantum cannot be cloned}},
  journal   = Nat,
  volume    = {299},
  number    = {5886},
  pages     = {802--803},
  month     = oct,
  year      = {1982},
  doi       = {10.1038/299802a0},
  url       = {https://doi.org/10.1038/299802a0}
}

@article{A:02:IEEE_J_JSAC,
  author  = {Siavash M. Alamouti},
  title   = {{A Simple Transmit Diversity Technique for Wireless Communications}},
  journal = {IEEE J. Sel. Areas Commun.},
  volume  = {16},
  number  = {8},
  pages   = {1451--1458},
  year    = {1998},
  doi     = {10.1109/49.730453},
  url     = {https://doi.org/10.1109/49.730453}
}

@article{PLOB:17:NC,
  author  = {Stefano Pirandola and Riccardo Laurenza and Carlo Ottaviani and Leonardo Banchi},
  title   = {{Fundamental Limits of Repeaterless Quantum Communications}},
  journal = {Nature Communications},
  volume  = {8},
  pages   = {15043},
  year    = {2017},
  doi     = {10.1038/ncomms15043},
  url     = {https://doi.org/10.1038/ncomms15043}
}

@article{WEH:18:Sci,
  author  = {Stephanie Wehner and David Elkouss and Ronald Hanson},
  title   = {{Quantum Internet: A Vision for the Road Ahead}},
  journal = {Science},
  volume  = {362},
  number  = {6412},
  pages   = {eaam9288},
  year    = {2018},
  doi     = {10.1126/science.aam9288},
  url     = {https://doi.org/10.1126/science.aam9288}
}

@article{GM:97:PRL,
  author    = {Nicolas Gisin and Serge Massar},
  title     = {{Optimal Quantum Cloning Machines}},
  journal   = PRL,
  volume    = {79},
  number    = {11},
  pages     = {2153--2156},
  month     = sep,
  year      = {1997},
  doi       = {10.1103/PhysRevLett.79.2153},
  url       = {https://doi.org/10.1103/PhysRevLett.79.2153}
}

@article{Wer:98:PRA,
  author    = {R. F. Werner},
  title     = {{Optimal cloning of pure states}},
  journal   = PRA,
  volume    = {58},
  number    = {3},
  pages     = {1827--1832},
  month     = sep,
  year      = {1998},
  doi       = {10.1103/PhysRevA.58.1827},
  url       = {https://doi.org/10.1103/PhysRevA.58.1827}
}

@article{kay_optimal_2016,
  author    = {Alastair Kay},
  title     = {{Optimal universal quantum cloning: asymmetries and fidelity measures}},
  journal   = QIC,
  volume    = {16},
  number    = {11--12},
  pages     = {991--1028},
  month     = sep,
  year      = {2016},
  doi       = {10.26421/QIC16.11-12-5},
  url       = {https://doi.org/10.26421/QIC16.11-12-5}
}

@article{abbott_communication_2020,
  author    = {Alastair A. Abbott and Julian Wechs and Dominic Horsman and
               Mehdi Mhalla and Cyril Branciard},
  title     = {{Communication through coherent control of quantum channels}},
  journal   = Qu,
  volume    = {4},
  pages     = {333},
  month     = sep,
  year      = {2020},
  doi       = {10.22331/q-2020-09-24-333},
  url       = {https://doi.org/10.22331/q-2020-09-24-333}
}

@article{rubino_experimental_2021,
  author    = {Giulia Rubino and Lee A. Rozema and Daniel Ebler and
               Hl{\'e}r Kristj{\'a}nsson and Sina Salek and
               Philippe Allard Gu{\'e}rin and Alastair A. Abbott and
               Cyril Branciard and {\v{C}}aslav Brukner and
               Giulio Chiribella and Philip Walther},
  title     = {{Experimental quantum communication enhancement by superposing
               trajectories}},
  journal   = PRR,
  volume    = {3},
  number    = {1},
  pages     = {013093},
  month     = jan,
  year      = {2021},
  doi       = {10.1103/PhysRevResearch.3.013093},
  url       = {https://doi.org/10.1103/PhysRevResearch.3.013093}
}

@article{CEM:99:PRL,
  author    = {J. I. Cirac and A. K. Ekert and C. Macchiavello},
  title     = {{Optimal Purification of Single Qubits}},
  journal   = PRL,
  volume    = {82},
  number    = {21},
  pages     = {4344--4347},
  month     = may,
  year      = {1999},
  doi       = {10.1103/PhysRevLett.82.4344},
  url       = {https://doi.org/10.1103/PhysRevLett.82.4344}
}

@article{yao_protocols_2025,
  author    = {Hongshun Yao and Yu-Ao Chen and Erdong Huang and Kaichu Chen and
               Honghao Fu and Xin Wang},
  title     = {{Protocols and trade-offs of quantum state purification}},
  journal   = QST,
  volume    = {10},
  number    = {3},
  pages     = {035020},
  month     = oct,
  year      = {2025},
  doi       = {10.1088/2058-9565/add17e},
  url       = {https://doi.org/10.1088/2058-9565/add17e}
}

@article{KO:26:NQI,
  author    = {Kao-Yueh Kuo and Yingkai Ouyang},
  title     = {{Degenerate quantum erasure decoding}},
  journal   = NQI,
  volume    = {12},
  number    = {1},
  pages     = {75},
  publisher = {Nature Publishing Group},
  month     = mar,
  year      = {2026},
  doi       = {10.1038/s41534-026-01212-3},
  url       = {https://doi.org/10.1038/s41534-026-01212-3}
}

@article{KL:97:PR,
  author    = {Emanuel Knill and Raymond Laflamme},
  title     = {{Theory of quantum error-correcting codes}},
  journal   = PRA,
  volume    = {55},
  number    = {2},
  pages     = {900--911},
  publisher = {American Physical Society},
  month     = feb,
  year      = {1997},
  doi       = {10.1103/PhysRevA.55.900},
  url       = {https://doi.org/10.1103/PhysRevA.55.900}
}

@article{KL:09:QIP_2,
  author    = {Robert L. Kosut and Daniel A. Lidar},
  title     = {{Quantum error correction via convex optimization}},
  journal   = QIP,
  volume    = {8},
  number    = {5},
  pages     = {443--459},
  month     = oct,
  year      = {2009},
  doi       = {10.1007/s11128-009-0120-2},
  url       = {https://doi.org/10.1007/s11128-009-0120-2}
}

@article{FSW:08:PRA,
  author    = {Andrew S. Fletcher and Peter W. Shor and Moe Z. Win},
  title     = {{Structured near-optimal channel-adapted quantum error correction}},
  journal   = PRA,
  volume    = {77},
  number    = {1},
  pages     = {012320},
  publisher = {American Physical Society},
  month     = jan,
  year      = {2008},
  doi       = {10.1103/PhysRevA.77.012320},
  url       = {https://doi.org/10.1103/PhysRevA.77.012320}
}

@article{TKL:10:IEEE_J_IT,
  author    = {Soraya Taghavi and Robert L. Kosut and Daniel A. Lidar},
  title     = {{Channel-Optimized Quantum Error Correction}},
  journal   = IEEE_J_IT,
  volume    = {56},
  number    = {3},
  pages     = {1461--1473},
  month     = mar,
  year      = {2010},
  doi       = {10.1109/TIT.2009.2039162},
  url       = {https://doi.org/10.1109/TIT.2009.2039162}
}

@article{L:50:JCP,
  author  = {L{\"o}wdin, Per-Olov},
  title   = {On the Non-Orthogonality Problem Connected with the Use of Atomic
             Wave Functions in the Theory of Molecules and Crystals},
  journal = {The Journal of Chemical Physics},
  volume  = {18},
  number  = {3},
  pages   = {365--375},
  year    = {1950},
  doi     = {10.1063/1.1747632}
}

@article{CB:02:JMC,
  author  = {Carlson, B. C. and Keller, Joseph M.},
  title   = {Orthogonalization Procedures and the Localization of Wannier
             Functions},
  journal = {Physical Review},
  volume  = {105},
  number  = {1},
  pages   = {102--103},
  year    = {1957},
  doi     = {10.1103/PhysRev.105.102}
}

@article{MRL:08:PRA,
  author    = {M. Mohseni and A. T. Rezakhani and D. A. Lidar},
  title     = {{Quantum-process tomography: Resource analysis of different strategies}},
  journal   = PRA,
  volume    = {77},
  number    = {3},
  pages     = {032322},
  publisher = {American Physical Society},
  month     = mar,
  year      = {2008},
  doi       = {10.1103/PhysRevA.77.032322},
  url       = {https://doi.org/10.1103/PhysRevA.77.032322}
}

@article{BNWK:09:NJP,
  author    = {M. P. A. Branderhorst and J. Nunn and I. A. Walmsley and R. L. Kosut},
  title     = {{Simplified quantum process tomography}},
  journal   = NJP,
  volume    = {11},
  number    = {11},
  pages     = {115010},
  publisher = {IOP Publishing},
  month     = nov,
  year      = {2009},
  doi       = {10.1088/1367-2630/11/11/115010},
  url       = {https://doi.org/10.1088/1367-2630/11/11/115010}
}

@article{GRRWK:20:NC,
  author    = {L. C. G. Govia and G. J. Ribeill and D. Rist{\`e} and M. Ware and H. Krovi},
  title     = {{Bootstrapping quantum process tomography via a perturbative ansatz}},
  journal   = NC,
  volume    = {11},
  number    = {1},
  pages     = {1084},
  publisher = {Springer Nature},
  month     = feb,
  year      = {2020},
  doi       = {10.1038/s41467-020-14873-1},
  url       = {https://doi.org/10.1038/s41467-020-14873-1}
}

@article{TWACCEtAl:23:NC,
  author    = {Giacomo Torlai and Christopher J. Wood and Atithi Acharya and
               Giuseppe Carleo and Juan Carrasquilla and Leandro Aolita},
  title     = {{Quantum process tomography with unsupervised learning and tensor networks}},
  journal   = NC,
  volume    = {14},
  number    = {1},
  pages     = {2858},
  publisher = {Springer Nature},
  month     = may,
  year      = {2023},
  doi       = {10.1038/s41467-023-38332-9},
  url       = {https://doi.org/10.1038/s41467-023-38332-9}
}

@article{CK:19:PRSA,
  author  = {Chiribella, Giulio and Kristj{\'a}nsson, Hl{\'e}r},
  title   = {Quantum {S}hannon theory with superpositions of trajectories},
  journal = {Proceedings of the Royal Society A},
  volume  = {475},
  number  = {2225},
  pages   = {20180903},
  year    = {2019},
  doi     = {10.1098/rspa.2018.0903}
}

@article{KCSEW:20:NJP,
  author  = {Kristj{\'a}nsson, Hl{\'e}r and Chiribella, Giulio and Salek, Sina
             and Ebler, Daniel and Wilson, Matthew},
  title   = {Resource theories of communication},
  journal = {New Journal of Physics},
  volume  = {22},
  number  = {7},
  pages   = {073014},
  year    = {2020},
  doi     = {10.1088/1367-2630/ab8ef7}
}

@article{TRS:18:PRX,
  author    = {Nora Tischler and Carsten Rockstuhl and Karolina S{\l}owik},
  title     = {{Quantum Optical Realization of Arbitrary Linear Transformations Allowing for Loss and Gain}},
  journal   = {Physical Review X},
  volume    = {8},
  number    = {2},
  pages     = {021017},
  publisher = {American Physical Society},
  year      = {2018},
  doi       = {10.1103/PhysRevX.8.021017},
  url       = {https://doi.org/10.1103/PhysRevX.8.021017}
}

@article{RZBB:94:PRL,
  author    = {Michael Reck and Anton Zeilinger and Herbert J. Bernstein and Philip Bertani},
  title     = {{Experimental Realization of Any Discrete Unitary Operator}},
  journal   = PRL,
  volume    = {73},
  number    = {1},
  pages     = {58--61},
  publisher = {American Physical Society},
  year      = {1994},
  doi       = {10.1103/PhysRevLett.73.58},
  url       = {https://doi.org/10.1103/PhysRevLett.73.58}
}

@article{CHMKW:16:Optica,
  author    = {William R. Clements and Peter C. Humphreys and Benjamin J. Metcalf and W. Steven Kolthammer and Ian A. Walmsley},
  title     = {{Optimal Design for Universal Multiport Interferometers}},
  journal   = {Optica},
  volume    = {3},
  number    = {12},
  pages     = {1460--1465},
  publisher = {Optica Publishing Group},
  year      = {2016},
  doi       = {10.1364/OPTICA.3.001460},
  url       = {https://doi.org/10.1364/OPTICA.3.001460}
}

@article{E:96:PRL,
  author    = {Berthold-Georg Englert},
  title     = {{Fringe Visibility and Which-Way Information: An Inequality}},
  journal   = PRL,
  volume    = {77},
  number    = {11},
  pages     = {2154--2157},
  publisher = {American Physical Society},
  year      = {1996},
  doi       = {10.1103/PhysRevLett.77.2154},
  url       = {https://doi.org/10.1103/PhysRevLett.77.2154}
}

@article{KL:66:AO,
  author    = {H. Kogelnik and T. Li},
  title     = {{Laser Beams and Resonators}},
  journal   = {Applied Optics},
  volume    = {5},
  number    = {10},
  pages     = {1550--1567},
  year      = {1966},
  doi       = {10.1364/AO.5.001550},
  url       = {https://doi.org/10.1364/AO.5.001550}
}

@article{PLCFBG:10:PRL,
  author    = {S. M. Popoff and G. Lerosey and R. Carminati and M. Fink and A. C. Boccara and S. Gigan},
  title     = {{Measuring the Transmission Matrix in Optics: An Approach to the Study and Control of Light Propagation in Disordered Media}},
  journal   = PRL,
  volume    = {104},
  number    = {10},
  pages     = {100601},
  year      = {2010},
  doi       = {10.1103/PhysRevLett.104.100601},
  url       = {https://doi.org/10.1103/PhysRevLett.104.100601}
}

@article{RBFFRW:13:OE,
  author    = {Saleh Rahimi-Keshari and Matthew A. Broome and Robert Fickler and Alessandro Fedrizzi and Timothy C. Ralph and Andrew G. White},
  title     = {{Direct Characterization of Linear-Optical Networks}},
  journal   = {Optics Express},
  volume    = {21},
  number    = {11},
  pages     = {13450--13458},
  year      = {2013},
  doi       = {10.1364/OE.21.013450},
  url       = {https://doi.org/10.1364/OE.21.013450}
}

@article{MS:14:PRA,
  author    = {Iman Marvian and Robert W. Spekkens},
  title     = {{Modes of Asymmetry: The Application of Harmonic Analysis to Symmetric Quantum Dynamics and Quantum Reference Frames}},
  journal   = PRA,
  volume    = {90},
  number    = {6},
  pages     = {062110},
  year      = {2014},
  doi       = {10.1103/PhysRevA.90.062110},
  url       = {https://doi.org/10.1103/PhysRevA.90.062110}
}

@article{PGN:04:PoIEEE,
  author = {PAULRAJ, A.J. and GORE, D.A. and NABAR, R.U. and BOLCSKEI, H.},
  doi = {10.1109/JPROC.2003.821915},
  journal = {Proceedings of the IEEE},
  month = feb,
  number = {2},
  pages = {198--218},
  title = {An overview of {MIMO} communications - a key to gigabit wireless},
  volume = {92},
  year = {2004}
}

@article{C:20:JOSA,
  author  = {Mikhail Charnotskii},
  title   = {{Comparison of Four Techniques for Turbulent Phase Screens Simulation}},
  journal = JOSA,
  volume  = {37},
  number  = {5},
  pages   = {738--747},
  year    = {2020},
  doi     = {10.1364/JOSAA.385754},
  url     = {https://doi.org/10.1364/JOSAA.385754}
}

@inproceedings{PHRDKB:18:SPIE,
  author    = {Emiel H. Por and Sebastiaan Y. Haffert and Vikram M. Radhakrishnan and David S. Doelman and Maaike van Kooten and Steven P. Bos},
  title     = {{High Contrast Imaging for Python ({HCIPy}): An Open-Source Adaptive Optics and Coronagraph Simulator}},
  booktitle = {Adaptive Optics Systems VI},
  series    = {Proc. SPIE},
  volume    = {10703},
  year      = {2018},
  doi       = {10.1117/12.2314407},
  url       = {https://doi.org/10.1117/12.2314407}
}

\begin{IEEEbiography}[{\includegraphics[width=1in,height=1.25in,clip,keepaspectratio]{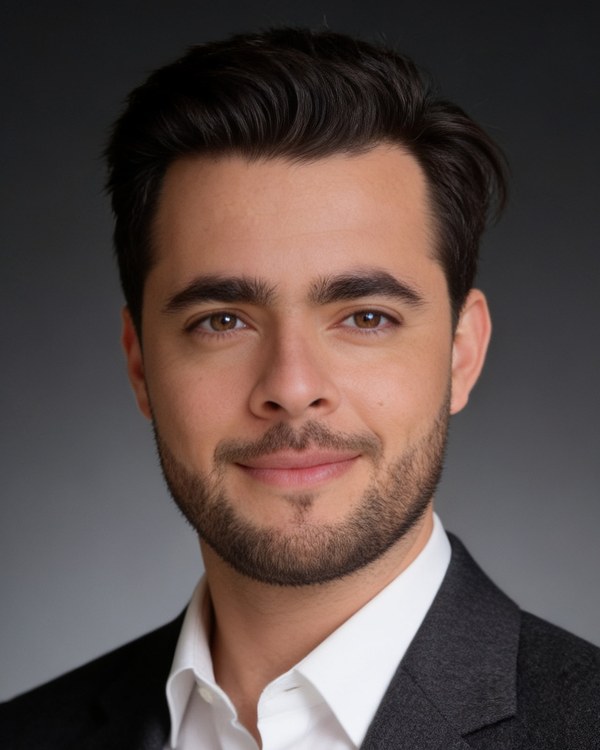}}]{Shehbaz Tariq}
received the B.S. degree in Electrical Engineering from the University of
Engineering and Technology, Peshawar, Pakistan, in 2020, and the Ph.D. degree
in Electronics and Information Convergence Engineering from Kyung Hee
University, Seoul, South Korea, in 2025. Since 2025, he has been a
Postdoctoral Fellow with the Interdisciplinary Centre for Security, Reliability
and Trust (SnT), University of Luxembourg, Luxembourg. His research interests
include quantum machine learning, quantum information theory, and quantum
communications.
\end{IEEEbiography}
\vfill

\begin{IEEEbiography}[{\includegraphics[width=1in,height=1.25in,clip,keepaspectratio]{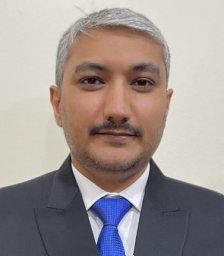}}]{Junaid ur Rehman}
received the B.S. degree in Electrical Engineering from National University of
Sciences and Technology (NUST), Islamabad, Pakistan, in 2013, and the Ph.D.
degree in Electronic Engineering from Kyung Hee University (KHU), Yongin-si,
Korea, in 2019. He worked as a researcher (Postdoctoral Fellow/Research
Associate/Research Scientist) with various research groups including groups at
KHU, Korea Institute of Science and Technology (KIST), and Signal Processing
and Communications (SIGCOM) group of the University of Luxembourg. Currently,
he is an assistant professor at the Department of Electrical Engineering at
King Fahd University of Petroleum and Minerals, Saudi Arabia. His research
interests include quantum information science, particularly quantum
communications, quantum computing, and quantum sensing. Dr. ur Rehman has
served as a technical program committee member of multiple IEEE conferences
including IEEE International Conference on Communications (ICC) and IEEE
Global Communications Conference (Globecom). He has reviewed for numerous IEEE
and non-IEEE conferences and journals. He was an Exemplary Reviewer for the
IEEE Wireless Communications Letters in 2022.
\end{IEEEbiography}
\vfill

\begin{IEEEbiography}[{\includegraphics[width=1in,height=1.25in,clip,keepaspectratio]{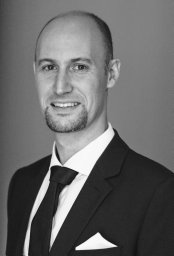}}]{Symeon Chatzinotas}
(S'06, M'09, SM'13, F'23) is Full Professor and Head of the SIGCOM Research
Group at SnT, University of Luxembourg. He is coordinating the research
activities on communications and networking across a group of 80 researchers,
acting as a PI for more than 40 projects and main representative for 3GPP,
ETSI, DVB. He is currently serving in the editorial board of the IEEE
Transactions on Communications, IEEE Open Journal of Vehicular Technology and
the International Journal of Satellite Communications and Networking. In the
past, he has been a Visiting Professor at the University of Parma, Italy and
was involved in numerous R\&D projects for NCSR Demokritos, CERTH Hellas and
CCSR, University of Surrey. He was the co-recipient of the 2014 IEEE
Distinguished Contributions to Satellite Communications Award and Best Paper
Awards at WCNC, 5GWF, EURASIP JWCN, CROWNCOM, ICSSC. He has (co-)authored more
than 700 technical papers in refereed international journals, conferences and
scientific books.
\end{IEEEbiography}
\vfill

\end{document}